\documentclass[twocolumn]{aastex63}
\pdfoutput=1 
\usepackage{amsmath,amstext}
\usepackage[T1]{fontenc}
\usepackage{apjfonts} 
\usepackage[figure,figure*]{hypcap}
\usepackage{upgreek}
\usepackage{float}
\usepackage{rotating}
\usepackage{boldline}
\usepackage{listings}
\usepackage{hyperref}

\submitjournal{ApJ}
\received{November 2, 2018}
\accepted{September 27, 2019}

\shorttitle{TRAPPIST-1}
\shortauthors{Gonzales et al.}
\begin{document}

\title{A Reanalysis of the Fundamental Parameters and Age of TRAPPIST-1\footnote{This paper includes data gathered with the 6.5 meter Magellan Telescopes located at Las Campanas Observatory, Chile.}}

\author[0000-0003-4636-6676]{Eileen C. Gonzales}
\altaffiliation{LSSTC Data Science Fellow}
\affiliation{Department of Astrophysics, American Museum of Natural History, New York, NY 10024, USA}
\affiliation{The Graduate Center, City University of New York, New York, NY 10016, USA}
\affiliation{Department of Physics and Astronomy, Hunter College, City University of New York, New York, NY 10065, USA}

\author[0000-0001-6251-0573]{Jacqueline K. Faherty}
\affiliation{Department of Astrophysics, American Museum of Natural History, New York, NY 10024, USA}

\author[0000-0002-2592-9612]{Jonathan Gagn\'e}
\affiliation{Department of Terrestrial Magnetism, Carnegie Institution of Washington, Washington, DC 20015, USA}
\affil{Institute for Research on Exoplanets, Universit\'e de Montr\'eal, D\'epartement de Physique, C.P.~6128 Succ. Centre-ville, Montr\'eal, QC H3C~3J7, Canada}

\author{Johanna Teske}
\altaffiliation{Hubble Fellow}
\affiliation{Observatories of the Carnegie Institution for Science, 813 Santa Barbara St., Pasadena, CA 91101}
\affiliation{Department of Terrestrial Magnetism, Carnegie Institution of Washington, Washington, DC 20015, USA}

\author{Andrew McWilliam}
\affiliation{Observatories of the Carnegie Institution for Science, 813 Santa Barbara St., Pasadena, CA 91101}

\author[0000-0002-1821-0650]{Kelle Cruz}
\affiliation{Department of Astrophysics, American Museum of Natural History, New York, NY 10024, USA}
\affiliation{The Graduate Center, City University of New York, New York, NY 10016, USA}
\affiliation{Department of Physics and Astronomy, Hunter College, City University of New York, New York, NY 10065, USA}

\correspondingauthor{Eileen Gonzales}
\email{egonzales@amnh.org}

\begin{abstract}  
We present the distance-calibrated spectral energy distribution (SED) of TRAPPIST-1 using a new medium resolution (R$\sim$6000) near-infrared FIRE spectrum and its \textit{Gaia} parallax. We report an updated bolometric luminosity ($L_\mathrm{bol}$) of $-3.216 \pm 0.016$, along with semi-empirical fundamental parameters: effective temperature $T_\mathrm{eff}=2628 \pm 42$~K, mass=$90 \pm 8~M_\mathrm{Jup}$, radius=$1.16 \pm 0.03~R_\mathrm{Jup}$, and log\, $g$=$5.21 \pm 0.06$ dex. It's kinematics point toward an older age while spectral indices indicate youth therefore, we compare the overall SED and near-infrared bands of TRAPPIST-1 to field-age, low-gravity, and low-metallicity dwarfs of similar $T_\mathrm{eff}$ and $L_\mathrm{bol}$. We find field dwarfs of similar $T_\mathrm{eff}$ and $L_\mathrm{bol}$ best fit the overall and band-by-band features of TRAPPIST-1. Additionally, we present new \cite{Alle13} spectral indices for the SpeX SXD and FIRE spectra of TRAPPIST-1, both classifying it as intermediate gravity. Examining $T_\mathrm{eff}$, $L_\mathrm{bol}$, and absolute $JHKW1W2$ magnitudes versus optical spectral type places TRAPPIST-1 in an ambiguous location containing both field- and intermediate-gravity sources. Kinematics place TRAPPIST-1 within a subpopulation of intermediate-gravity sources lacking bonafide membership in a moving group with higher tangential and $UVW$ velocities. We conclude that TRAPPIST-1 is a field-age source with subtle spectral features reminiscent of a low surface gravity object. To resolve the cause of TRAPPIST-1's intermediate gravity indicators we speculate two avenues which might be correlated to inflate the radius: (1) magnetic activity or (2) tidal interactions from planets. We find the M8 dwarf LHS 132 is an excellent match to TRAPPIST-1's spectral peculiarities along with the M9\,$\beta$ dwarf 2MASS J10220489$+$0200477, the L1\,$\beta$ 2MASS J10224821$+$5825453, and the L0\,$\beta$ 2MASS J23224684$-$3133231 which have distinct kinematics making all three intriguing targets for future exoplanet studies.
\end{abstract}

\keywords{stars: individual (2MASS J23062928$-$0502285) --- stars: fundamental parameters --- stars: low mass --- brown dwarfs}

\section{Introduction}
The majority of stars in our galaxy are low mass, M dwarfs being the most numerous with the longest main sequence lifetime \citep{Boch10}. Their low mass and abundance in the solar neighborhood make M dwarfs favorable targets for exoplanet observations. Their small radii enable easier detection of Earth-sized planets using transit and radial velocity methods, therefore they are prime targets when searching for rocky planets within a star's habitable zone.

With numerous searches for exoplanets- such as \textit{Kepler} (aimed at detecting planets around Sun-like stars \citealt{Boru10}) and \textit{TESS} (an all sky survey searching for planets smaller then Neptune around nearby stars, \citealt{Rick15})- understanding stellar properties of M dwarfs as exoplanet host stars is extremely pertinent to understanding planet habitability. \cite{Kane16} found 40\% of all \textit{Kepler} planet candidates (1) with radii less than 2 $R_\oplus$ and (2) lying within an optimistically-sized habitable zone, orbit stars cooler than 4000 K. This is despite cool dwarfs being less than 5\% of initial \textit{Kepler} targets \citep{Bata10}. \cite{Muld15c} and \cite{Gaid16} determined planets with radii of 1--3 $R_\oplus$ occur 2--4 times higher around M dwarfs than FGK stars. Furthermore, \cite{Dres15} estimate the frequency of habitable Earths around M dwarfs to be $2.5\pm0.2$ planets. Such objects would have radii of $1-4$~$R_\oplus$ and periods shorter than 200 days. They calculate an occurrence rate of $0.56\substack{+0.06 \\ -0.05}$ Earth-sized planets with periods shorter than 50 days and $0.46\substack{+0.07 \\ -0.05}$ super-Earths ($1-1.5\,R_\oplus$) with periods shorter than 50 days per early-type M dwarf. \cite{Ball18} predict that \textit{TESS} will detect $900 \pm 350$ planets around $715\pm255$ M dwarfs spectral typed M1V--M4V. 

Four nearby mid- to late-type M dwarfs with habitable zone planets are: 2MASS J23062928$-$0502285 (hereafter TRAPPIST-1), Proxima Centauri, LHS 1140, and Teegarden's Star. TRAPPIST-1, a M7.5 dwarf at a distance of $\sim12.4$ parsecs, hosts a system of seven rocky Earth-sized exoplanets \citep{Gill16,Gill17} with four lying in the habitable zone. Proxima Centauri, a moderately active M5.5 dwarf, is our nearest stellar neighbor 1.295 parsecs away. It hosts an earth-sized planet ($1.3\,M_\oplus$) that could have liquid water on its surface \citep{Angl16}. LHS 1140, a metal-poor M4.5 dwarf older than 5 Gyr and about 15 parsecs away, hosts a super-Earth ($\sim7M_\oplus$) and an Earth-sized planet ($\sim2M_\oplus$) with an Earth-like composition \citep{Ditt17,Ment18}. Teegarden's Star, an old magnetically quiet M6.5 dwarf located 3.8 parsecs away, was classified as intermediate gravity by \cite{Gagn15b} and hosts two Earth-sized planets both within the conservative habitable zone \citep{Zech19}. Having the most precise stellar parameters for these systems is critical for understanding planet habitability because stellar size and temperature drive the habitable zone boundaries.

While most M dwarfs are stars, late-type M dwarfs can be either stars or brown dwarfs depending on age. Brown dwarfs are low-mass, low-temperature objects unable to sustain stable hydrogen burning in their cores and thus cool throughout their lifetime. With masses $<75\ M_\mathrm{Jup}$, brown dwarfs lie between the boundary of low mass stars and planets \citep{Saum96,Chab97}. They typically fall into three main age subpopulations: field dwarfs, low surface gravity dwarfs, and subdwarfs. Field dwarfs anchor the brown dwarf spectral classification system, while the low-gravity and subdwarfs show differences in their observed spectra deviating them from the field classification. Red infrared colors, enhanced metal oxide in the optical, and weak alkali lines differentiate low-gravity dwarfs from the field sources \citep{Kirk06, Kirk10, Cruz09, Alle10}. High-likelihood or candidate membership in young nearby moving groups has been seen for many low-gravity sources \citep{Liu_13,Liu_16, Fahe16, Kell16,Schn16,Gagn17}. Low-gravity dwarfs are further classified into two gravity groups- (1) very low gravity, designated by $\gamma$ in the optical or by VL-G in the infrared, and (2) intermediate gravity, designated by $\beta$ in the optical or INT-G in the infrared. Subdwarfs are low-luminosity metal-poor sources that exhibit blue near-infrared colors \citep{Burg03c, Burg09a}, kinematics consistent with halo-membership \citep{Burg08a, Dahn08, Cush09}, enhanced metal-hydride absorption bands along with weak or absent metal oxides, and enhanced collision-induced H$_2$ absorption (\citealt{Burg03c} and references therein).

In this paper we examine the fundamental parameters of TRAPPIST-1 to determine if it is a typical field M-dwarf host star or if it is more akin to low gravity or subdwarf equivalents. Previously published data on TRAPPIST-1 is presented in Section~\ref{PubTrap}. New FIRE and SpeX SXD spectra observations are discussed in Section~\ref{Obs}. Section~\ref{motivation} presents the motivation of our chosen samples and analysis process. Section~\ref{FundParamTrap} discusses how we derive the fundamental parameters of TRAPPIST-1 using its distanced-calibrated SED and the \cite{Fili15} method. Section~\ref{CompTrap} discusses comparative samples for TRAPPIST-1. Sections~\ref{FieldAgeAssumption} and \ref{YoungAgeAssumption} examine the full SED of TRAPPIST-1 and the $Y$, $J$, $H$ and $K$ near-infrared (NIR) spectra of comparative objects. Section~\ref{subdwarfNaIKIcomps} examines whether an extreme subsolar metallicity for TRAPPIST-1 might explain anomalous spectral features. Section~\ref{FinalThoughts} makes concluding remarks on the age of TRAPPIST-1 after examining all samples. Section~\ref{LHS132} examines LHS 132 and other sources from \cite{Burg17} comparing their overall SEDs to TRAPPIST-1. Lastly, Section~\ref{Discussion} presents \cite{Alle13} indices and gravity classifications for the entire comparative sample, absolute magnitude and fundamental parameters versus spectral type comparisons, and a comparison of the kinematics of TRAPPIST-1 to $\beta$-gravity sources. We also present possible reasons for the gravity classification that TRAPPIST-1 receives.

\section{Published Data on TRAPPIST-1}\label{PubTrap}

\subsection{The discovery of a M-dwarf}
TRAPPIST-1 was discovered by \cite{Gizi00} as part of a search using optical and near-infrared sky survey data from the Second Palomar Sky Survey (POSS-II) and the Two Micron All Sky Survey (2MASS). \cite{Gizi00} spectral typed TRAPPIST-1 as an M7.5 based on its optical spectrum taken by the RC spectrograph at Kitt Peak on the 4 m telescope. Additional optical spectra of TRAPPIST-1 are presented in \cite{Cruz07}, \cite{Schm07}, \cite{Rein09} and \cite{Burg15}. NIR spectra are presented in \cite{Tann12}, \cite{Bard14}, \cite{Cruz18}, and this paper (FIRE). Thus, there are currently 5 optical and 4 NIR spectra of TRAPPIST-1.

\textit{Gaia} DR2 (\citealt{GaiaDR1,GaiaDR2,Lind18}) provides the most precise proper motion and parallax values for TRAPPIST-1 although it was also observed by numerous other surveys (\citealt{Cost06}, \citealt{Schm07}, \citealt{Wein16}, and \citealt{Boss17}, \citealt{vanG18}). Radial velocity measurements of TRAPPIST-1 were reported in \cite{Rein09}, \cite{Tann12} (using the NIRSPEC spectrum), and \cite{Burg15} (using the MagE spectrum). In this paper we present updated $UVW$ velocities using updated position and parallax values from \textit{Gaia} DR2 paired with the \cite{Tann12} radial velocity.

The equivalent width of the 6563\;\AA\, H$\alpha$ line, an indicator of activity for M-dwarfs, has been measured in many papers \citep{Gizi00,Schm07,Rein10,Barn14,Burg15}. The width measurements vary between $\sim2.3-7.7$\;\AA. Ratios of log($L_\mathrm{\alpha}/L_\mathrm{bol}$) in \cite{Gizi00}, \cite{Schm07} and \cite{Rein10} indicate that TRAPPIST-1 is a moderately active M dwarf. 

Bolometric luminosity, effective temperature, radius, mass, gravity and age for TRAPPIST-1 were initially determined in \cite{Fili15} via distance-scaled SED fitting. After the discovery of TRAPPIST-1 as an exoplanet host star with seven rocky planets, \cite{Gill17} presented fundamental parameters from a Markov chain Monte Carlo (MCMC) analysis with \textit{a priori} knowledge of stellar properties from \cite{Fili15}. Using their improved parallax measurement, \cite{vanG18} determined a bolometric luminosity almost two times as precise as the \cite{Fili15} value and used that to determine more accurate values of $T_\mathrm{eff}$, radius, and mass for TRAPPIST-1. Values for published data from the literature are listed in Table \ref{tab:Trapdata}.

\begin{deluxetable}{l c c }
\tabletypesize{\small}
\tablecaption{Properties of TRAPPIST-1 \label{tab:Trapdata}}
\tablehead{\colhead{Property} & \colhead{Value} & \colhead{Reference}} 
  \startdata
  R.A. & $23^h 06^m 29.36^s$ & 1 \\ 
  Decl. & $-05 ^\circ 02' 29''.2$ & 1 \\
  R.A. (epoch 2015.0) & $346.63 \pm 0.11$ & 2,3 \\ 
  Decl.(epoch 2015.0) & $-5.043 \pm 0.093$ & 2,3 \\
  Spectral type & M7.5 & 4\\
  $\pi$ (mas) & $80.45 \pm 0.12$  & 2,3 \\ 
  $Gaia$ BP (mag) & $18.998 \pm 0.048$ & 2,5\\
  $Gaia$ RP (mag) & $14.1 \pm 0.01$ & 2,5\\
  PS $g$ (mag) & $19.35 \pm 0.02$ & 6\\
  PS $r$ (mag) & $17.87 \pm 0.01$ & 6\\
  PS $i$ (mag) & $15.13 \pm 0.01$ & 6\\
  PS $z$ (mag) & $13.73 \pm 0.01$ & 6\\
  PS $y$ (mag) & $12.97 \pm 0.01$ & 6\\
  2MASS $J$ (mag) & $11.354 \pm 0.022$ & 1\\
  2MASS $H$ (mag) & $10.718 \pm 0.021$ & 1\\
  2MASS $K_{s}$ (mag) & $10.296 \pm 0.023$ & 1\\
  WISE $W1$ (mag) & $10.042 \pm 0.023$ & 7\\
  WISE $W2$ (mag) & $9.80 \pm 0.02$ & 7\\
  WISE $W3$ (mag) & $9.528 \pm 0.041$ & 7\\
  WISE $W4$ (mag) & $<8.397$ & 7\\
  $\mu_\alpha$ (mas yr$^{-1}$) & $930.88\pm0.25$ & 2,3\\                  
  $\mu_\delta$ (mas yr$^{-1}$) & $-479.40\pm0.17$ & 2,3 \\ 
  $V_{r}$ (km s$^{-1}$) & $-52.8 \pm 0.16$ & 8 \\ 
  $V_\mathrm{tan}$ (km s$^{-1}$)\tablenotemark{a} & $61.69 \pm 0.10$ & 9\\   
  $U$ (km s$^{-1}$)\tablenotemark{a} & $-44.1\pm0.1$ & 9\\ 
  $V$ (km s$^{-1}$)\tablenotemark{a} & $-67.2\pm0.3$ & 9\\  
  $W$ (km s$^{-1}$)\tablenotemark{a} & $11.7\pm0.4$ & 9\\ 
  $X$ (pc) & $2.369 \pm 0.004$ & 9\\ 
  $Y$ (pc) & $6.41\pm0.01$ & 9\\
  $Z$ (pc) & $-10.38\pm0.02$ & 9\\
  $L_\mathrm{bol}$ log($L_*/L_{\odot}$) & $-3.216 \pm 0.016$ & 9\\ 
  $T_\mathrm{eff}$ (K) & $2628 \pm 42$ & 9\\
  Radius ($R_\mathrm{Jup}$) & $1.16 \pm 0.03$ & 9\\ 
  Mass ($M_\mathrm{Jup}$ )& $90 \pm 8$ & 9\\
  log $g$ (dex)\tablenotemark{b} & $5.21 \pm 0.06$ & 9\\                                                 
  Age (Gyr) & $0.5-10$ & 9\\
  Distance (pc) & $12.43 \pm 0.02$ & 9\\
  $[\mathrm{Fe/H}]$ (dex) & $+0.04 \pm 0.08$ & 10\\ 
  H$\alpha$ EW (\AA) & $2.34-4.17$ & 11\\ 
  log($L_\mathrm{H\alpha}/L_\mathrm{bol}$) & min:$-4.85$, max:$-4.60$ & 11\\
  \enddata
\tablenotetext{a}{Not in LSR frame.}
\tablenotetext{b}{We account for gravitational reddening assuming 0.5 km s$^{-1}$.}
\tablerefs{(1) \cite{Cutr03}, (2) \cite{GaiaDR1,GaiaDR2}, (3) \cite{Lind18}, (4) \cite{Gizi00}, (5) \cite{Riel18, Evan18} , (6) \cite{Cham16}, (7) \cite{Cutr12}, (8) \cite{Tann12}, (9) This paper, (10) \cite{Gill16}, (11) \cite{Barn14}}
\end{deluxetable}

\subsection{Discovery as an exoplanet host star}
TRAPPIST (the TRansiting Planet and PlanetIsimals Small Telescope) monitored TRAPPIST-1 in the near-infrared (NIR) from mid September 2015 to the end of December 2015. Follow up photometry in the optical on the Himalayan Chandra 2-meter Telescope and in the NIR with the Very Large Telescope and UK Infrared Telescope helped to confirm signatures of three of exoplanets: TRAPPIST-1b, TRAPPIST-1c, and TRAPPIST-1d \citep{Gill16}. TRAPPIST-1d was later identified by follow up observations to be the signature of four planets: TRAPPIST-1d, TRAPPIST-1e, TRAPPIST-1f, and TRAPPIST-1g, along with the discovery of TRAPPIST-1h \citep{Gill17}. All seven of the planets have Earth-sized radii, ranging from 0.755 to 1.127 $R_\mathrm{Earth}$, with four of the planets TRAPPIST-1d,e,f, and g, lying in the habitable zone \citep{Gill17}. These planets are in a zone where temperatures are cool enough to potentially have long-lived liquid water present on the surface.

\section{Observations}\label{Obs}
\subsection{FIRE}

We used the Folded-port InfraRed Echellette (FIRE; \citealt{Simcoe13}) spectrograph on the 6.5m Baade \textit{Magellan} telescope to obtain near-infrared spectra of TRAPPIST-1. Observations were taken on 2017 July 28 using echellette mode and the 0.$\arcsec$45 slit, high gain (1.2 e-/DN), and sample-up-the-ramp (SUTR) readout covering the full 0.8 - 2.5 $\micron$ band. Each exposure was 600s long with ABBA nodding, and are bracketed by quartz lamp, ThAr lamp, and telluric standards. Airmass ranged from 1.7-1.0 and seeing from 0.7-1.1 over the night. The night sky emission lines were significantly larger than the internal Th-Ar comparison lines, even though the system was in best focus; this suggests that FIRE was not properly collimated at the time of observation. The result is that the resolving power we obtained was $\lambda$/$\Delta \lambda \sim$ 6000, not the R$\sim$8,000 that should have been obtained. This causes lines in our data to be shallower but wider with the equivalent widths preserved. All FIRE exposures of TRAPPIST-1 obtained on 2017 UT July 27 outside of transit were combined and reduced with the Interactive Data Language (IDL) FireHose v2 package\footnote{Available at \url{https://github.com/jgagneastro/FireHose_v2/}} \citep{Boch09,zenodofirehose}, as described in \cite{Gagn15b}.

\subsection{SpeX}
The following new SpeX spectra were obtained for objects we use in a comparative analysis to TRAPPIST-1.

\subsubsection{SXD}
A SpeX SXD spectrum was obtained for 2MASS J06085283$-$2753583 on 2007 November 13 across the $0.7-2.55\,\upmu$m region. The $0.\arcsec5$ slit was used providing a resolving power of $\lambda$/$\Delta \lambda \sim1200$. We obtained a two 200s exposures and four 300s exposures for a total integration time of 20 minutes using ABBA nodding. The slit was aligned to the parallactic angle and we observed at an airmass of~1.49. The spectrum was telluric corrected and flux calibrated using the spectrum of the A0~V standard HD~52487 taken at a similar airmass. Internal flat-field and Ar arc lamps exposures were taken for pixel response and wavelength calibration. The data was then reduced using standard procedures and the SpeXtool Package \cite{Cush04}.

\section{Is TRAPPIST-1 Young?}\label{motivation}
\cite{Burg17} noted TRAPPIST-1 exhibited weaker FeH absorption and a more triangular $H$ band, features that are typically associated with youth (see \citealt{Fahe16} for details). Using the \cite{Alle13} surface gravity indices, they determined the low-resolution SpeX spectrum of TRAPPIST-1 displayed signs of an intermediate gravity object, suggesting a young age. However, from examination of the kinematics and the lack of enhanced VO absorption in the SpeX prism spectrum, \cite{Burg17} concluded that the $\beta$ gravity classification may have arisen from other physical factors and thus is unrelated to youth. Indeed, all previous studies \citep{Gizi00,Fili15,Fahe09} indicated that TRAPPIST-1 was a field object. 

The low gravity indicators in the spectrum of TRAPPIST-1 motivate the remainder of the paper where we create comparison subsamples to examine whether youth or some other physical factor drive the observed parameters. Throughout our comparisons we examine whether the overall SED shape or specific features in the new FIRE medium-resolution NIR spectrum of TRAPPIST-1 show signs of low surface gravity. We also determine gravity indices for all objects in our sample as another aspect of our comparison to TRAPPIST-1, which is discussed in depth in Section \ref{AL13Indicescomparison}.

\section{Fundamental Parameters of TRAPPIST-1}\label{FundParamTrap}
\begin{figure*}
  \hspace{-0.25cm}
   \includegraphics[scale=.7]{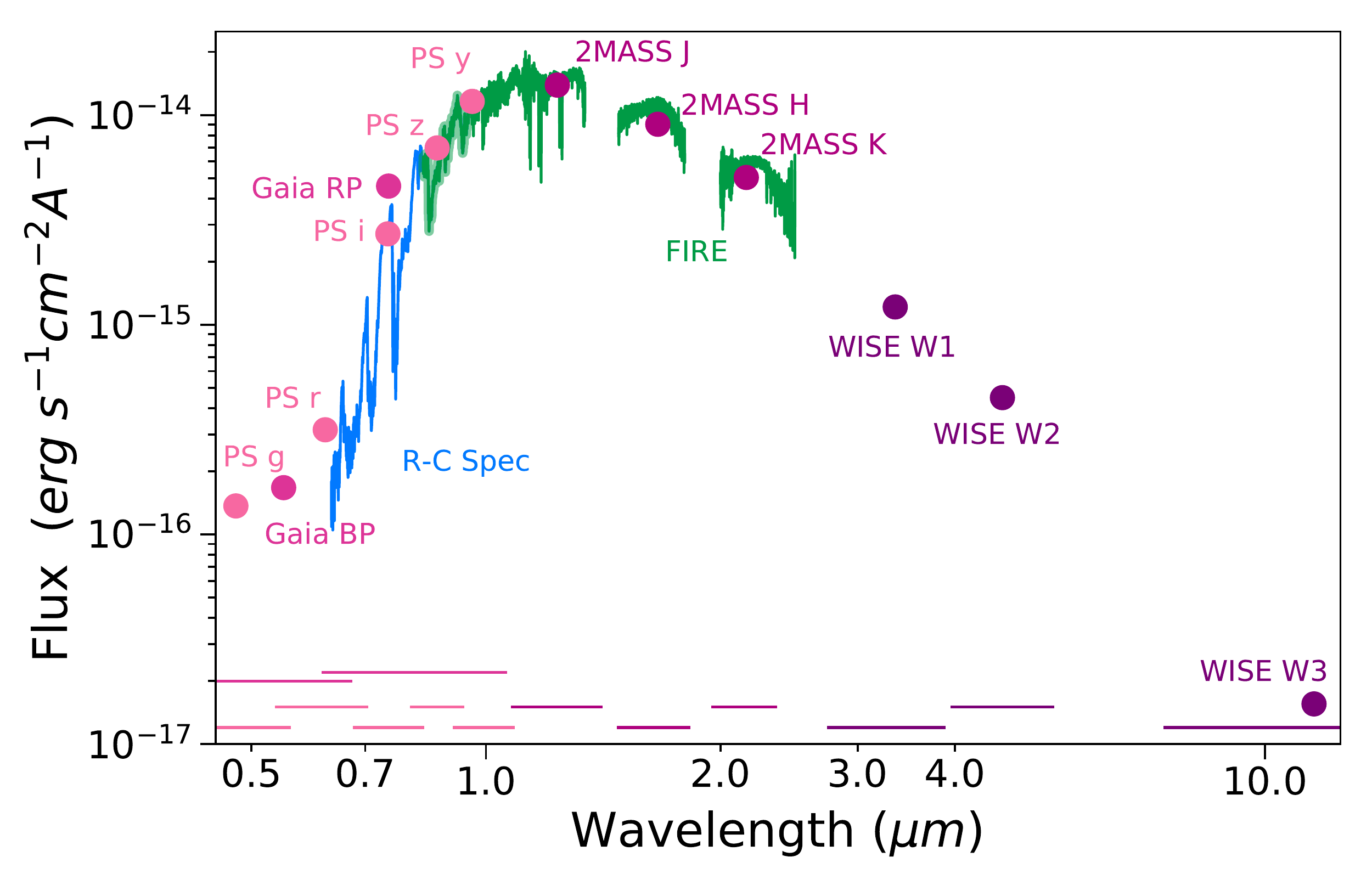}
\caption{Distance-calibrated SED of TRAPPIST-1. The spectra (optical in blue, NIR in green) and photometry (shades of pink and purple) are labeled by instrument or filter system. The horizontal lines at the bottom show the wavelength coverage for the corresponding photometric measurement. Error bars on the photometric points are smaller than the point size. The overlapping region for the optical and NIR is the fuzzy blue-green portion of the SED. Observation references can be found in Tables \ref{tab:Trapdata} and \ref{tab:SpectraReferences}.}
\label{fig:Regimes}
\vspace{0.5cm} 
\end{figure*}

Fundamental parameters for TRAPPIST-1 were determined from its distance-calibrated SED using the technique of \cite{Fili15}. Parameter values were determined using SEDkit\footnote{SEDkit is available on GitHub at \url{https://github.com/hover2pi/SEDkit}}, which requires spectra, photometry, and a parallax to create the distance-scaled SED and to determine the bolometric luminosity. The spectra, photometry, and parallax used in the generation of the SED of TRAPPIST-1 can be found in Tables \ref{tab:Trapdata} and \ref{tab:opticalphot}--\ref{tab:SpectraReferences}.

Using the optical and NIR spectra, we first construct a composite spectrum of TRAPPIST-1. The overlapping region from $0.8305-0.95\, \upmu$m (shown as a fuzzy blue-green line in Figure \ref{fig:Regimes}) was combined as an average. The composite spectrum is then scaled to the absolute magnitudes in each filter. We do not create synthetic magnitudes, those determined through empirical relations in the absence of photometric data as in \cite{Fili15}, instead we only use observed photometric data. The SED of TRAPPIST-1 is shown in Figure \ref{fig:Regimes}, with the various components labeled. 

The bolometric luminosity ($L_\mathrm{bol}$) was determined by integrating under the distance-calibrated SED from 0--1000~$\upmu$m, using a distance of $12.43\pm0.02$ pc. To obtain a radius estimate and the effective temperature ($T_\mathrm{eff}$), we used the DMEstar models \citep{Feid12} to extend the brown dwarf \cite{Saum08} hybrid cloud evolutionary models into the low-mass stellar range. They were connected via a cubic spline interpolation in the regions with no evolutionary model coverage. The same was done using the \citealt{Chab00} evolutionary models. For all three models we use an age range of $0.5-10$ Gyr, corresponding to the field ultracool dwarf age range \citep{Fili15}. The final radius range was set as the maximum and minimum from all model predictions as done in \citealt{Fili15}. The effective temperature was calculated using the inferred radius along with the bolometric luminosity using the Stefan-Boltzmann law. The uncertainty on the $T_\mathrm{eff}$ comes primarily from the uncertainty in the age of the system (leading to the range of radii) and the SED flux measurement. However, as noted in \cite{Dupu13} slight differences in radii do not have a large effect on the calculated $T_\mathrm{eff}$. The range of masses was determined using the \citealt{Saum08} and \citealt{Chab00} evolutionary models.

Using this approach, we derived the following parameters assuming a field age: $L_\mathrm{bol} = -3.216\pm0.016$, $T_\mathrm{eff} = 2628\pm42$\;K, $R = 1.16\pm0.03$\;$R_\mathrm{Jup}$, $M = 90\pm8 \;M_\mathrm{Jup}$, $\mathrm{log}\,g = 5.21\pm0.06$\;dex. Given the speculation on the age of TRAPPIST-1, we also repeat this process and assume an age range of $0-0.5$~Gyr to address the intermediate gravity classification. The fundamental parameters derived for TRAPPIST-1 for both age assumptions are also listed in Table \ref{tab:LitFunParams}, which compares our values to literature values.

\begin{deluxetable*}{l c c c c c c c c c}
\tablecaption{Comparison of fundamental parameters from the literature for TRAPPIST-1\label{tab:LitFunParams}} \tablehead{\colhead{Parameter} & \colhead{This Paper} & \colhead{This Paper} &\colhead{Burg17} & \colhead{Gill17} & \colhead{Fili15} & \colhead{vanG18}} 
  \startdata
  log $L_*/L_{\odot}$ & $-3.216 \pm 0.016$ & $-3.216 \pm 0.016$  & $\cdots$ & $-3.281\pm 0.028$ & $-3.28 \pm 0.03$ & $-3.28 \pm 0.016$\\
  $T_\mathrm{eff}$ (K) & $2628 \pm 42$& $2310 \pm 230$   & $\cdots$ & $2559 \pm 50$ & $2557 \pm 64$ & $2516 \pm 41$\\
  Radius ($R_\mathrm{Jup}$) & $1.16 \pm 0.03$ & $1.5 \pm 0.3$ & $1.177 \pm 0.029$& $1.138 \pm 0.035$ & $1.14 \pm 0.04$ & $1.177 \pm 0.029$\\
  Mass ($M_\mathrm{Jup}$) & $90 \pm 8$ & $49 \pm 34$ & $83.8 \pm 7.3$ & $84 \pm 7.6$ & $86.07 \pm 9.28$ & $93.2 \pm 6.3$\\
  log$g$ & $5.21 \pm 0.06$ & $4.61 \pm 0.54$ & $\cdots$ & $\cdots$ & $5.22 \pm 0.08$ & $\cdots$ \\
  Age (Gyr) & $0.5-10$ & $<0.5$ & $7.6 \pm 2.2$& $>0.5\tablenotemark{a}$ & $0.5-10$ & $\cdots$ \\
  Parallax (mas) & $80.45\pm0.12$ & $80.45\pm0.12$ & $\cdots$& $82.58\pm2.58$\tablenotemark{b} & $82.58\pm2.58$\tablenotemark{b} & $82.4 \pm 0.8$ \\
  \enddata
  \tablenotetext{a}{Value from \cite{Gill16}, where the constraint of >0.5 Gyr is from \cite{Fili15}.}
  \tablenotetext{b}{Parallax from \cite{Cost06}}
  \tablecomments{The effective temperature in this paper and \cite{Fili15} are determined from measured $L_\mathrm{bol}$ combined with an assumed age and radii ranges obtained from evolutionary models. Values not noted are either not used or derived in that work.} 
\end{deluxetable*}

\subsection{Fundamental parameter comparison to the literature} 
Table \ref{tab:LitFunParams} contains our calculated fundamental parameters assuming two age ranges (1) $0.5-10$~Gyr, the field age constraint from \cite{Fili15} and (2) $<0.5$~Gyr, to address the intermediate gravity classification. Also listed in Table \ref{tab:LitFunParams} are literature values for derived fundamental parameters of TRAPPIST-1 from \cite{Fili15}, \cite{Gill17}, \cite{Burg17}, and \cite{vanG18}. Our bolometric luminosity value differs from previous measurements by up to 2 $\sigma$, which can be accounted for by our use of the more precise \textit{Gaia} DR2 parallax measurement as well as replacing the wider Johnson-Cousins band measurements with the narrower band Pan-STARRS values. However, our radius, $T_\mathrm{eff}$, and mass values agree with literature values within 1$\sigma$ with the exception of our $0.5-10$~Gyr $T_\mathrm{eff}$ value which differs by 1.4~$\sigma$ to the \cite{vanG18} value.

\section{A Comparative Sample for TRAPPIST-1}\label{CompTrap}

\subsection{Sample Selection and Properties}
\begin{deluxetable*}{l c c c c c c c c c c c c}
\tablecaption{Comparative Samples\label{tab:Sample}}
\tabletypesize{\scriptsize}
\tablehead{\colhead{R.A.} & \colhead{Decl.} & \colhead{Designation} & \colhead{Shortname} & \colhead{Disc. Ref.} & \colhead{Opt. SpT} & \colhead{SpT Ref.} & \colhead{NIR SpT\tablenotemark{a}} & \colhead{SpT Ref.} & \colhead{$\pi$ (mas)} & \colhead{Pi Ref.}} 
  \startdata
  23 06 29.36 & $-$5 02 29.2 & 2MASS J23062928$-$0502285 & TRAPPIST-1 & 1 & M7.5 & 1 & $\cdots$ &$\cdots$ & $80.45 \pm 0.12$ & 2 \\ \hline  
  &&&\multicolumn{4}{c}{\textbf{Assuming Field Age for TRAPPIST-1}} &&&& \\ \hline \hline
  &&&\multicolumn{4}{c}{Field and Subdwarfs with similar $T_\mathrm{eff}$ and $L_\mathrm{bol}$} &&&& \\ \hline
  03 20 59.65 & $+$18 54 23.3 & 2MASS J03205965$+$1854233 & J0320$+$1854 & 3 & M8 & 3 & $\cdots$ & $\cdots$ &$68.28 \pm 0.15$ & 2\\   
  10 13 07.35 & $-$13 56 20.4 & SSSPM J1013$-$1356 & J1013$-$1356 & 4 & sdM9.5 & 3 &$\cdots$ & $\cdots$ & $18.37 \pm 0.39$ & 2 \\ 
  14 56 38.31 & $-$28 09 47.3 & LHS 3003 & $\cdots$ & 5 & M7 & 3 & M7 & 6 & $141.69\pm 0.11$ & 2 \\                             
  16 55 35.29 & $-$8 23 40.1 & vB 8 & $\cdots$ & 7 & M7 & 8 & M7 & 9 & $153.81 \pm 0.11 $ & 2\\ 
  19 16 57.62 & $+$05 09 02.2 & vB 10 & $\cdots$ & 7 & M8 & 10 & M8 & 9 & $168.96 \pm 0.13$ & 2\\ \hline   
  &&&\multicolumn{4}{c}{Low gravity similar $T_\mathrm{eff}$ sources} &&&& \\ \hline
  04 36 27.88 & $-$41 14 46.5 & 2MASS J04362788$-$4114465 & J0436$-$4114& 11 & M8\,$\beta$ & 12 & M9 VL-G & 13 & $25.30 \pm 0.12$ & 2\\                               
  06 08 52.83 & $-$27 53 58.3 & 2MASS J06085283$-$2753583\tablenotemark{b} & J0608$-$2753 & 14 & M8.5\,$\gamma$ & 15 & L0 VL-G & 13 & $22.65 \pm 0.19$ & 2\\ 
  12 47 44.28 & $-$38 16 46.4& 2MASS J12474428$-$3816464 & J1247$-$3816 & 16 & $\cdots$ &$\cdots$ & M9 VL-G & 16 & $11.81 \pm 0.48$ & 2\\ \hline
  &&&\multicolumn{4}{c}{Low gravity, similar $L_\mathrm{bol}$ sources} &&&& \\ \hline
  04 43 37.61 & $+$00 02 05.1 & 2MASS J04433761$+$0002051 & J0443$+$0002 & 17 & M9\,$\gamma$ & 18 & L0 VL-G & 13 & $47.41 \pm 0.19$ & 2 \\ 
  05 18 46.16 & $-$27 56 45.7 & 2MASS J05184616$-$2756457 & J0518$-$2756 & 19 & L1\,$\gamma$ & 20 & L1 VL-G & 13 & $17.27 \pm 0.81$ & 2\\ 
  12 07 48.36 & $-$39 00 04.3 & 2MASS J12074836$-$3900043 & J1207$-$3900 & 16 & L0\,$\gamma$ & 16 & L1 VL-G & 16 & $14.92 \pm 0.90$ & 2\\ \hline
  &&&\multicolumn{4}{c}{\textbf{Assuming Age <0.5 Gyr for TRAPPIST-1}} &&&& \\ \hline \hline
  08 53 36.19 & $-$03 29 32.1 & 2MASS J08533619$-$0329321 & J0853$-$0329 & 21 & M9 & 22 & M9 & 6 & $115.30 \pm 0.11$ & 2\\
  10 48 14.64 & $-$39 56 06.2 & DENIS--P J1048.0$-$3956 & J1048$-$3956 & 23 & M9 & 23 &$\cdots$ & $\cdots$ & $247.22 \pm 0.12$ & 2 \\
  14 44 20.67 & $-$20 19 22.3 & SSSPM J1444$-$2019 & J1444$-$2019 & 24 & sdM9 & 24 & $\cdots$ &$\cdots$ &$57.80 \pm 0.56$ & 25 \\
  17 12 51.21 & $-$5 07 24.9 & GJ 660.1B & $\cdots$ & 26 & $\cdots$ & $\cdots$ & d/sdM7 & 27 & $50.1 \pm 3.62$ & 28\\  
  18 35 37.90 & $+$32 59 54.5 & 2MASS J18353790+3259545 & J1835$+$3259 & 29 & M8.5 & 30 & $\cdots$ & $\cdots$ & $175.82 \pm 0.09$ & 2\\
  20 00 48.41& $-$75 23 07.0 & 2MASS J20004841$-$7523070 & J2000$-$7523 & 31 & M9\,$\gamma$ & 32 & M9\,$\gamma$ & 32 & $33.95 \pm 0.15$ & 2\\ \hline
  &&&\multicolumn{4}{c}{\textbf{Exploring Low-z: Subdwarf J band Comparison sample}} &&&& \\ \hline \hline
  05 32 53.46 & $+$82 46 46.5 & 2MASS J05325346$+$8246465 & J0532$+$8246 & 33 & sdL7 & 34 & $\cdots$ &$\cdots$ & $40.24 \pm 0.64$ & 2\\    
  12 56 37.13 & $-$02 24 52.4 & SDSS J125637.13$-$022452.4\tablenotemark{b} & J1256$-$0224& 35 & sdL3.5 & 36 & $\cdots$ &$\cdots$ & $12.55 \pm 0.72$ & 2\\  
  14 39 00.31 & $+$18 39 38.5 & LHS 377 & $\cdots$ & 37 & sdM7 & 38 & $\cdots$ &$\cdots$ & $25.75 \pm 0.10$ & 2 \\                
  16 10 29.00 & $-$00 40 53.0 & LSR J1610$-$0040 & J1610$-$0040 & 39 & sdM7\tablenotemark{c} & 34 & $\cdots$ &$\cdots$ & $30.73 \pm 0.34$ & 40\\ 
  20 36 21.65 & $+$51 00 05.2 & LSR J2036$+$5059 & J2036$+$5059 & 41 & sdM7.5 & 42 & $\cdots$ &$\cdots$ & $23.10 \pm 0.29$ & 25\\ \hline  
  &&&\multicolumn{4}{c}{\textbf{Burgasser \& Majamajek Sources}} &&&& \\ \hline \hline
  01 02 51.00 & $-$37 37 43.0 & LHS 132 & $\cdots$ & 5 & M8 & 43 & M8 & 44 & $87.87 \pm 0.12$ & 2 \\  
  23 41 28.68 & $-$11 33 35.6 & 2MASS J23412868$-$1133356 & J2341$-$1133 & 19 & M8 & 19 & $\cdots$ &$\cdots$ & $20.33 \pm 0.24$ & 2\\ 
  23 52 05.00 & $-$11 00 43.5 & 2MASS J23520507$-$1100435 & J2352$-$1100 & 19 & M7 & 19 & M8\,$\beta$ & 45 & $24.88 \pm 0.51$ & 2 \\ \hline
  \enddata
\tablenotetext{a}{We refer to any objects classified as VL-G as $\gamma$ and INT-G as $\beta$ sources throughout this work when making reference to the NIR spectral type.}
\tablenotetext{b}{Also part of the Youth Assumption sample.}
\tablenotetext{c}{Spectroscopic binary.}
\tablerefs{(1) \cite{Gizi00}, (2) \cite{GaiaDR1,GaiaDR2,Lind18}, (3) \cite{Kirk95}, (4) \cite{Scho04a}, (5) \cite{Luyt79a}, (6) \cite{Geba02}, (7) \cite{vanB61}, (8) \cite{Reid05}, (9) \cite{Kirk10}, (10) \cite{Henr90}, (11) \cite{Phan03}, (12) \cite{Fahe09}, (13) \cite{Alle13},(14) \cite{Cruz03}, (15) \cite{Fahe12},  (16) \cite{Gagn14b}, (17) \cite{Hawl02}, (18) \cite{Kirk08}, (19) \cite{Cruz07}, (20) \cite{Cruz18}, (21) \cite{Reid87}, (22) \cite{Kirk91}, (23) \cite{Mont01}, (24) \cite{Scho04c}, (25) \cite{Dahn17}, (26)  \cite{Schn11}, (27) \cite{Agan16}, (28) \cite{Dupu12a}, (29) \cite{Gizi02b}, (30)  \cite{Reid03b}, (31) \cite{Cost06}, (32) \cite{Fahe16}, (33) \cite{Burg03c}, (34) \cite{Burg07a}, (35) \cite{Siva09}, (36) \cite{Burg09a}, (37) \cite{Lieb79}, (38) \cite{Gizi97}, (39) \cite{Lepi03c}, (40) \cite{Kore16}, (41)  \cite{Lepi02}, (42) \cite{Lepi03a}, (43) \cite{Diet14} , (44)  \cite{Bard14}, (45) \cite{Gagn15b}}
\tabletypesize{\small}
\end{deluxetable*}

\begin{deluxetable*}{l c c c c c c c c c}
\tablecaption{Fundamental Parameters of Comparison Samples \label{tab:SampleFunParams}}
\tabletypesize{\footnotesize}
\tablehead{\colhead{Name} & \colhead{Opt. SpT} & \colhead{NIR SpT} &\colhead{log L$_\mathrm{bol}$} & \colhead{T$_\mathrm{eff}$} & \colhead{Radius} & \colhead{Mass} & \colhead{log(g)} &\colhead{Age} & \colhead{Distance} \vspace{-.1cm}\\ 
 & & & &\colhead{(K)} & \colhead{(R$_\mathrm{J}$)} & \colhead{(M$_\mathrm{J}$)} & \colhead{(dex)} & \colhead{(Gyr)} & \colhead{(pc)} }
  \startdata
  TRAPPIST-1 & M7.5 & $\cdots$ & $-3.216 \pm 0.016$ & $2628 \pm 42$ & $1.16 \pm 0.03$  & $90 \pm 8$ & $5.21 \pm 0.06$ & $0.5-10$ & $12.43 \pm 0.02$\\ 
  TRAPPIST-1 & M7.5 & $\cdots$ & $-3.216 \pm 0.016$ & $2310 \pm 230$ & $1.5 \pm 0.3$  & $49 \pm 34$ & $4.61 \pm 0.54$ & $0-0.5$ & $12.43 \pm 0.02$\\  \hline 
  LHS 132 & M8 & M8 & $-3.264 \pm 0.015$ & $2579 \pm 41$ & $1.14 \pm 0.03$ & $87 \pm 8$ & $5.21 \pm 0.07$ & $0.5-10$ & $11.38 \pm 0.02$\\ 
  J0320$+$1554 & M8 & $\cdots$ & $-3.226 \pm 0.007$ & $2613 \pm 35$ & $1.16 \pm 0.03$ & $89 \pm 8$ & $5.21 \pm 0.06$ & $0.5-10$ & $14.65 \pm 0.03$  \\ 
  J0436$-$4114 & M8\,$\beta$ & M9\,$\gamma$ & $-2.927 \pm 0.019$ & $2560 \pm 260$ & $1.7 \pm 0.34$ & $60 \pm 43$ & $4.58 \pm 0.56$ & $0-0.5$ & $39.52 \pm 0.19$\\
  J0443$+$0002 & M9\,$\gamma$ & L0\,$\gamma$ & $-3.194\pm 0.009$ & $2232 \pm 29$ & $1.65 \pm 0.04$ & $25 \pm 2$ & $4.34 \pm 0.06$ & $0.021-0.027$ &  $21.09 \pm 0.08$\\ 
  J0518$-$2756 & L1\,$\gamma$ & L1\,$\gamma$ &$-3.273 \pm 0.041$ & $2229 \pm 57$ & $1.51 \pm 0.03$ & $31 \pm 3$ &	$4.51 \pm 0.05$ & $0.038-0.048$ & $57.9 \pm 2.7$\\ 
  J0532$+$8246 & sdL7 & $\cdots$ &$-4.28 \pm 0.07$ & $1670 \pm 70$ & $0.84 \pm 0.02$ & $72 \pm 5 $ & $5.4 \pm 0.05$ & $5-10$ & $24.85 \pm 0.39$\\ 
  J0608$-$2753 & M8.5\,$\gamma$ & L0\,$\gamma$ &$-2.996 \pm 0.014$ & $2510 \pm 250$ & $1.64 \pm 0.33$ & $ 57 \pm 40 $ & $4.59 \pm 0.56$ & $0-0.5$ & $44.15 \pm 0.38$\\ 
  J0853$-$0329 & M9 & M9 & $-3.485 \pm 0.033$ & $2330 \pm 70$ & $1.08 \pm 0.05$ & $78 \pm 10$ & $5.21 \pm 0.1$ & $0.5-10$ & $8.67 \pm 0.01$\\ 
  J1013$-$1356 & sdM9.5 & $\cdots$ &$-3.303 \pm 0.027$ & $2628 \pm 43$ & $1.05 \pm 0.01$ & $92 \pm 1$ & $5.32 \pm 0.05$ & $5-10$ & $54.4 \pm 1.2 $\\ 
  J1048$-$3956 & M9 & $\cdots$ & $-3.485 \pm 0.019$ & $2330 \pm 60$ & $1.08 \pm 0.05$ &	$78 \pm 10$ & $5.21 \pm 0.1$ & $0.5-10$ & $4.045 \pm 0.002$ \\
  J1207$-$3900 & L0\,$\gamma$ & L1\,$\gamma$ &$-3.337 \pm 0.053$ & $2013 \pm 66$ &	$1.72 \pm 0.04$ & $15 \pm 1$ & $4.1 \pm 0.03$ & $0.007-0.013$ & $67 \pm 4$\\ 
  J1247$-$3816 & $\cdots$ & M9\,$\gamma$ & $-2.84 \pm 0.039$ & $2630 \pm 290$ & $1.79 \pm 0.39$ & $67 \pm 49$ & $4.56 \pm 0.57$ & $0-0.5$ & $84.6 \pm 3.5$\\ 
  J1256$-$0224 & sdL3.5 & $\cdots$ & $-3.63 \pm 0.05$ & $2301\pm71$ & $0.94\pm0.02$ & $83\pm2$ & $5.37\pm0.01$ & $0.5-10$ & $79.7\pm4.6$  \\
  LHS 377 & sdM7 & $\cdots$ & $ -3.041 \pm 0.008$ & $2840 \pm 60$ & $1.212 \pm 0.05$ & $106 \pm 1$ & $5.25 \pm 0.05$ & $5-10$ & $38.84 \pm 0.16$\\ 
  J1444$-$2019 & sdM9 & $\cdots$ & $-3.49\pm 0.02$ & $2442\pm68$ & $0.98\pm 0.05$ & $87\pm1$ & $5.35\pm0.05$ & $5-10$ & $17.3\pm0.17$ \\ 
  LHS 3003 & M7 & M7 & $-3.224 \pm 0.012$ & $2616 \pm 38$ &  $1.16 \pm 0.03$ & $89 \pm 8$ &	$5.21 \pm 0.06$ & $0.5-10$ & $7.06 \pm 0.01$\\ 
  J1610$-$0040 & sdM7 & $\cdots$ & $-2.98 \pm 0.01$ & $2880 \pm 20$ & $1.27 \pm 0.01$ & $111 \pm 1$ & $5.23 \pm 0.05$ & $5-10$ & $32.54 \pm 0.36$ \\ 
  vB 8 & M7 & M7 & $-3.192 \pm 0.006$ & $2642 \pm 35$ & $1.18 \pm 0.03$ & $91 \pm 7$ & $5.21 \pm 0.06$ & $0.5-10$ & $6.501\pm0.005$\\        
  GJ 660.1B & $\cdots$ & d/sdM7 & $-3.523 \pm 0.063$ & $2409 \pm 91$ & $0.97 \pm 0.02$ & $86 \pm 2$ & $5.36 \pm 0.01$ & $0.5-10$ & $20 \pm 1.4$\\ 
  J1835$+$3295 & M9 & M9& $-3.502\pm0.016$ & $2321\pm58$ & $1.07\pm0.05$ & $77\pm10$ & $5.21\pm 0.1$ & $0.5-10$ & $5.687\pm 0.003$\\ 
  vB 10 & M8 & M8 & $-3.298 \pm 0.018$ & $2540 \pm 52$ & $1.13 \pm 0.04$ & $86 \pm 8$ & $5.22 \pm 0.07$ & $0.5-10$ & $5.918 \pm 0.005$\\ 
  J2000$-$7523 & M9\,$\gamma$ & M9\,$\gamma$ & $-3.04\pm 0.02$	& $2389\pm 44$ & $1.72\pm0.05$ & $30\pm3$ & $4\pm0.07$ & $0.021-0.027$ & $29.45\pm0.13$\\ 
  J2036$+$5059 & sdM7.5 & $\cdots$ & $-3.057 \pm 0.035$ & $2832 \pm 58$ & $1.2 \pm 0.01$ & $105 \pm 1$ & $5.25 \pm 0.05$ & $0.5-10$ & $43.29 \pm 0.54$\\ 
  J2341$-$1133 & M8 & $\cdots$ & $-2.921\pm0.015$ & $2898\pm41$ & $1.34\pm0.03$& $109\pm7$ & $5.18\pm0.04$ & $0.5-10$& $49.18\pm0.58$\\ 
  J2352$-$1100 & M7 & M8\,$\beta$ & $-2.797\pm0.018$ & $2913\pm56$& $1.53\pm0.05$& $100\pm11$& $5.02\pm0.06$& $0.13-0.2$& $40.19\pm0.82$\\ 
  \enddata
  \tablecomments{The effective temperature is determined based off age estimates and evolutionary models.}
  \tabletypesize{\small}
\end{deluxetable*}

To determine if TRAPPIST-1 exhibits similar or different features compared to the average field M dwarf, we constructed three comparative samples.
\begin{itemize}
  \item Sample \#1: Assumes TRAPPIST-1 is a field age ($0.5-10$~Gyr) source and contains comparative field, very low-gravity ($\gamma$), and old dwarfs with similar effective temperatures (within $1\, \sigma$ of TRAPPIST-1) and/or bolometric luminosity (within up to $2.5\, \sigma$) of TRAPPIST-1
  \item Sample \#2: Assumes an age of $<0.5$~Gyr for TRAPPIST-1 and contains field, very low-gravity ($\gamma$), and old dwarfs with similar effective temperatures within $1\, \sigma$
  \item Sample \#3: Assumes TRAPPIST-1 is a field age and contains subdwarfs of varying $T_\mathrm{eff}$ and $L_\mathrm{bol}$ (sdM7 and later) with medium resolution data (R $\sim 5000$).
\end{itemize}

The effective temperature samples compare the overall SED shape and spectral features of TRAPPIST-1 to objects with similar atmospheric chemistry. The bolometric luminosity sample examines how the flux of TRAPPIST-1 is redistributed across different wavelengths compared to comparable $L_\mathrm{bol}$ sources. The subdwarf sample explores the \ion{Na}{1} and \ion{K}{1} doublets in the $J$ band to see if TRAPPIST-1 exhibits similarities with these metal poor old objects. 

Objects in each comparative sample were chosen from (1) \cite{Fili15}, which examined a large sample of field and low-gravity objects, (2) \cite{Fahe16}, which examined a large sample of low-gravity sources, or (3) \cite{Gonz18}, which examined subdwarfs later than sdM6 with parallax measurements. The bolometric luminosities in each sample were empirically derived, while their effective temperatures were semi-empirically derived using radii from evolutionary models depending on the age. Ages were determined by updated membership in moving groups (Faherty et al. in prep) for young sources from \cite{Fahe16}. For field sources from \cite{Fili15} we use their field dwarf age range of 0.5--10 Gyr, and for subdwarfs from \cite{Gonz18} we use their subdwarf age range of 5--10 Gyr. Exact evolutionary models used for radii values are listed in \cite{Fili15}, \cite{Fahe16}, and \cite{Gonz18}. 

In order to resolve features in a band-by-band comparative analysis of the spectra, we required objects to have medium-resolution NIR data ($\lambda/\Delta\lambda > 1000$ at $J$ band). Field dwarf comparison SEDs chosen from \cite{Fili15} were reconstructed with the same data used in that work, with the exception of replacing their low-resolution SpeX data with medium-resolution NIR data in this work. Low-gravity dwarf comparison SEDs from \cite{Fahe16} were constructed with new NIR FIRE data or re-reduced spectra. Subdwarf comparison SEDs were constructed with the same data from \cite{Gonz18}, with the exception of GJ 660.1B which was not part of that sample, SSSPM J1013$-$1356 which only uses the NIR spectrum in this work, and LSR 2036$+$5059 which includes synthetic WISE photometry points for proper appending of the Rayleigh-Jeans tail.

All SEDs constructed for this paper used measured photometric values alone with the exceptions listed above. We did not incorporate synthetic photometry as done by \cite{Fili15} and \cite{Fahe16}. We also used updated parallax measurements from \textit{Gaia} DR2 for objects chosen from \cite{Fili15} and \cite{Fahe16}. Therefore our values differ slightly but fall within $\sim1\, \sigma$ for all values except bolometric luminosity. 

Sample \#1 contains 5 field sources that have similar $T_\mathrm{eff}$ and $L_\mathrm{bol}$, three sources with similar $T_\mathrm{eff}$, and three with similar $L_\mathrm{bol}$ for a total of 8 sources when examining $T_\mathrm{eff}$ or $L_\mathrm{bol}$. Sample \#2 contains 8 sources- 3 field dwarfs, 3 subdwarfs, and 2 low-gravity sources. Sample \#3 contains the 5 subdwarfs from \cite{Gonz18} which have medium-resolution $J$-band data. Details for all comparative objects are listed in Table \ref{tab:Sample}. The corresponding parallaxes, photometry, and spectra used in the creation of the SEDs are listed in Tables \ref{tab:opticalphot}--\ref{tab:SpectraReferences}. Table \ref{tab:SampleFunParams} lists the derived fundamental parameters for TRAPPIST-1 along with the comparative objects.

\section{Spectral Analysis For Sample \#1: Assuming a Field Age for TRAPPIST-1}\label{FieldAgeAssumption}
\subsection{Full SED Comparisons}\label{FullSEDFeildAssumption}
\begin{figure*}
 \gridline{\fig{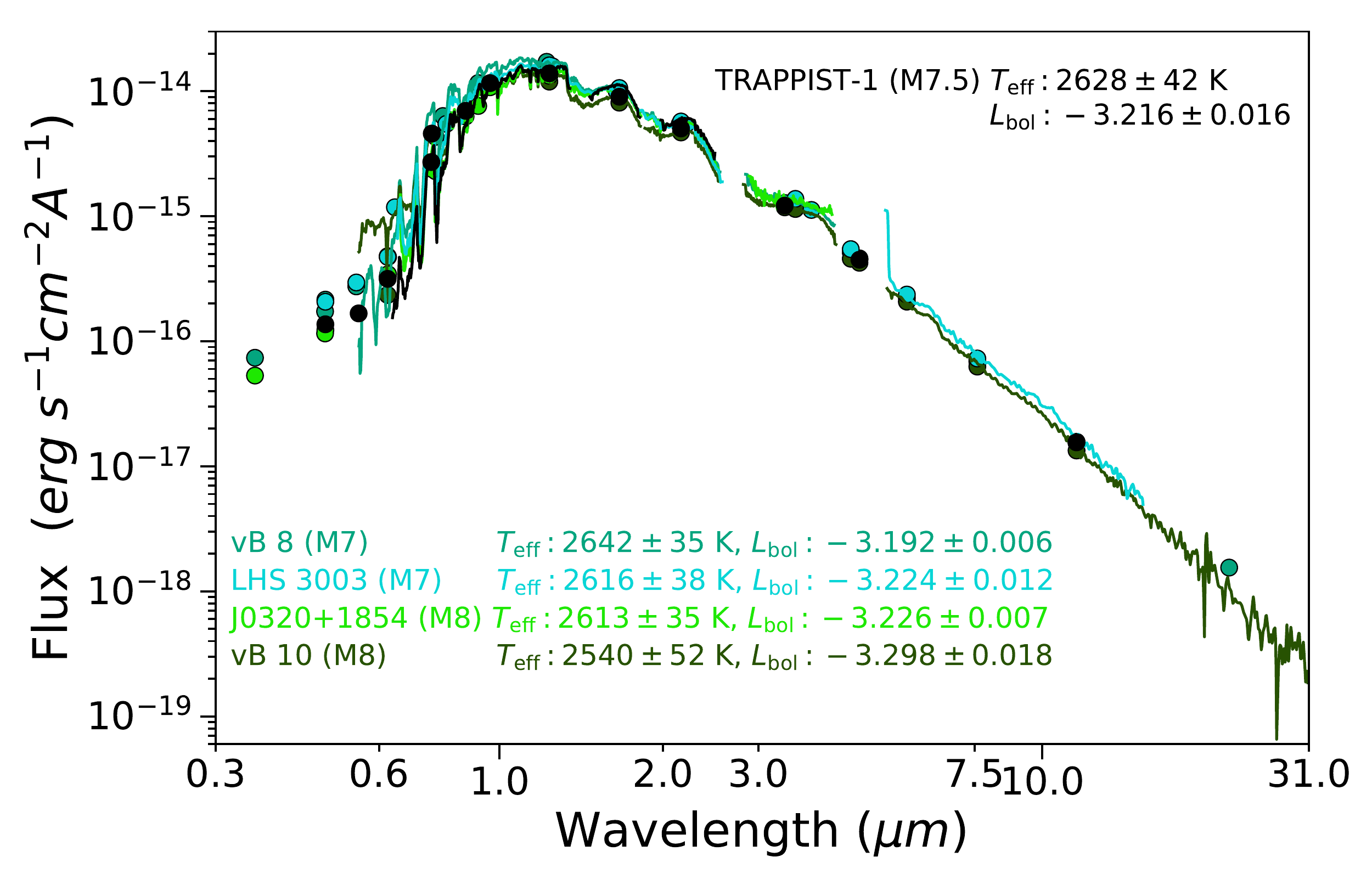}{0.48\textwidth}{\large(a)}
           \fig{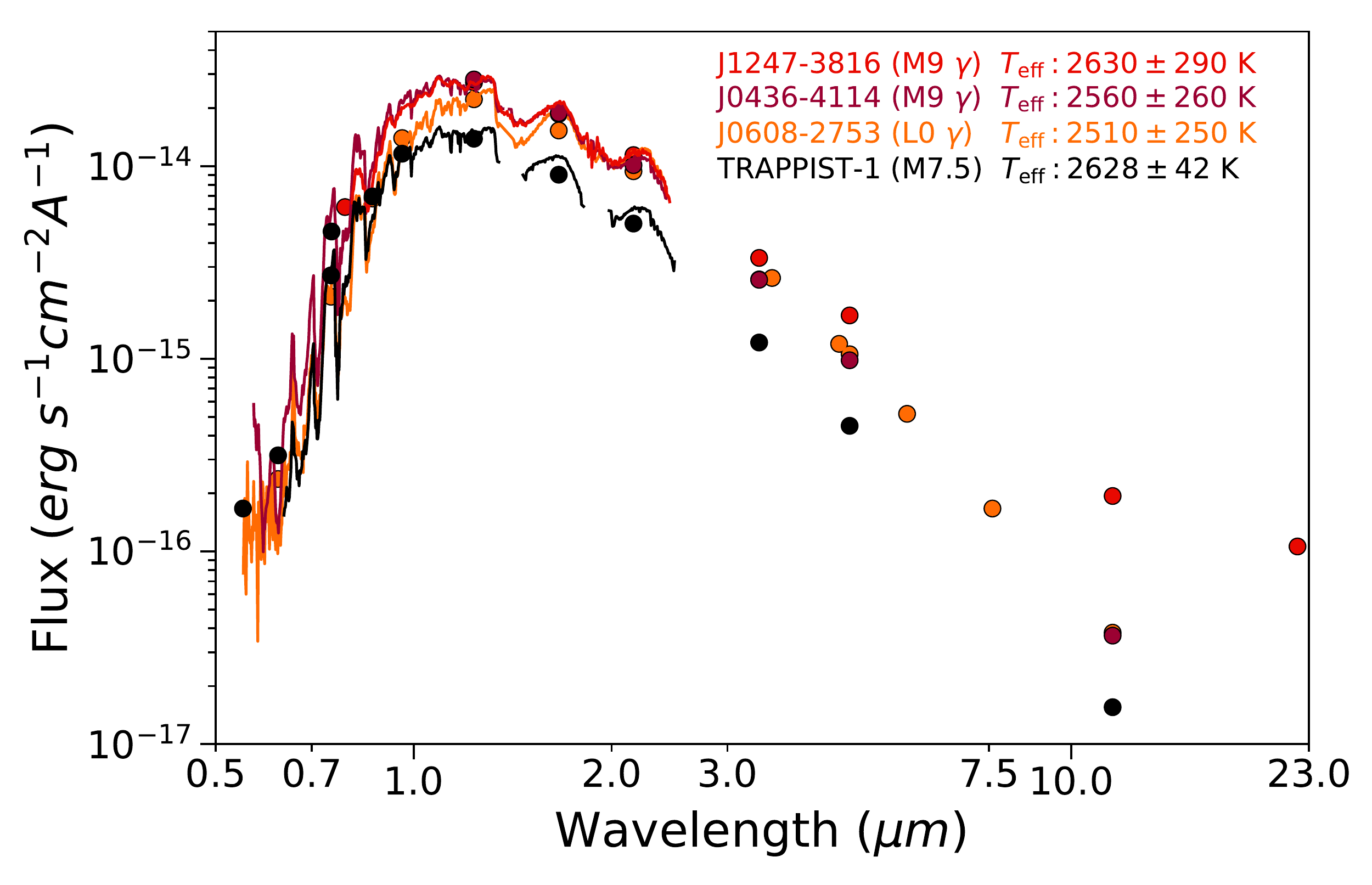}{0.48\textwidth}{\large(b)}} 
 \gridline{\fig{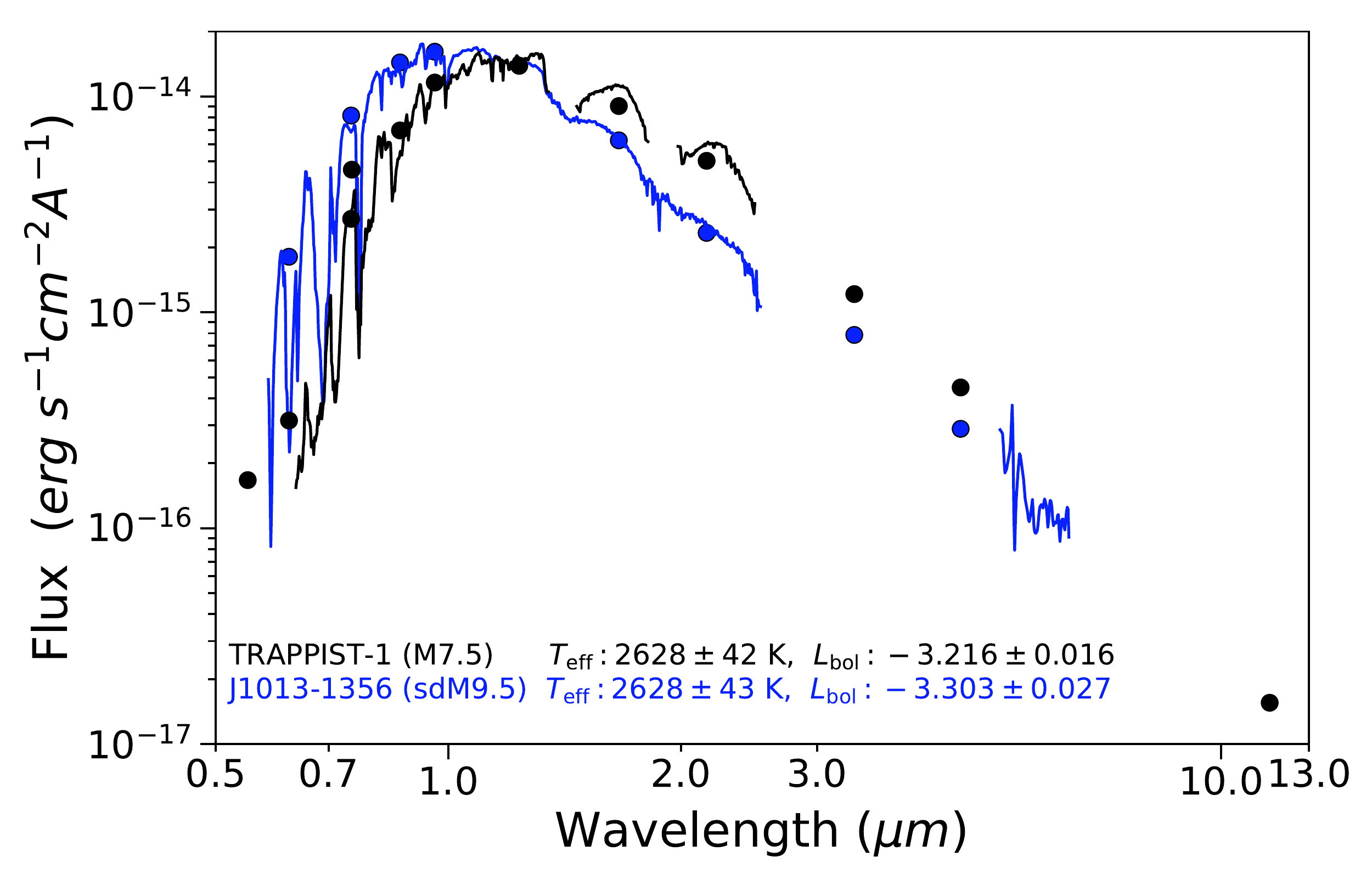}{0.48\textwidth}{\large(c)}
           \fig{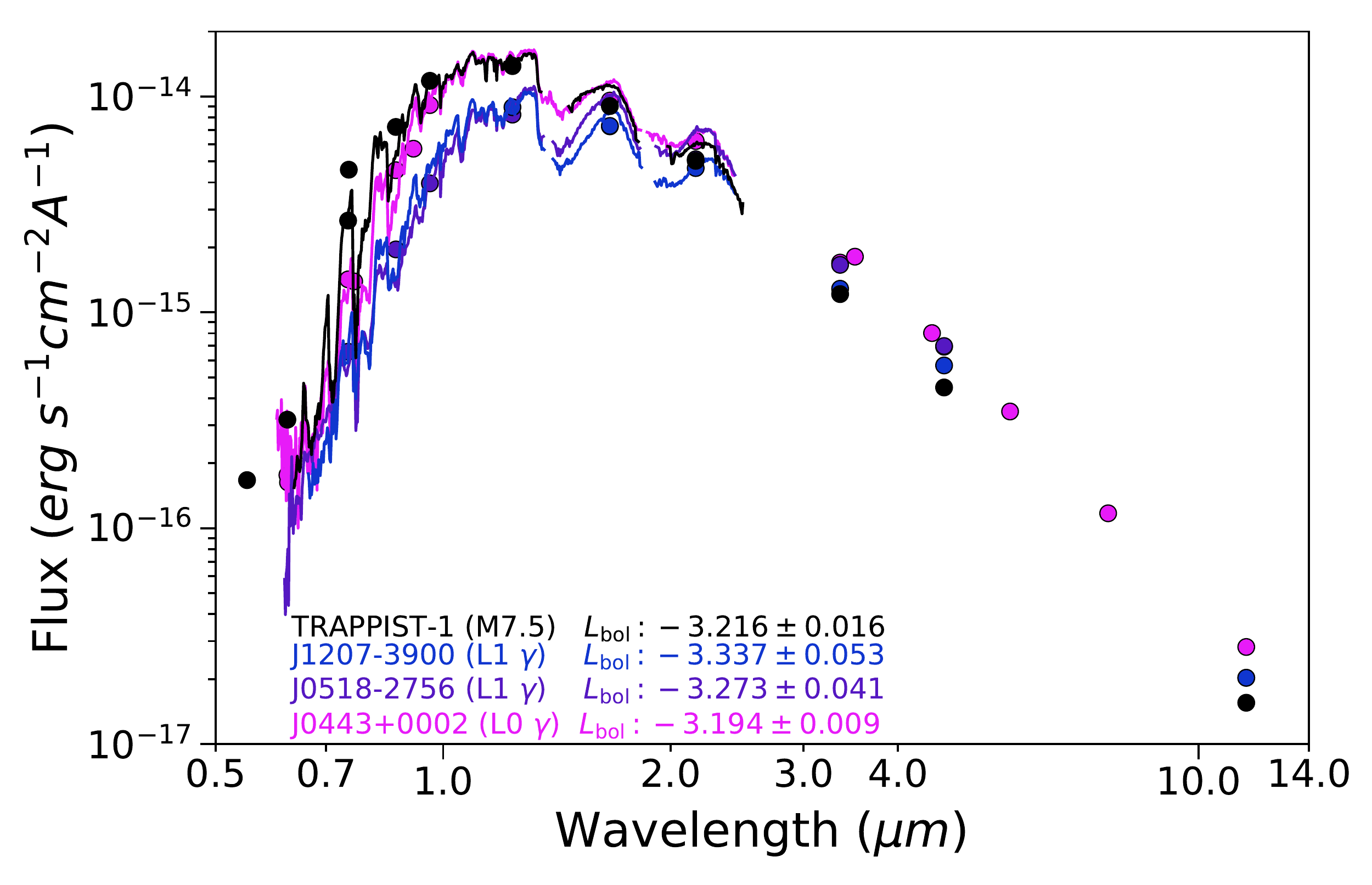}{0.48\textwidth}{\large(d)}}
 \caption{Distance-calibrated SEDs of field dwarfs, low-gravity dwarfs, and subdwarfs of approximately the same effective temperature and/or bolometric luminosity as TRAPPIST-1 (black). All spectra were resampled to the same dispersion relation using a wavelength-dependent Gaussian convolution. The optical and near infrared spectra are flux calibrated and scaled to the absolute magnitudes of the photometry shown. No normalization is applied. (a) Field dwarfs of similar $T_\mathrm{eff}$ and $L_\mathrm{bol}$ (various shades of green). (b) Low-gravity dwarfs of similar $T_\mathrm{eff}$ (shades of red and orange). (c) Subdwarfs of similar $T_\mathrm{eff}$ and $L_\mathrm{bol}$ (various shades of blue). (d) Low-gravity dwarfs of similar $L_\mathrm{bol}$ (pink, purple, and blue).}
 \label{fig:TeffLbolSEDs}
 \vspace{0.5cm} 
\end{figure*}

\begin{figure*}
 \gridline{\fig{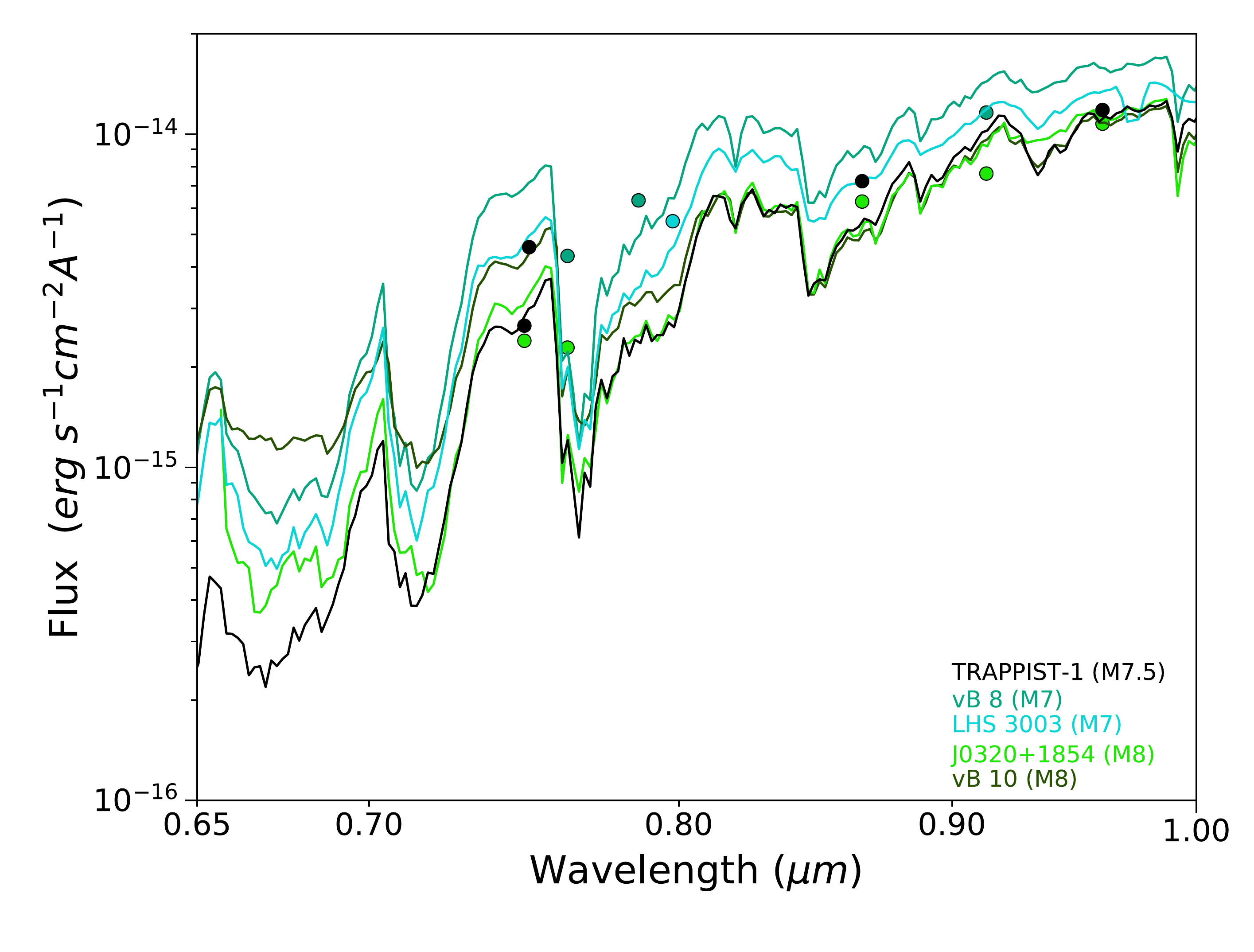}{0.5\textwidth}{\large(a)}
           \fig{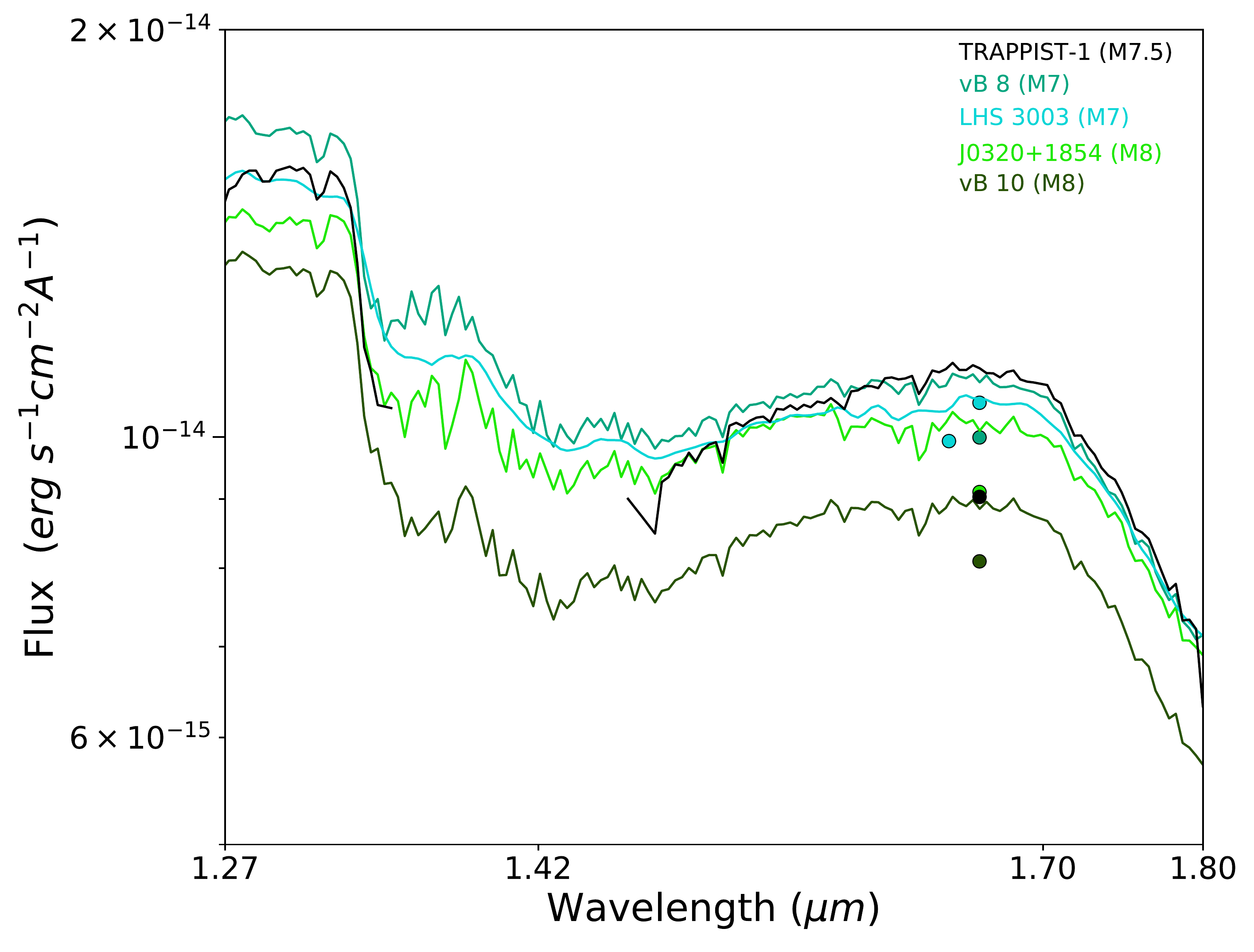}{0.5\textwidth}{\large(b)}} 
 \caption{Zoom of Figure \ref{fig:TeffLbolSEDs}a with colors corresponding to the same objects.  (a) Zoom in of the $0.65-1\, \upmu$m region. Objects fan out, but do not fan based on temperature or luminosity. (b) Zoom in of the $1.27-1.8\, \upmu$m region. Objects fan out in a temperature sequence, with the hottest object at the top and the coldest at the bottom.}
 \label{fig:Fieldzoom}
 \vspace{0.5cm} 
\end{figure*}

Under the assumption that TRAPPIST-1 is field age, we present the overall SEDs for objects of equivalent temperature and/or bolometric luminosity to that of TRAPPIST-1 in Figures \ref{fig:TeffLbolSEDs} and \ref{fig:Fieldzoom}. The sample is composed of three age subsets: field, low-gravity (young), and subdwarfs (old). All objects of similar $T_\mathrm{eff}$ are within $1\,\sigma$ of TRAPPIST-1 and of similar $L_\mathrm{bol}$ are within $2.5\,\sigma$.

Figure \ref{fig:TeffLbolSEDs}a compares the overall SED assuming an older age for TRAPPIST-1 with field dwarfs of similar $T_\mathrm{eff}$ and $L_\mathrm{bol}$ across $0.3-31\,\upmu$m. All comparative field dwarfs in the sample are within one spectral type of TRAPPIST-1. From $0.65-1\, \upmu$m there is a spread in the SEDs (zoomed version in Figure \ref{fig:Fieldzoom}a), however this spread does not appear to correspond to temperature or bolometric luminosity of the objects. There appears to be a tighter temperature-dependent sequence from $1.28-1.8\, \upmu$m, which can be seen in Figure \ref{fig:Fieldzoom}b most clearly from the start of the $H$ band out to $1.70\,\upmu$m.  Other than the large spread in the red optical, the overall SED shape of TRAPPIST-1 is similar to all sources in this sub-sample.

Figure \ref{fig:TeffLbolSEDs}b compares TRAPPIST-1 to very low-gravity ($\gamma$) dwarfs of similar $T_\mathrm{eff}$ sample across the $0.5-23\,\upmu$m region. The low-gravity comparison objects are not members of known moving groups so we can't give a definitive age bracket for them however, their spectra show strong indications of youth such as triangular $H$-band and weakened FeH (see \citealt{Fahe16}, Faherty et al. in prep). From $\sim 0.80-1\,\upmu$m, we see TRAPPIST-1 directly on top of J0608$-$2753, however from $J$ band onward all low-gravity sources are much brighter than TRAPPIST-1.

Figure \ref{fig:TeffLbolSEDs}c displays the overall SED of TRAPPIST-1 against an older subdwarf of similar $T_\mathrm{eff}$ and $L_\mathrm{bol}$ across the $0.5-13\,\upmu$m region. From $\sim0.7-1\, \upmu$m SSSPM J1013$-$1356 (hereafter J1013$-$1356) is brighter than TRAPPIST-1, while beyond $1\, \upmu$m TRAPPIST-1 is much brighter. TRAPPIST-1 is clearly poorly fit by the subdwarf in this sample.

SEDs of low-gravity objects with equivalent $L_\mathrm{bol}$ (within $2.5\, \sigma$) of TRAPPIST-1 across the $0.5-14\,\upmu$m region are shown in Figure \ref{fig:TeffLbolSEDs}d. 2MASS J04433761$+$0002051 (hereafter J0443$+$0002) is a member of the $21-27$ Myr \citep{Bell15} $\beta$ Pictoris moving group, 2MASS J05184616$-$2756457 (hereafter J0518$-$2756) is a member of the $38-48$ Myr \citep{Bell15} Columba moving group, and lastly 2MASS J12074836$-$3900043 (hereafter J1207$-$3900) is a member of the $7-13$ Myr \citep{Bell15} TW Hydra moving group (\citealt{Fahe16}, Faherty et al. in prep.). From $0.5\,\upmu$m to the start of the $J$ band, TRAPPIST-1 is brighter than all low-gravity sources, while in the $J$ and $H$ bands J0443$+$0002 and TRAPPIST-1 are of similar brightness. Beyond $W1$ TRAPPIST-1 is fainter than the low-gravity sources. 

Overall under the assumption of field age for TRAPPIST-1, comparing the shape to field, low-gravity, and subdwarfs leads to the strong conclusion that this is a field source.

\subsection{Band-by-Band Analysis}\label{bandbybandField}
While the overall SED comparisons give a general examination of where flux is matched between sources, the subtle but significant feature detail is lost. Detailed line analysis can be used to tease out signatures of gravity or metallicity therefore in this subsection we compare NIR band-by-band features.

\begin{figure*}
 \gridline{\fig{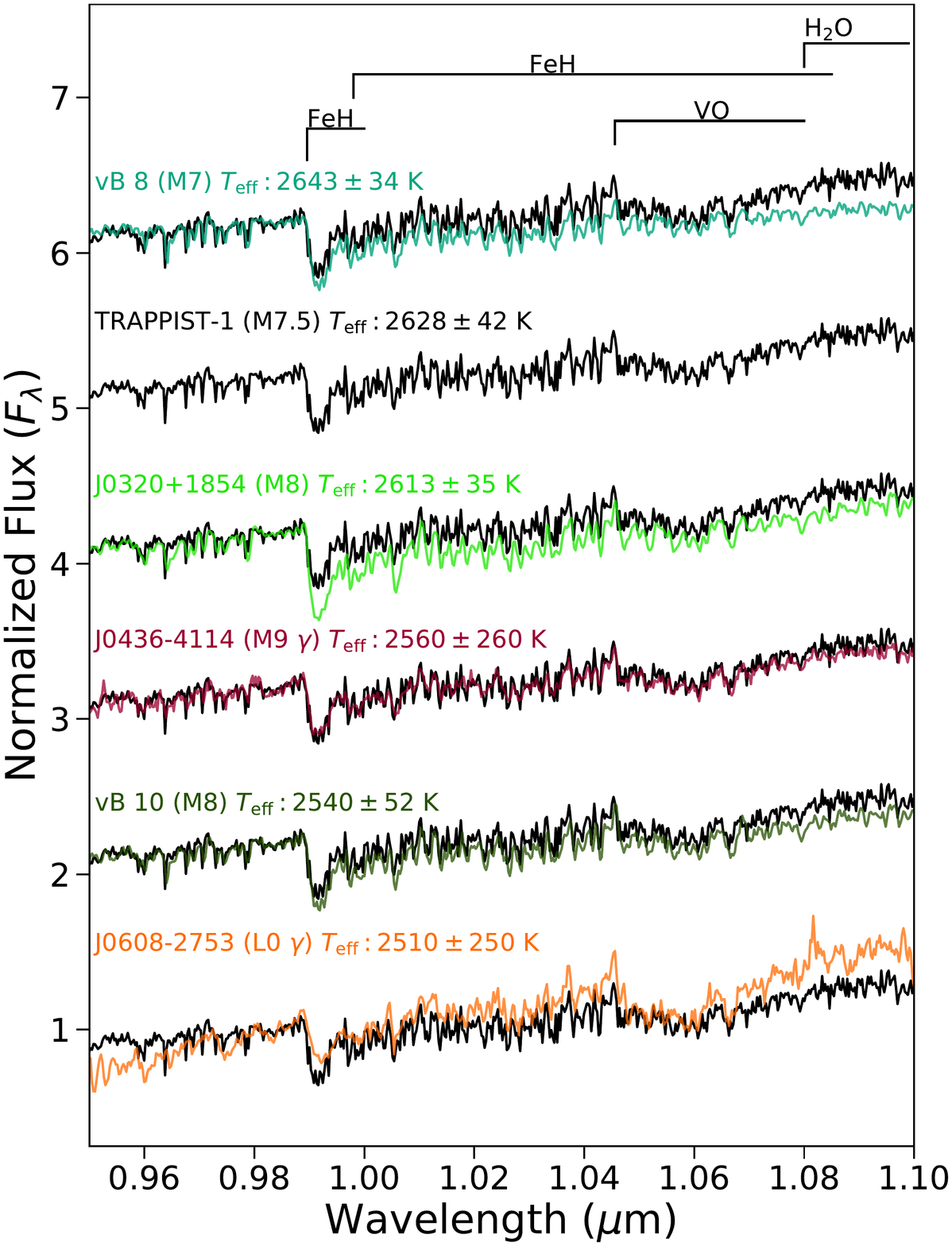}{0.45\textwidth}{\large(a)}
           \fig{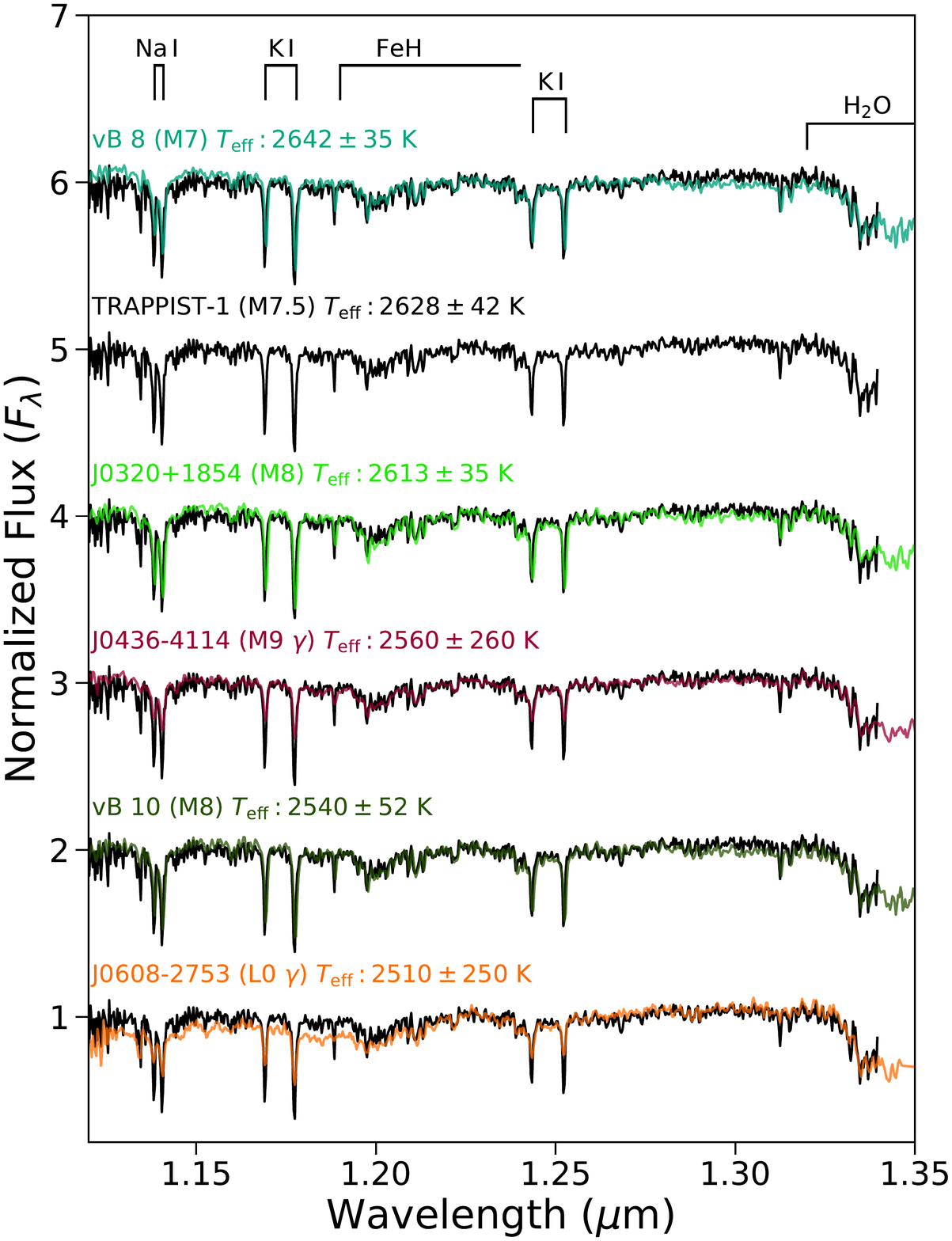}{0.45\textwidth}{\large(b)}} 
 \gridline{\fig{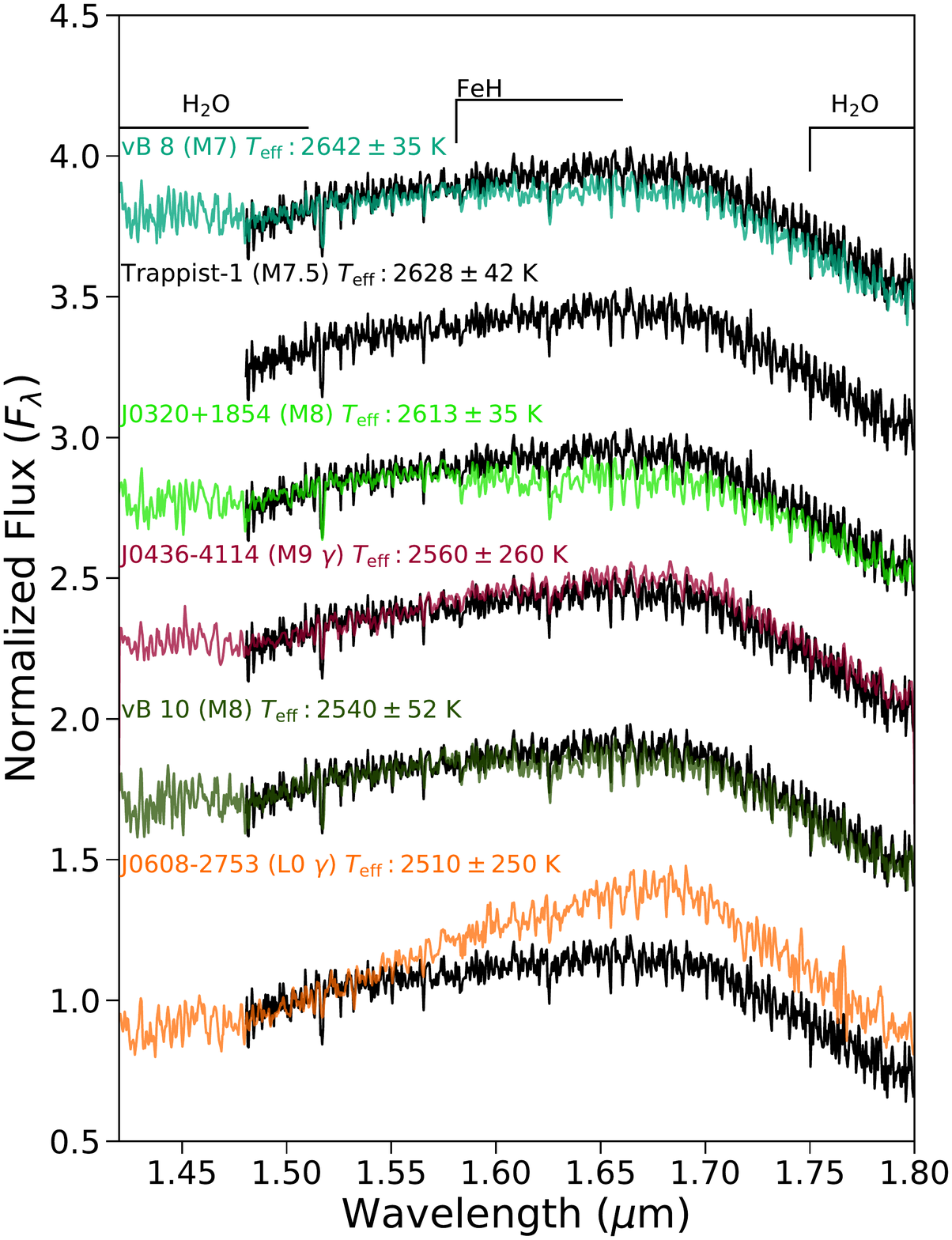}{0.45\textwidth}{\large(c)}
           \fig{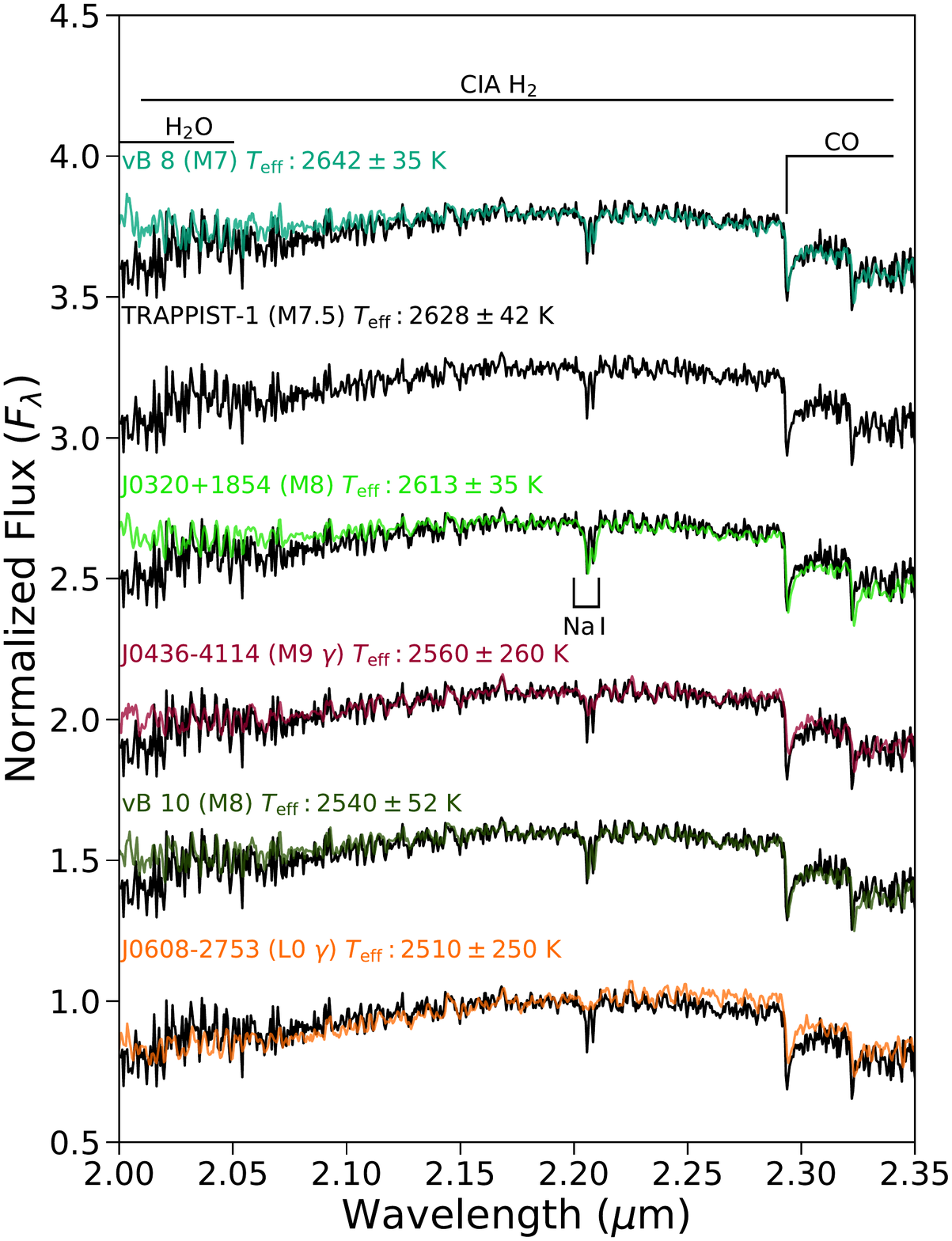}{0.45\textwidth}{\large(d)}}
 \caption{Band-by-band comparison of field (shades of green) and low-gravity (burgundy and orange) dwarfs of similar $T_\mathrm{eff}$ as TRAPPIST-1 (black). All spectra were resampled to the same dispersion relation using a wavelength-dependent Gaussian convolution and are offset by a constant. NIR spectral types as displayed. (a) $Y$ band. Spectra are normalized by the average flux taken across $0.98-0.988\, \upmu$m. (b) $J$ band. Spectra are normalized by the average flux taken across $1.29-1.31\, \upmu$m. (c) $H$ band. Spectra are normalized by the average flux taken across $1.5-1.52\, \upmu$m. (d) $K$ band. Spectra are normalized by the average flux taken across $2.16-2.20\, \upmu$m.}
 \label{fig:Teffbandbyband}
\end{figure*} 

\subsubsection{Similar $T_\mathrm{eff}$}
Figure \ref{fig:Teffbandbyband}a shows the $0.95-1.10\, \upmu$m $Y$-band data with FeH, VO, and H$_2$O features labeled. The spectra for the $Y$ band were normalized by averaging the flux taken across the relatively featureless $0.98-0.988\, \upmu$m region. From first glance, we see that TRAPPIST-1 appears to be most similar in shape in the $Y$ band to the low-gravity dwarf J0436$-$4114 and the field dwarf vb10. The slope from $0.95 - 0.99\, \upmu$m of TRAPPIST-1 is most similar to those of the field dwarfs however, it also matches the slope of the low-gravity dwarf J0436$-$4114. The Wing-Ford FeH band-head of TRAPPIST-1 appears to be of similar depth to J0436$-$4114, but only slightly shallower than vB 8 and vB 10. From $1-1.05\,\upmu$m the spectrum of TRAPPIST-1 overlaps that of J0436$-$4114, but shares the shape of the field dwarfs more than that of the low-gravity dwarfs. There is an indication of slight VO absorption in the spectrum of TRAPPIST-1, similar in depth to the low-gravity dwarf J0436$-$4114. From all of these features we therefore conclude the TRAPPIST-1 $Y$-band spectrum exhibits a blend of both field and low-gravity dwarf features.

Figure \ref{fig:Teffbandbyband}b shows the $1.12-1.35\, \upmu$m $J$-band data with FeH and H$_2$O molecular features, as well as the \ion{Na}{1} and \ion{K}{1} alkali doublets labeled. The $J$-band spectra were normalized by the average flux over the featureless $1.29-1.31\, \upmu$m region. The depth of the \ion{Na}{1} and \ion{K}{1} doublets of TRAPPIST-1 are slightly deeper than the field dwarfs and much deeper than the low-gravity dwarfs. The FeH absorption of TRAPPIST-1 is similar to the field dwarfs, while beyond the $1.25\,\upmu$m \ion{K}{1} doublet the shape is similar for all objects in the sample. TRAPPIST-1 has some $J$-band spectral features similar to field dwarfs, while others differ from both the field and low-gravity dwarfs.

Figure \ref{fig:Teffbandbyband}c shows the $1.42-1.80\, \upmu$m $H$-band data with FeH and H$_2$O molecular features labeled. The $H$-band spectra were normalized by the average flux over the featureless $1.5-1.52\, \upmu$m region. The $H$-band shape of TRAPPIST-1 is similar to the field objects, however its overall shape is also similar to J0436$-$4114 but slightly less triangular. There are no features in the $H$-band spectrum of TRAPPIST-1 that match those of the low-gravity dwarf J0608$-$2753. In the $H$ band, TRAPPIST-1 is more like a field object.

Figure \ref{fig:Teffbandbyband}d shows the $2.0-2.35\, \upmu$m $K$-band data with H$_2$O, CO, and collision-induced H$_2$ absorption features labeled. The $K$-band spectra were normalized by the average flux over the $2.16-2.20\, \upmu$m region due to the relatively flat spectral region. TRAPPIST-1 has a visible \ion{Na}{1} doublet like the field dwarfs, while the low-gravity dwarfs don't display this feature significantly or at all. The depth of TRAPPIST-1's CO lines are similar to the field dwarfs, while the low-gravity dwarfs have shallower CO absorption lines.  Thus in the $K$ band TRAPPIST-1 best matches the field dwarfs. 

\subsubsection{Similar $L_\mathrm{bol}$}
\begin{figure*}
 \gridline{\fig{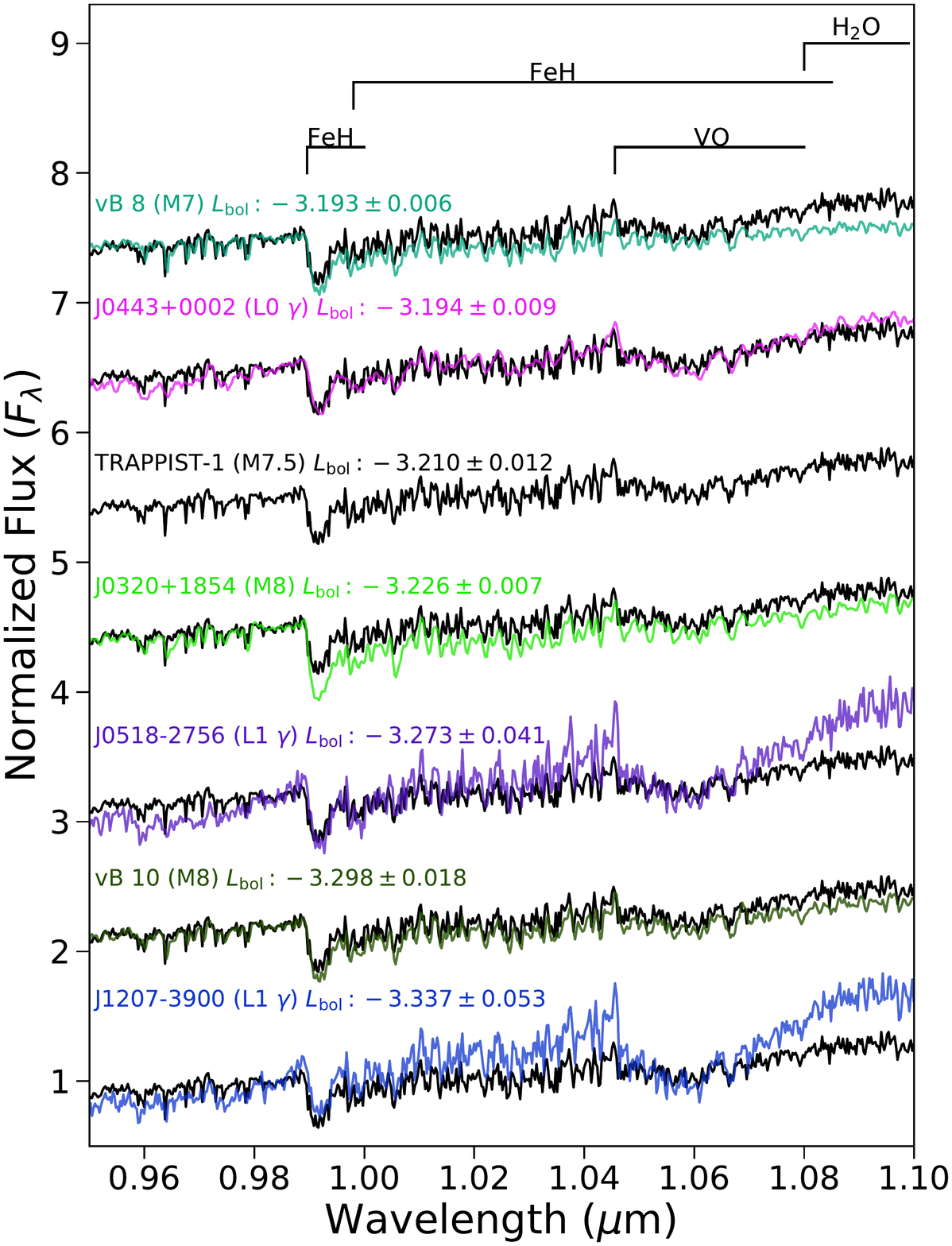}{0.45\textwidth}{\large(a)}
           \fig{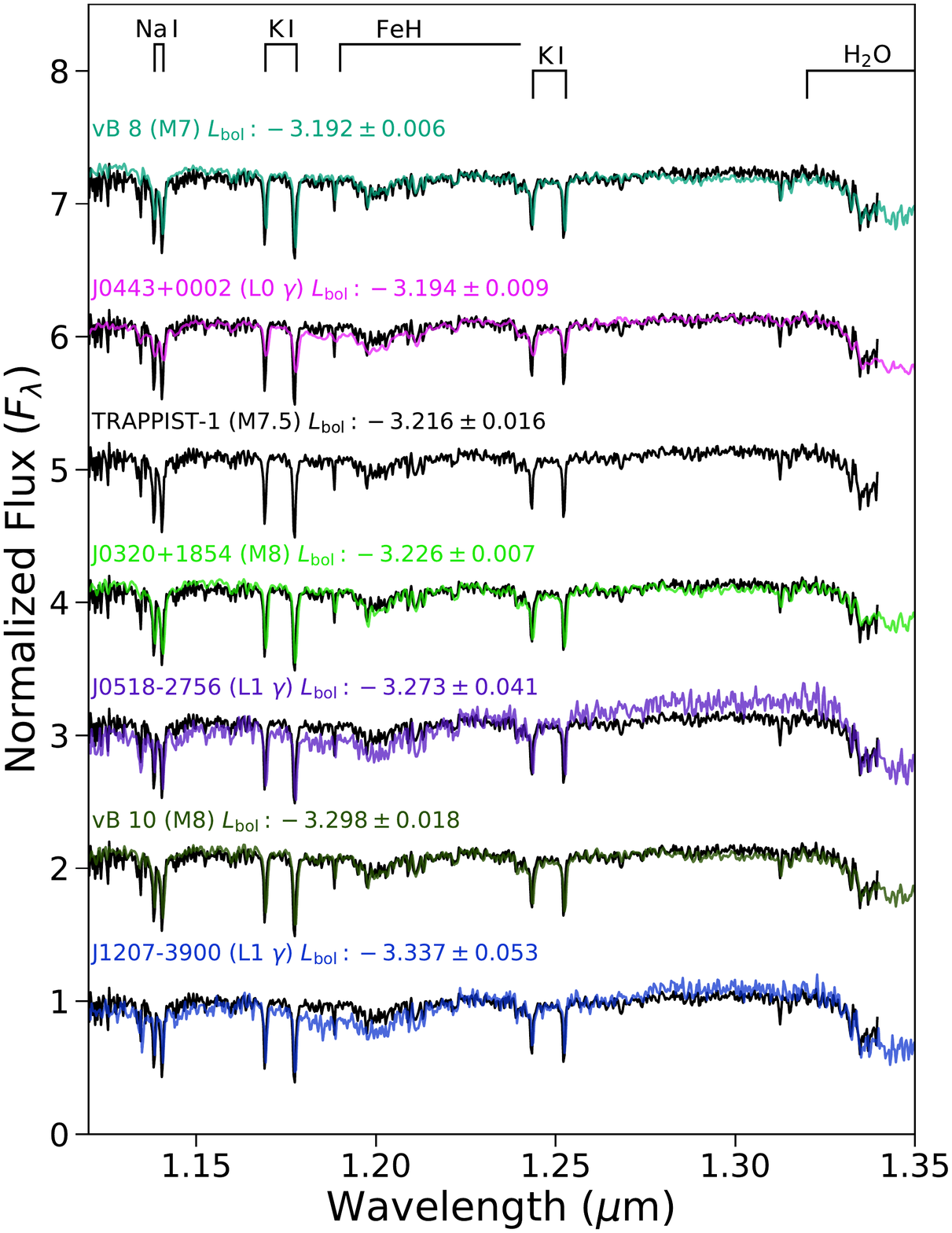}{0.45\textwidth}{\large(b)}} 
 \gridline{\fig{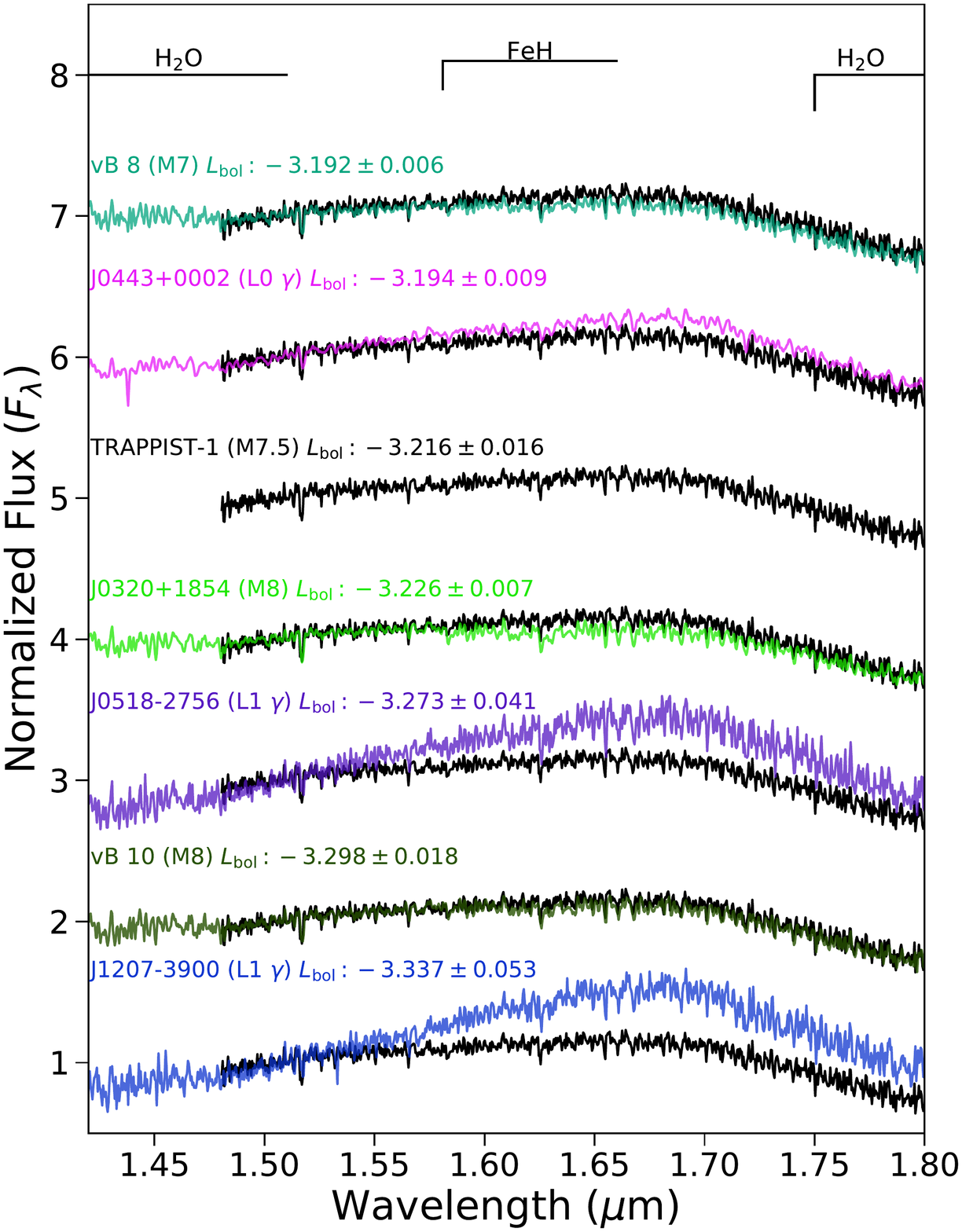}{0.45\textwidth}{\large(c)}
           \fig{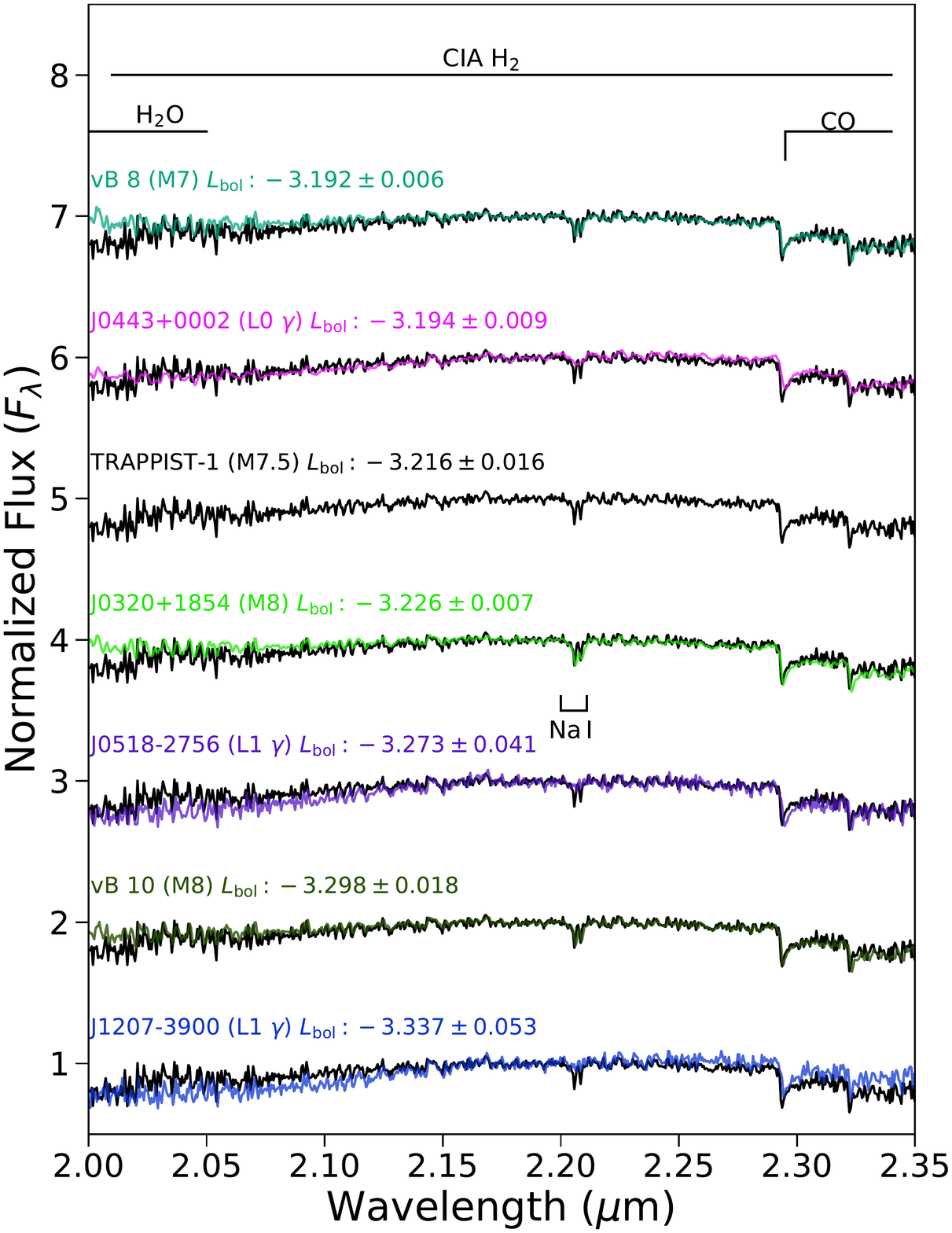}{0.45\textwidth}{\large(d)}}
 \caption{Band-by-band comparison of objects of similar $L_\mathrm{bol}$ with field objects (greens), low-gravity sources (pink, purple, and blue) and TRAPPIST-1 (black). Re-sampling and normalization same as Figure \ref{fig:Teffbandbyband}. (a) $Y$ band.  (b) $J$ band. (c) $H$ band. (d) $K$ band.}
 \label{fig:Lbolbandbyband}
\end{figure*}

In the $Y$ band, Figure \ref{fig:Lbolbandbyband}a, the depth of the Wing-Ford FeH band head for TRAPPIST-1 is similar to all comparative sources except J0320$+$1854 and J1207$-$3900 which are deeper than TRAPPIST-1. The shape of the spectrum carved out by the longer FeH band ($\sim 1-1.04\,\upmu$m) for TRAPPIST-1 is most similar first to J0443$+$0002 and second to the field dwarfs. 

The \ion{Na}{1} and \ion{K}{1} doublets of TRAPPIST-1 appears to be deeper than all comparative objects in Figure \ref{fig:Lbolbandbyband}b. Interestingly, the low-gravity dwarfs J0518$-$2756 and J1207$-$3900 have relativity deep alkali lines for low-gravity sources. The shape of the FeH feature, particularly near $1.20\,\upmu$m, for TRAPPIST-1 is shallower than the low-gravity dwarfs, more similar to the field dwarfs. Between the second \ion{K}{1} and the H$_2$O band we see that J0518$-$2756 and J1207$-$3900 slope slightly redward, while TRAPPIST-1 has a flat slope like the other field objects and J0443$+$0002. Thus in the $J$ band we see TRAPPIST-1 showing a hybrid of features both similar to field and low-gravity dwarfs like J0443$+$0002. 

Figure \ref{fig:Lbolbandbyband}c shows the $H$ band, where the low-gravity dwarfs are more triangular in shape compared to TRAPPIST-1. Again, we see that the $H$ band of TRAPPIST-1 clearly resembles that of a field object. The overall $K$ band shape of TRAPPIST-1, in Figure \ref{fig:Lbolbandbyband}d, is similar to the field dwarfs and J0443$+$0002. TRAPPIST-1 shows \ion{Na}{1} absorption and CO depths matching the field dwarfs.

When compared to sources of similar $T_\mathrm{eff}$ or $L_\mathrm{bol}$, the band-by-band fits show under the assumption of field age TRAPPIST-1 exhibits a blend of field and low-gravity spectral features.


\section{Spectral Analysis For Sample \#2: Assuming an Age of $<0.5$~Gyr for TRAPPIST-1}\label{YoungAgeAssumption}
Despite the above conclusion that the overall SED of TRAPPIST-1 is well fit as a field source, the SpeX SXD, prism, and FIRE spectra all show subtle signatures deviating from normal leading to an intermediate gravity classification. If we assume that TRAPPIST-1 is not a field-aged source due to its $\beta$ gravity classification, then we should treat it as we have other $\beta$ gravity sources and assume an age range of $<0.5$~Gyr. With this age assumption, the $T_\mathrm{eff}$ comparison sample we presented in Section \ref{FieldAgeAssumption} would no longer be suitable. Here we present the new temperature samples of field, young and old age sources compared to TRAPPIST-1 assuming a younger age leading to a cooler temperature and larger radius.

\subsection{Full SEDs}

\begin{figure*}
 \gridline{\fig{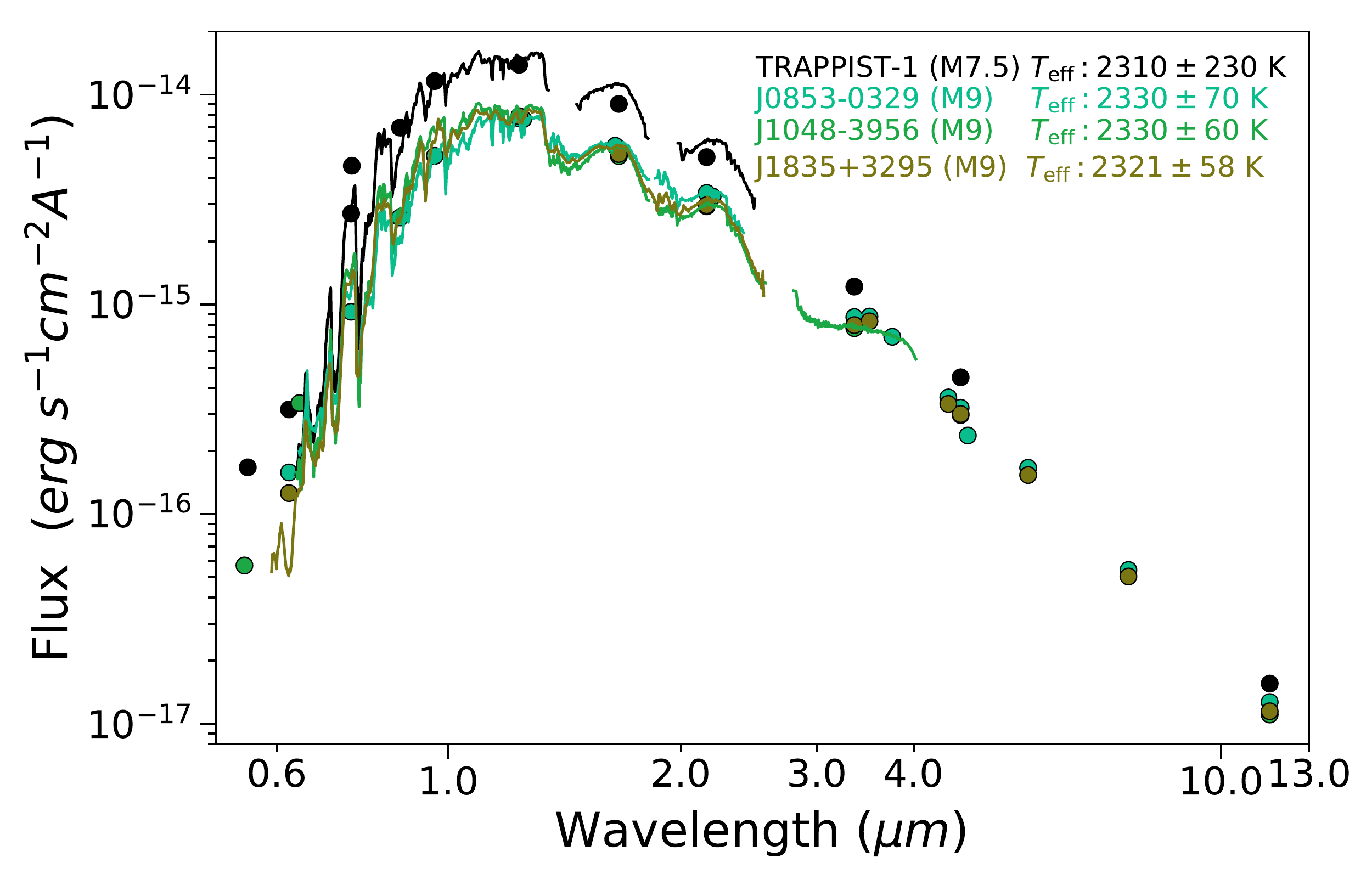}{0.5\textwidth}{\large(a)}
           \fig{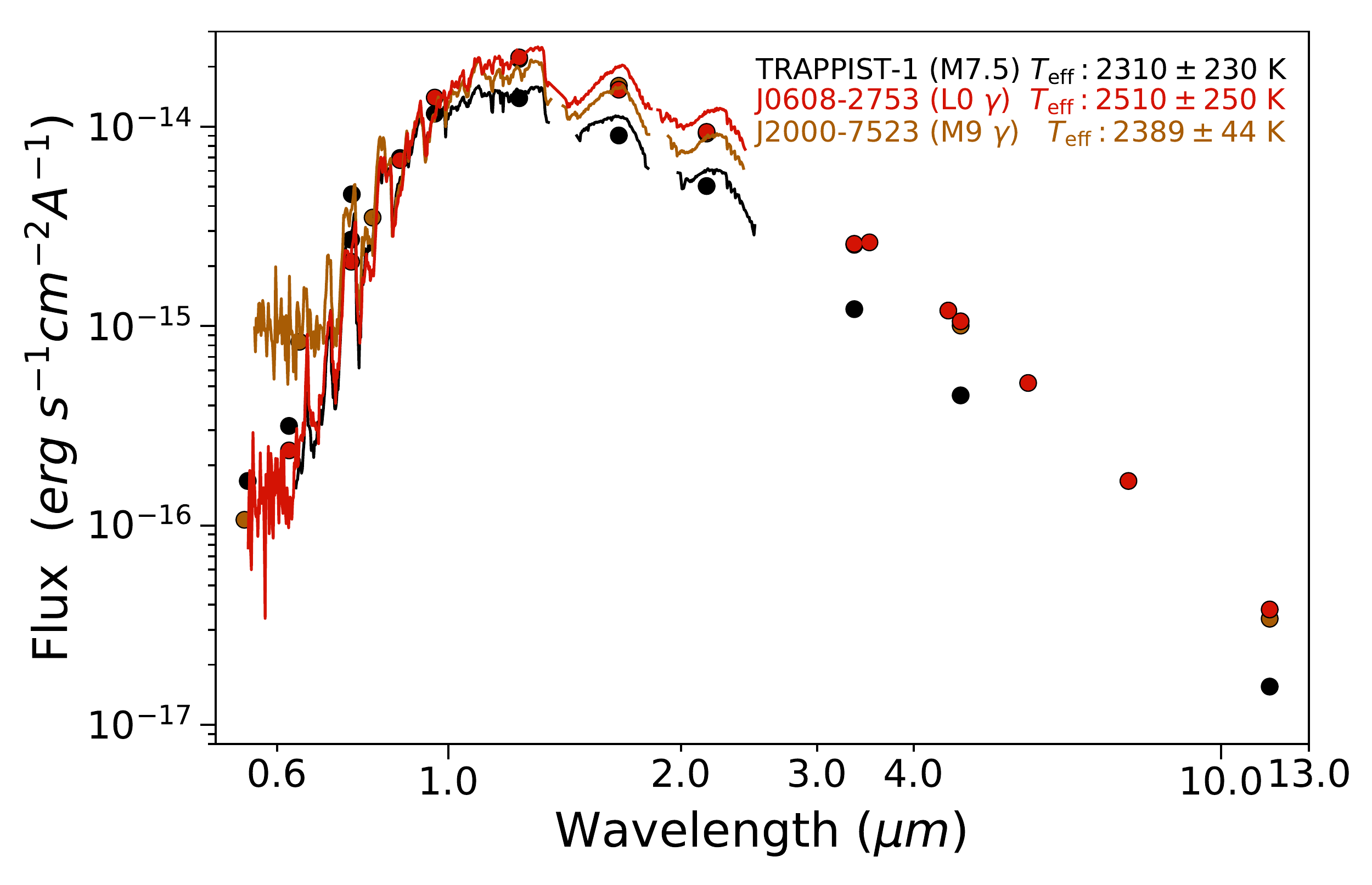}{0.5\textwidth}{\large(b)}} 
 \gridline{\fig{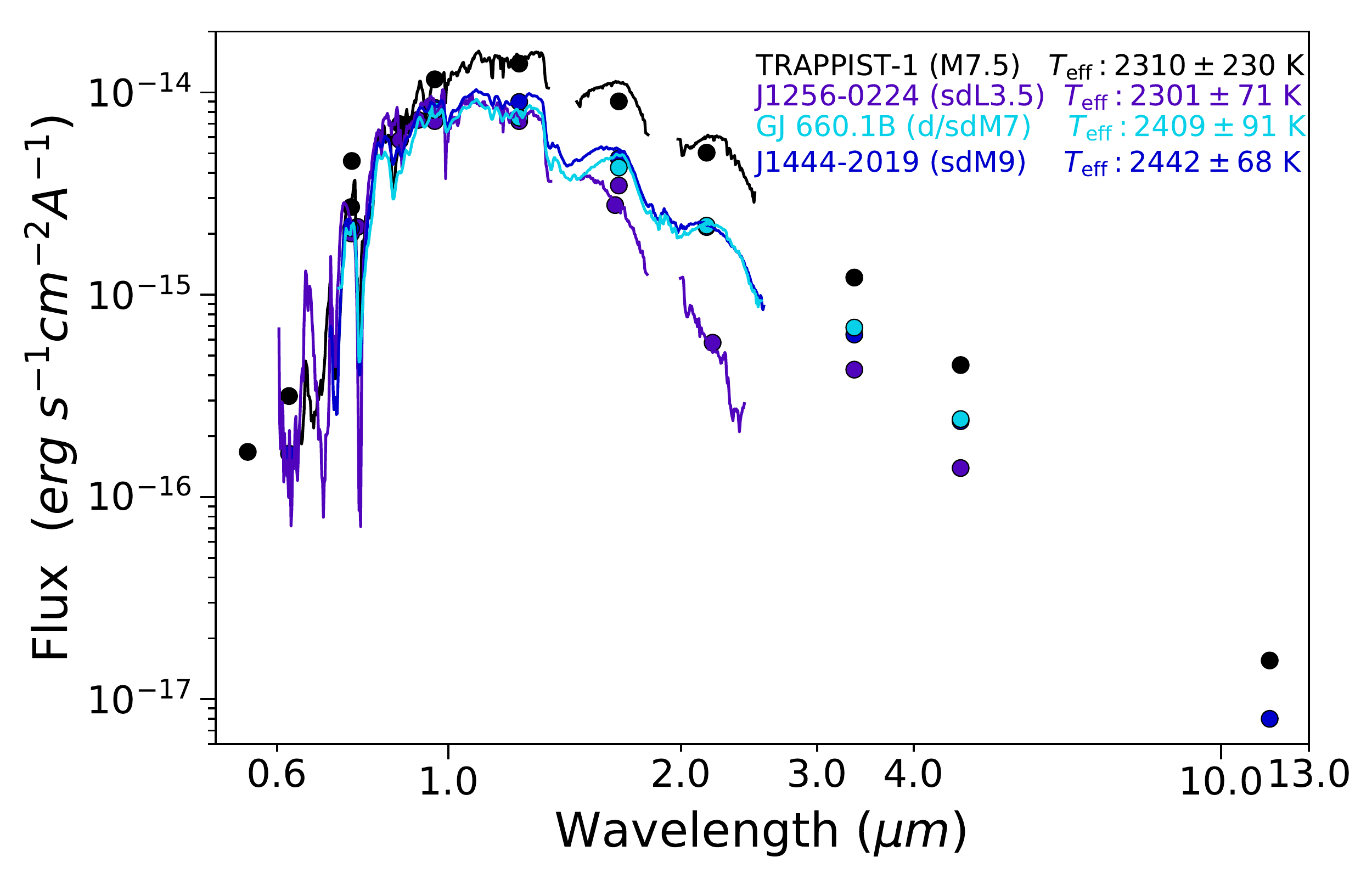}{0.5\textwidth}{\large(c)}}
 \caption{Distance-calibrated SEDs of field dwarfs, low-gravity dwarfs, and subdwarfs of approximately the same effective temperature as TRAPPIST-1 (black) assuming an age of $<500$~Myr. SEDs are displayed as described in Figure \ref{fig:TeffLbolSEDs}. (a) Field dwarfs (various shades of green). (b) Low-gravity dwarfs  (shades of red and orange). (c) Subdwarfs (various shades of blue).}
\label{fig:YoungAgeSample}
\vspace{0.5cm}
\end{figure*}

Figures \ref{fig:YoungAgeSample}a, b, and c compare TRAPPIST-1 to field-age, low-gravity, and subdwarfs of similar temperatures over the $0.5-13\,\upmu$m range. Compared to the field-age sources in Figure \ref{fig:YoungAgeSample}a, TRAPPIST-1 is brighter than all sources across the entire range, having only a similar brightness from $0.63-0.72\,\upmu$m. In Figure \ref{fig:YoungAgeSample}b, TRAPPIST-1 overlaps with the comparative sample from $\sim0.6-1\,\upmu$m, however beyond $1\,\upmu$m TRAPPIST-1 is fainter. This trend is similar to what is seen in Figure \ref{fig:TeffLbolSEDs}b, showing no matter the assumed age TRAPPIST-1 does not resemble the very-low-gravity sources. TRAPPIST-1 overlaps with the comparative sample in Figure \ref{fig:YoungAgeSample}c from $\sim0.6-1\,\upmu$m, however spectral features in this region are not well fit by the subdwarfs. Beyond $\sim1\,\upmu$m, TRAPPIST-1 is brighter than the comparative sample, however it displays a more triangular $H$ band similar to GJ 660.1B. Again, as seen with the low-gravity sources, no matter the assumed age range TRAPPIST-1's overall SED is poorly fit by subdwarfs. 

\subsection{Band-by-Band Analysis}\label{bandbybandYoung}
\begin{figure*}
 \gridline{\fig{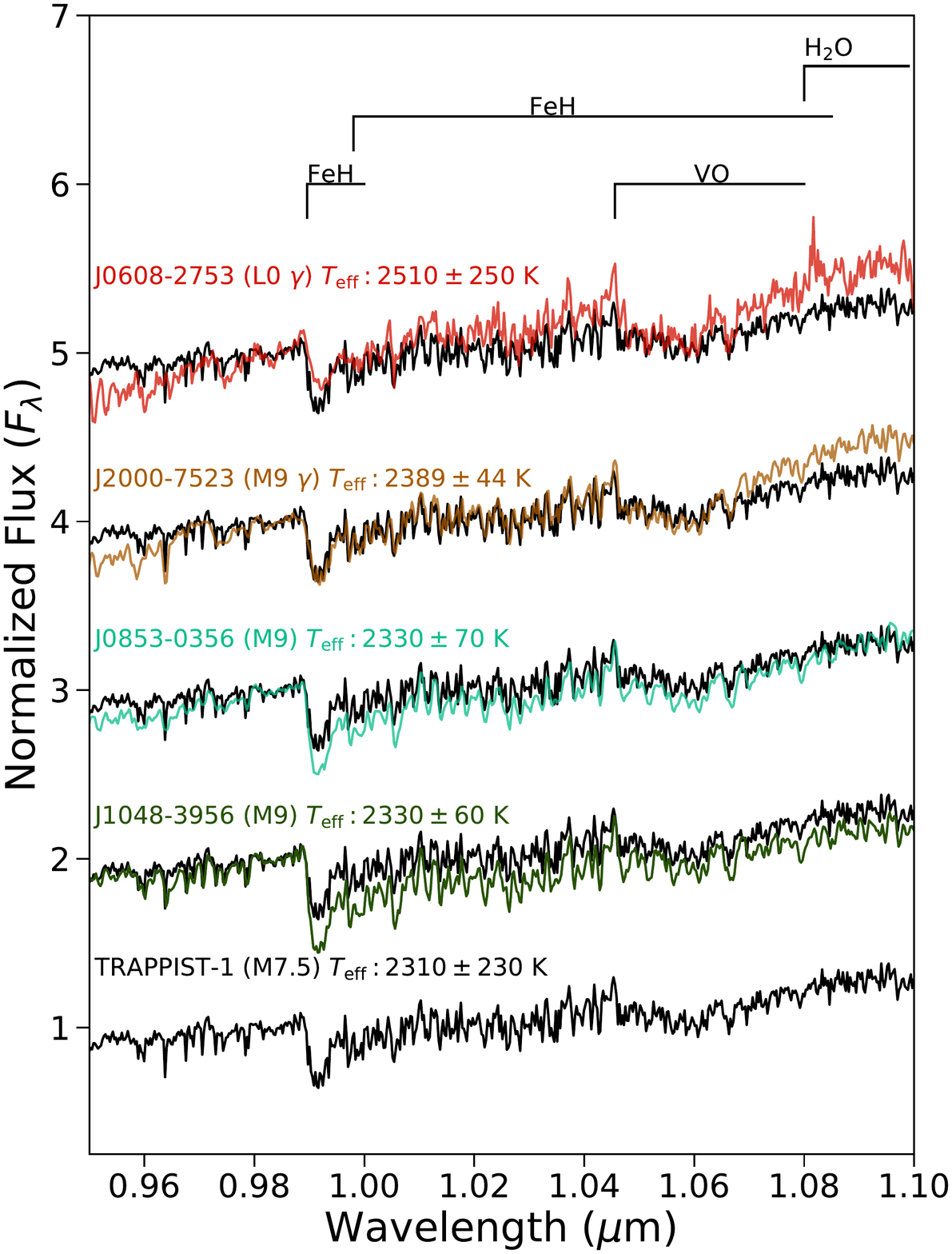}{0.45\textwidth}{\large(a)}
           \fig{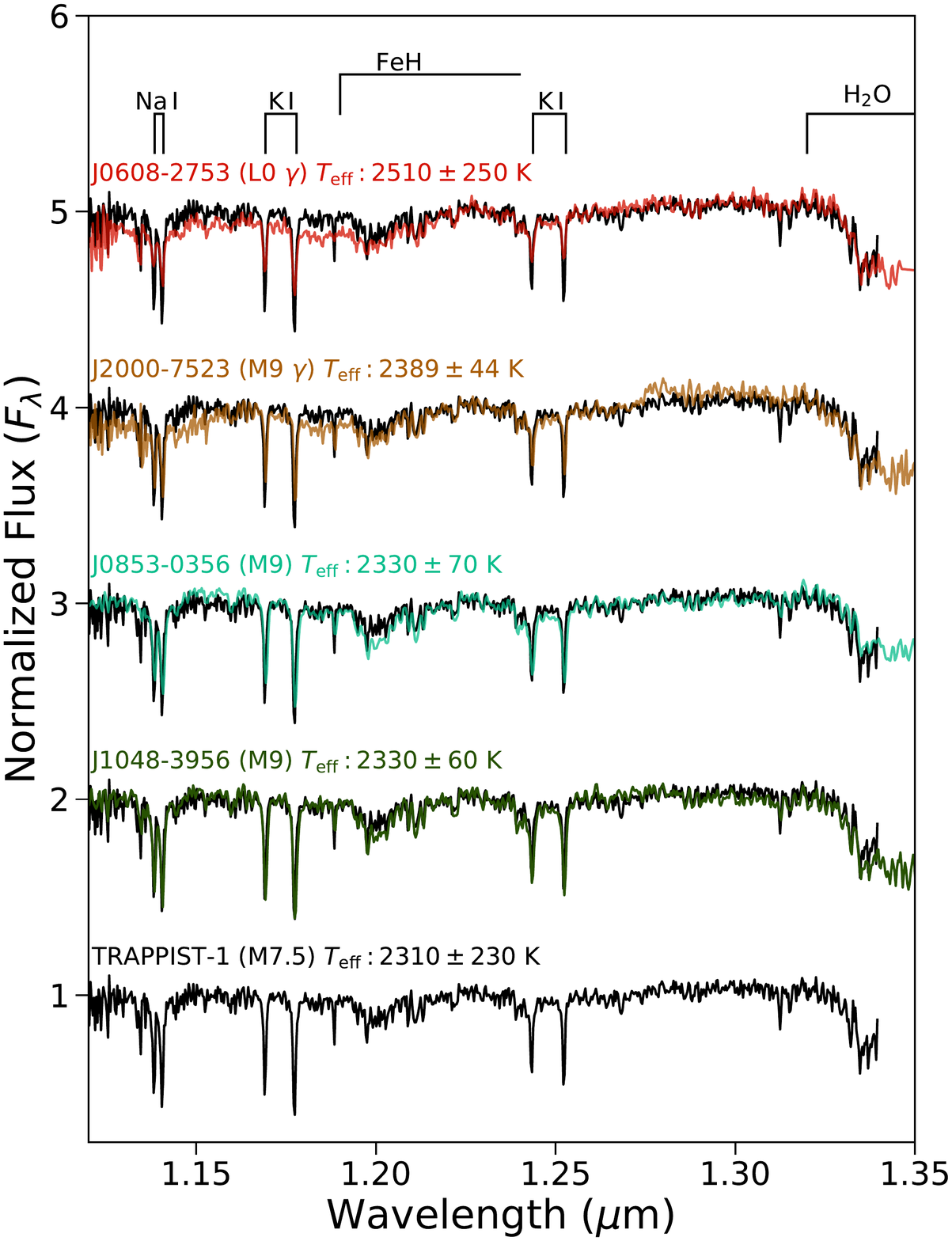}{0.45\textwidth}{\large(b)}} 
 \gridline{\fig{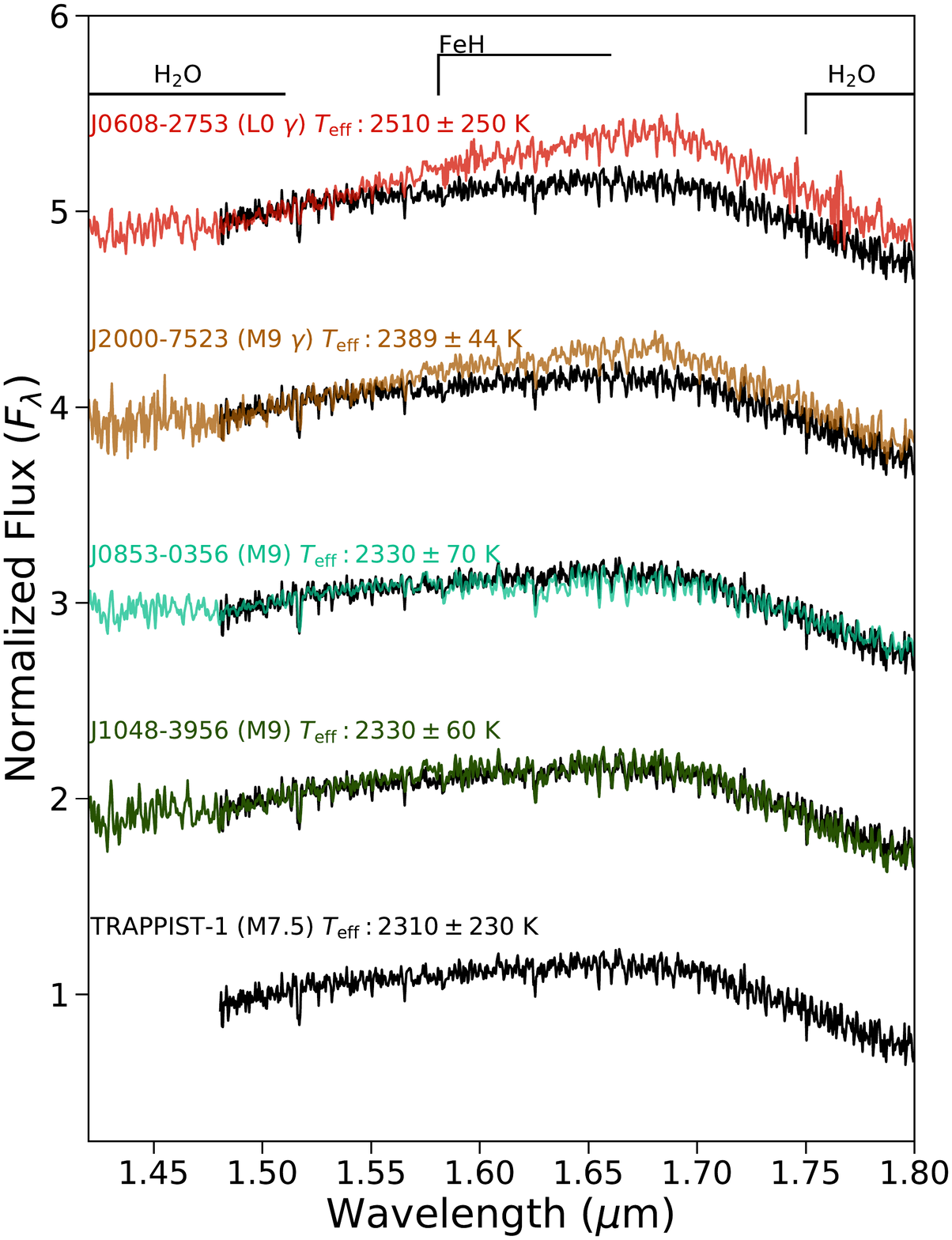}{0.45\textwidth}{\large(c)}
           \fig{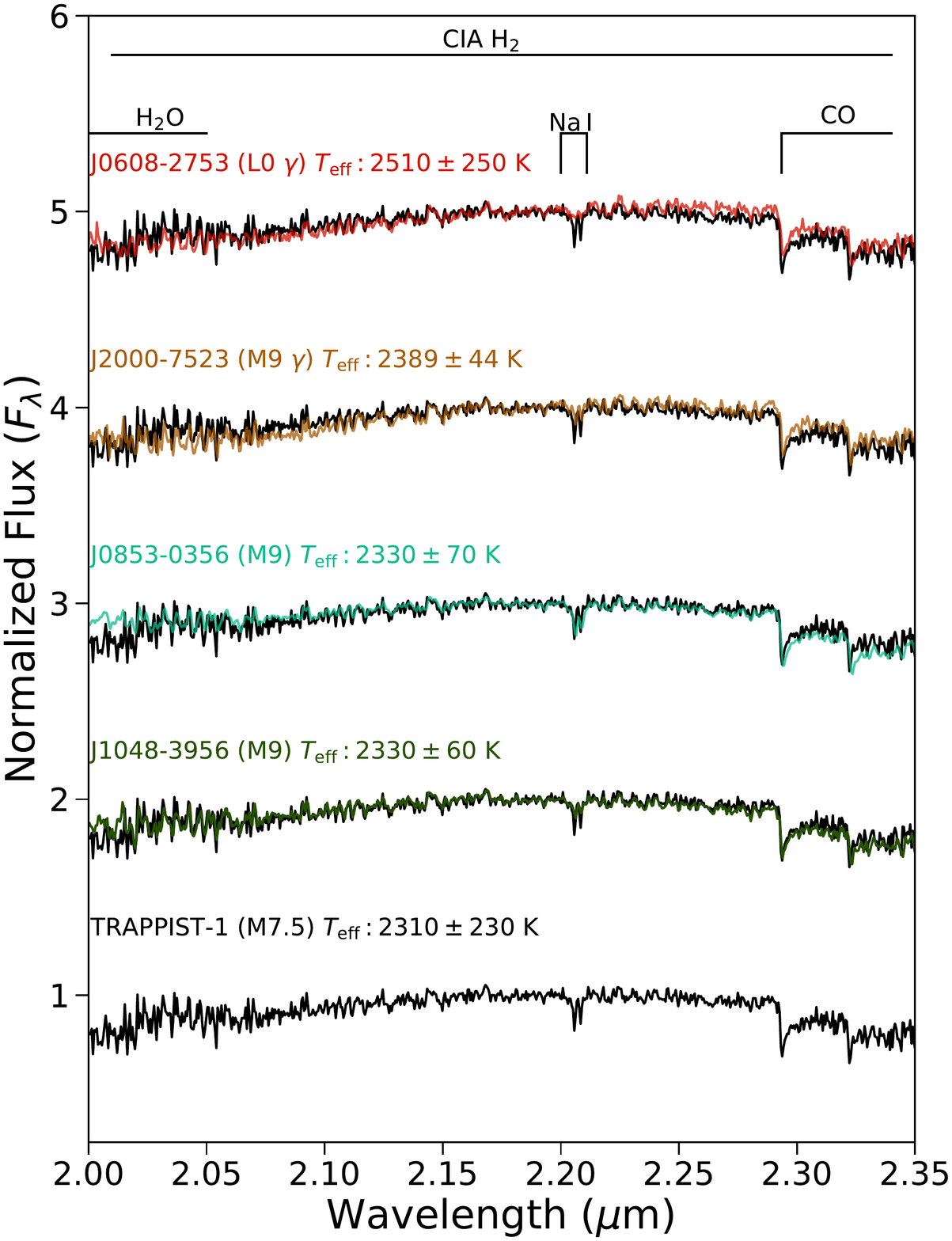}{0.45\textwidth}{\large(d)}}
 \caption{Band-by-band comparison of field-age (greens) and low-gravity (red/brown) sources of similar $T_\mathrm{eff}$ to assuming a younger age for TRAPPIST-1 (black). Re-sampling and normalization same as Figure \ref{fig:Teffbandbyband}. (a) $Y$ band. (b) $J$ band.  (c) $H$ band.  (d) $K$ band.}
\label{fig:YoungAgeBands}
\vspace{0.5cm}
\end{figure*}

Assuming a younger age for TRAPPIST-1 we take a closer look at the comparative sample described in detail in Tables \ref{tab:Sample} and \ref{tab:SampleFunParams}. At the younger age, TRAPPIST-1 is 319~K cooler hence the sample changes from an assumed field age. All panels in Figure \ref{fig:YoungAgeBands} show how TRAPPIST-1 fits to field and low-gravity equivalent sources in the $Y$, $J$, $H$, and $K$ bands. 

In Figure \ref{fig:YoungAgeBands}a TRAPPIST-1 is best fit by the field dwarf DENIS--P J1048.0$-$3956 (hereafter J1048$-$3956), from $0.95-0.99\,\upmu$m, while from $0.99-1.06\,\upmu$m TRAPPIST-1 is best fit by low-gravity dwarf 2MASS J20004841$-$7523070 (hereafter J2000$-$7523). The majority of the $Y$ band is poorly fit by both comparative field dwarfs and the low-gravity dwarf J0608$-$2753. While TRAPPIST-1 has regions that are matched well to J1048$-$3956 and J2000$-$7523, the field- and low-gravity dwarfs when assuming an older age more closely match TRAPPIST-1's spectrum in the $Y$ band (see Section \ref{bandbybandField}).  

TRAPPIST-1 shares alkali depth and spectral shape features most similar to the field dwarfs in Figure \ref{fig:YoungAgeBands}b, similar $H$-band shape as the field dwarfs in \ref{fig:YoungAgeBands}c, and similar Na\,I depth to J0853$-$0356 and overall $K$-band shape to both field dwarfs in \ref{fig:YoungAgeBands}d. These band-by-band similarities to the field dwarfs clearly anchor the classification of TRAPPIST-1 as a field aged source.

\section{Sample \#3: Could TRAPPIST-1's spectral features be low metallicity mimicing low gravity?}\label{subdwarfNaIKIcomps} 

\begin{figure*}
 \gridline{\fig{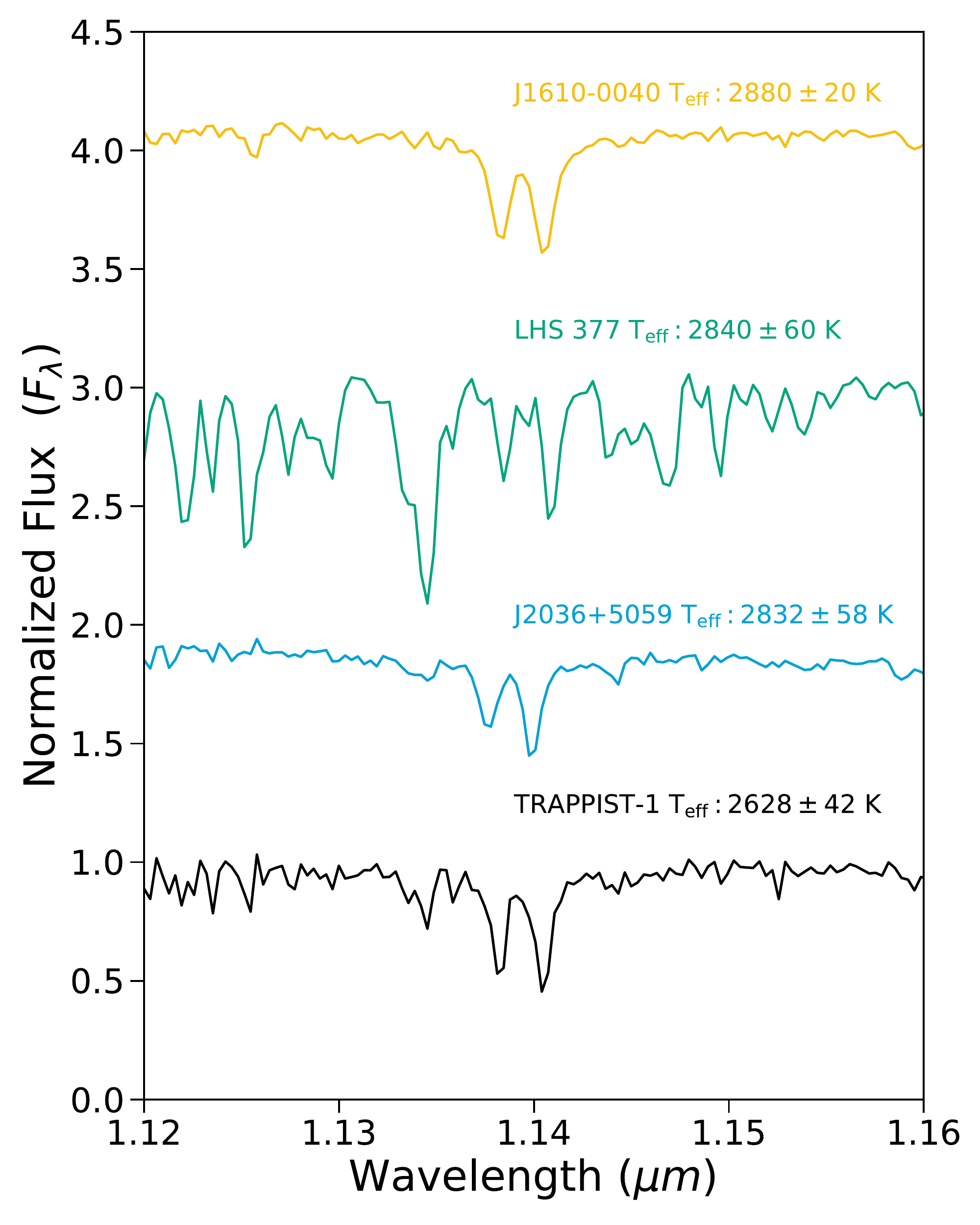}{0.5\textwidth}{\large(a)}
           \fig{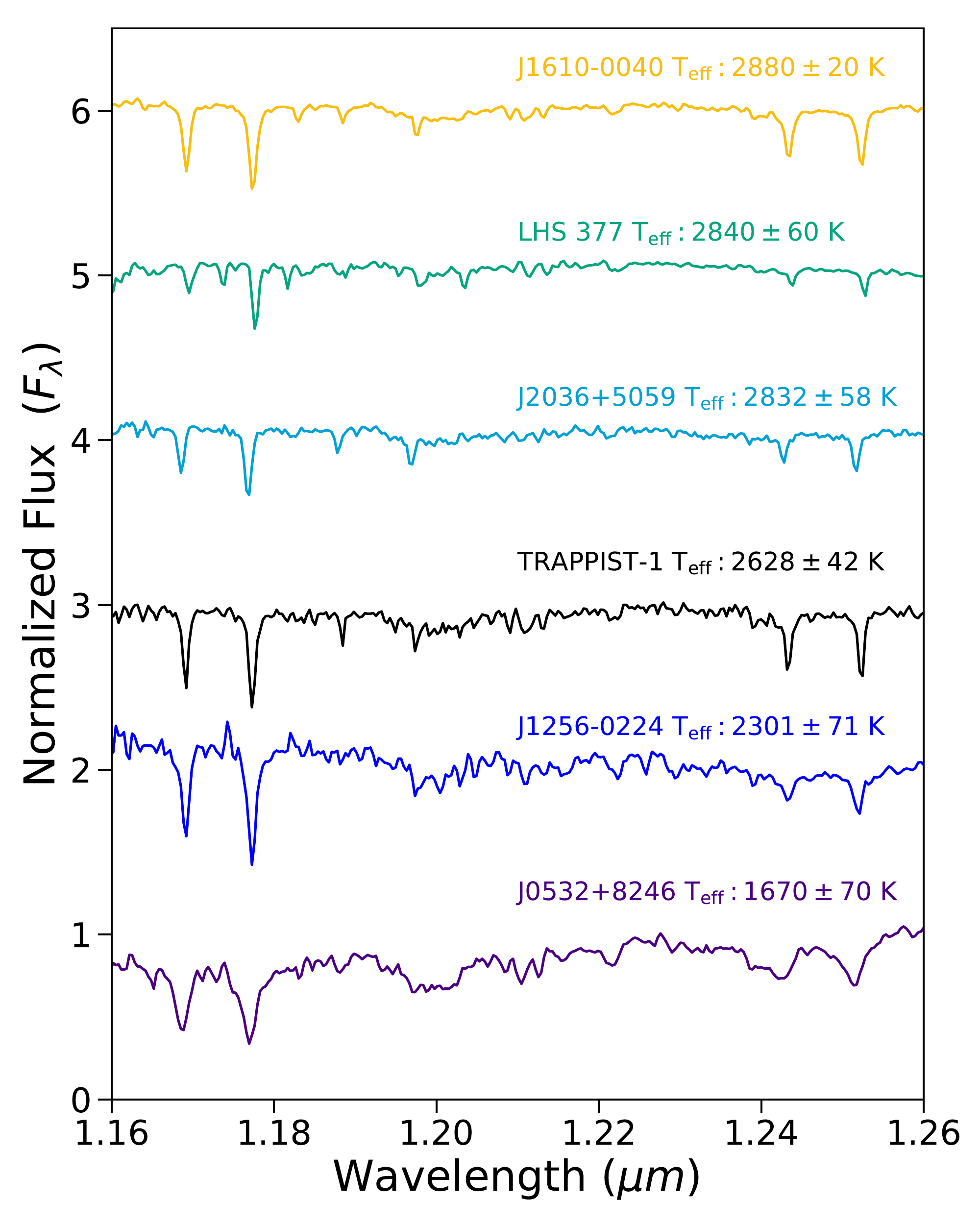}{0.5\textwidth}{\large(b)}} 
 \caption{Comparison of gravity sensitive spectral lines of \ion{Na}{1} and \ion{K}{1} in the $J$ band of TRAPPIST-1 to subdwarfs with medium resolution data from \cite{Gonz18}. All spectra were resampled to the same dispersion relation using a wavelength-dependent Gaussian convolution. Spectra are normalized by the average flux taken across $1.29-1.31\, \upmu$m. (a) $1.14\, \upmu$m \ion{Na}{1} doublet.  (b) 1.17 and 1.25 $\upmu$m \ion{K}{1} doublets}
 \label{fig:SubdwarfJband}
\end{figure*}

The low-metallicity d/sd GJ 660.1B discussed in \cite{Agan16} also showed signatures of youth similar to TRAPPIST-1 with its triangular $H$ band. Previous work by \cite{Kirk10} also saw this feature in spectra of late-M dwarfs with high proper motion and with no evidence of youth. \cite{Agan16} measured the \cite{Alle13} indices for GJ 660.1B assuming two different spectral types, M7 and M9.5. When typed as an M9.5, two of the indices received intermediate gravity scores, while when typed as M7 (the final decided upon spectral type) all scores were field gravity. Thus \cite{Agan16} (and references therein) state that additional opacity from the $1.55-1.6\, \upmu$m FeH absorption band and stronger H$_2$O, due to reduced condensate opacity of low-metallicity subdwarfs, may help shape the $H$-band continuum of mild subdwarfs and therefore potentially skew gravity index-based classifications. 

Since the $H$-band shape of TRAPPIST-1 is similar to GJ 660.1B, we investigate if the intermediate gravity classification could be due to a low metallicity. However, there is no medium-resolution spectral data of GJ 660.1B or J1013$-$1356 and J1444$-$2019, two of the subdwarfs from the field and younger age assumption samples, therefore we instead compare the $J$-band \ion{Na}{1} and \ion{K}{1} doublets of TRAPPIST-1 to subdwarfs with medium resolution spectra from \cite{Gonz18}. Figure \ref{fig:SubdwarfJband} displays these sources in a decreasing effective temperature sequence. Both lines in the \ion{Na}{1} doublet of TRAPPIST-1 in Figure \ref{fig:SubdwarfJband}a appear to be of similar depth, unlike the subdwarfs, and are deeper than the subdwarf doublets. In Figure \ref{fig:SubdwarfJband}b the $1.17\,\upmu$m \ion{K}{1} doublet of TRAPPIST-1 is narrower than the subdwarfs and has a depth in between that of the binary LSR J1610$-$0040 and SDSS J125637.13$-$022452.4 (hereafter J1256$-$0224), while the $1.25\,\upmu$m \ion{K}{1} doublet is deeper and narrower than the subdwarfs. Thus the $J$-band alkali lines of TRAPPIST-1 are not similar to those of subdwarfs. In conclusion while low metallicity may mimic some low-gravity features, we find no evidence for that in the case of TRAPPIST-1.

\section{Final Thoughts on TRAPPIST-1's Age}\label{FinalThoughts}
From comparing the overall SEDs in the various samples, we see that TRAPPIST-1 most resembles those of the field dwarfs no matter what age we assume. When comparing the NIR band-by-band fits, TRAPPIST-1 is most similar to the field sources in some areas, while it has some similarities to aspects of the low-gravity sources. From our examination with the subdwarfs in the $J$ band, we see no evidence for the low-gravity spectral features to be caused by low-metallicity. Thus we conclude that the spectral features of TRAPPIST-1 are a blend of the field dwarfs of equivalent $T_\mathrm{eff}$ and $L_\mathrm{bol}$ and that of low-gravity sources of similar $L_\mathrm{bol}$. TRAPPIST-1 is likely a field age source with these spectral features originating from some cause other than youth.

\section{Examining LHS 132 as an ideal comparative source}\label{LHS132}

\begin{figure*}
\centering
 \includegraphics[scale=.57]{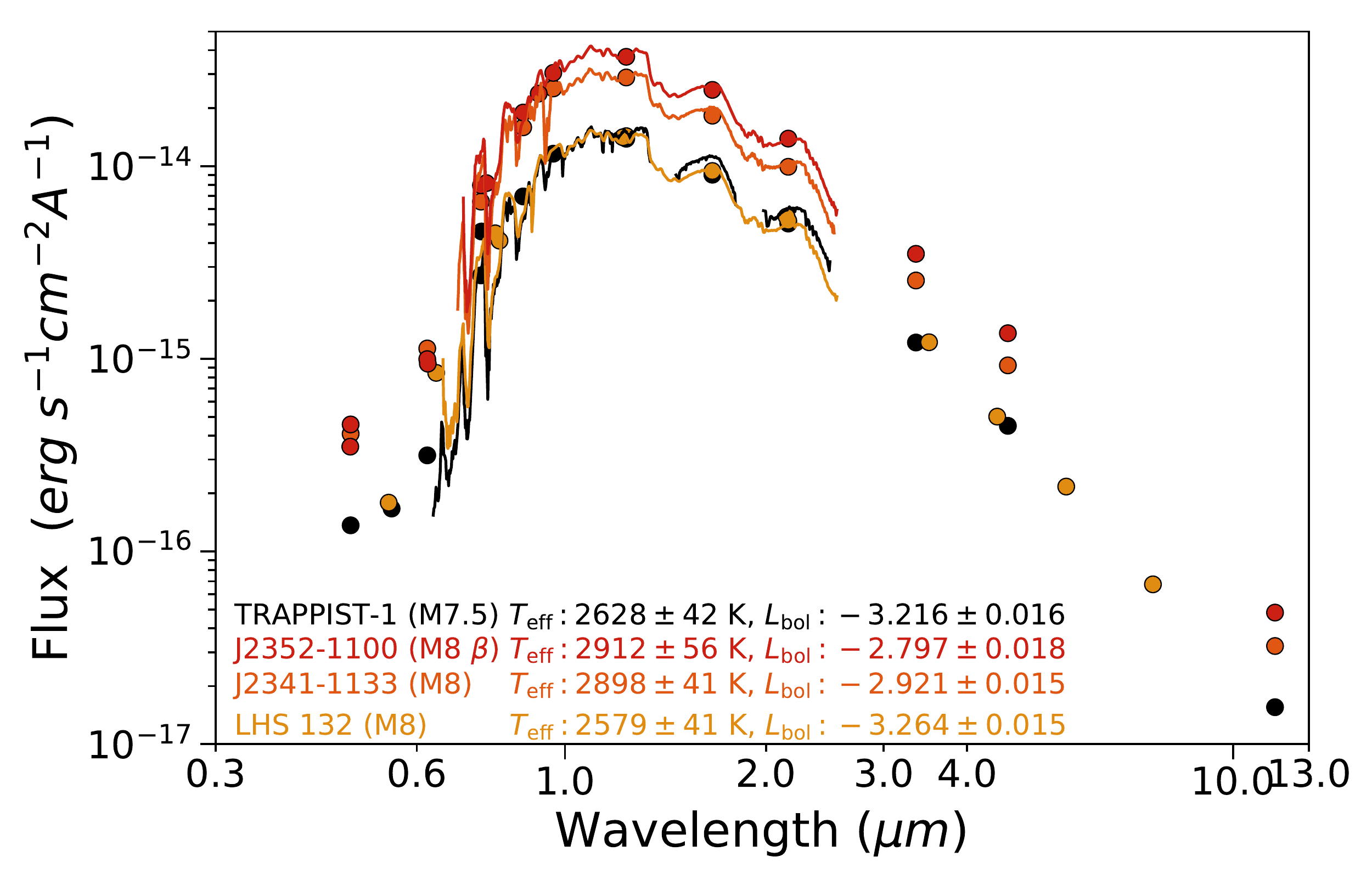}
\caption{Distance-calibrated SEDs of the comparative sources from \cite{Burg17} (colors) compared to  TRAPPIST-1 (black). All spectra were resampled to the same dispersion relation using a wavelength-dependent Gaussian convolution. SEDs are displayed as described in Figure \ref{fig:TeffLbolSEDs}.}
\label{fig:RebuttalSED}
\vspace{0.5cm}
\end{figure*}

When determining the age of TRAPPIST-1 \cite{Burg17} used the M7 dwarf 2MASS J23520507$-$1100435 (hereafter J2352$-$1100) and the M8 dwarfs LHS 132 and 2MASS 23412868$-$1133356 (hereafter J2341$-$1133) to constrain the age based on surface gravity features. J2352$-$1100 is a member of the $130-200$ Myr old \citep{Bell15} AB Doradus moving group (Faherty et al. in prep.), while both LHS 132 and J2341$-$1133 are field age sources \citep{Fili15,Cruz07}. We have created full SEDs of these sources to compare to TRAPPIST-1 in more detail than done in \cite{Burg17}. 

In Figure \ref{fig:RebuttalSED} we see that LHS 132 fits the overall SED shape of TRAPPIST-1 very well, while J2352$-$1100 and J2341$-$1133 both have significantly more flux over the entire region. LHS 132 has a similar $T_\mathrm{eff}$ and $L_\mathrm{bol}$ as TRAPPIST-1, while J2352$-$1100 and J2341$-$1133 are much hotter, brighter, and have larger radii. Looking at the SEDs alone, J2352$-$1100 and J2341$-$1133 are both poor comparison sources to TRAPPIST-1 since these objects are fundamentally different, while LHS 132 is still a fair comparative source. LHS 132 also receives an intermediate gravity classification which is discussed further in Section \ref{AL13Indicescomparison}. Therefore, LHS 132 is a target to explore in more detail to see if other aspects match those of TRAPPIST-1.

\section{Discussion}\label{Discussion}
\subsection{Comparison of the \cite{Alle13} Gravity Indices}\label{AL13Indicescomparison}

Because of the intermediate gravity (INT-G or $\beta$) classification that \cite{Burg17} found for TRAPPIST-1 when using the \cite{Alle13} gravity-sensitive indices, we calculated the indices for our entire comparative sample using our modified version of the \texttt{ALLERS13\_INDEX} IDL code on both low- and medium-resolution spectra when available. The index values and final gravity scores are listed in Tables \ref{tab:gravityindiceslow} and \ref{tab:gravityindicesmed}. The equivalent width measurements for the medium resolution gravity score are listed in Table \ref{tab:eqw}. All objects receive the same gravity class using both the low- and medium-resolution indices, with the exception of vB 10 which receives a FLD-G classification with low-resolution, but a $\beta$ classification with medium resolution due to the FeH$_J$ score. We note that while our FeH$_J$ score is a 2, the FeH$_J$ feature does not appear to look different from the other field sources in our sample. The FeH$_J$ score from \cite{Martin17} was a 0, therefore we may be getting a spurious measurement. 

The best fit source from the \cite{Burg17} sample, LHS 132, also receives an INT-G gravity classification. LHS 132 and TRAPPIST-1 have similar radii, mass, log $g$, $T_\mathrm{eff}$, and $L_\mathrm{bol}$ as well as the same scores for each of the gravity indices. Therefore, whatever physical factor is causing TRAPPIST-1 to receive an $\beta$ classification may also be the same for LHS 132.

All subdwarfs receive a $\beta$ gravity class, with the exception of the spectroscopic binary J1610$-$0400, which received a FLD-G classification, and LHS377 which did not receive a gravity class due to a lack of measurement for the FEH$_z$ index. Therefore as stated in \cite{Alle13ConfP, Agan16, Burg17, Martin17} there is some aspect of the spectrum that fools the indices into classifying objects with older ages as $\beta$, which could be due to metallicity. However, as stated in \cite{Burg17} and from our comparison to subdwarfs, TRAPPIST-1 is not low metallicity nor does it show $J$-band features similar to subdwarfs and thus this is unlikely to be the cause of the $\beta$ classification for TRAPPIST-1. 

\begin{figure}
\gridline{\hspace{-0.1cm}\fig{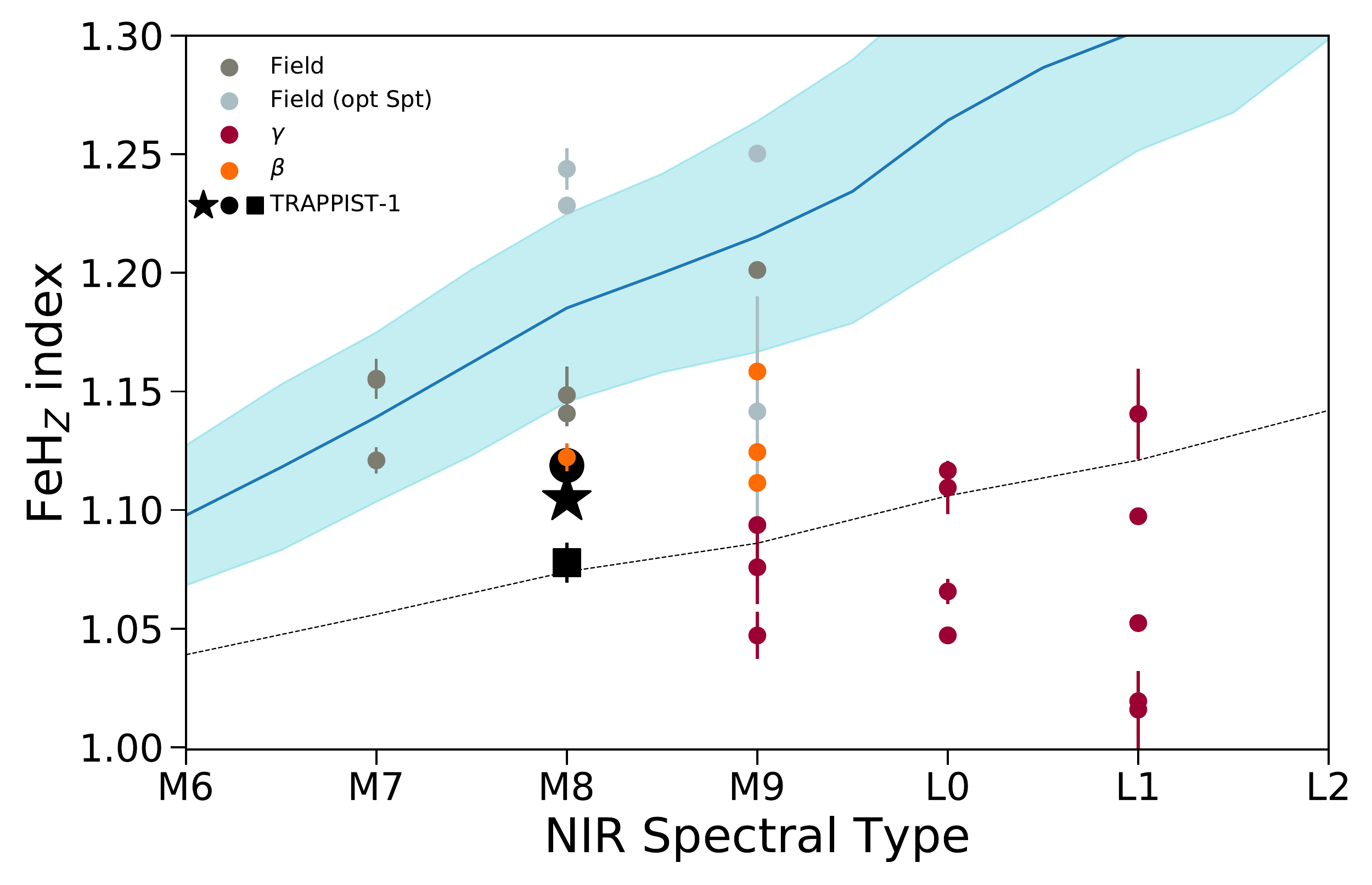}{0.48\textwidth}{\large(a)}}
\gridline{\hspace{-0.1cm}\fig{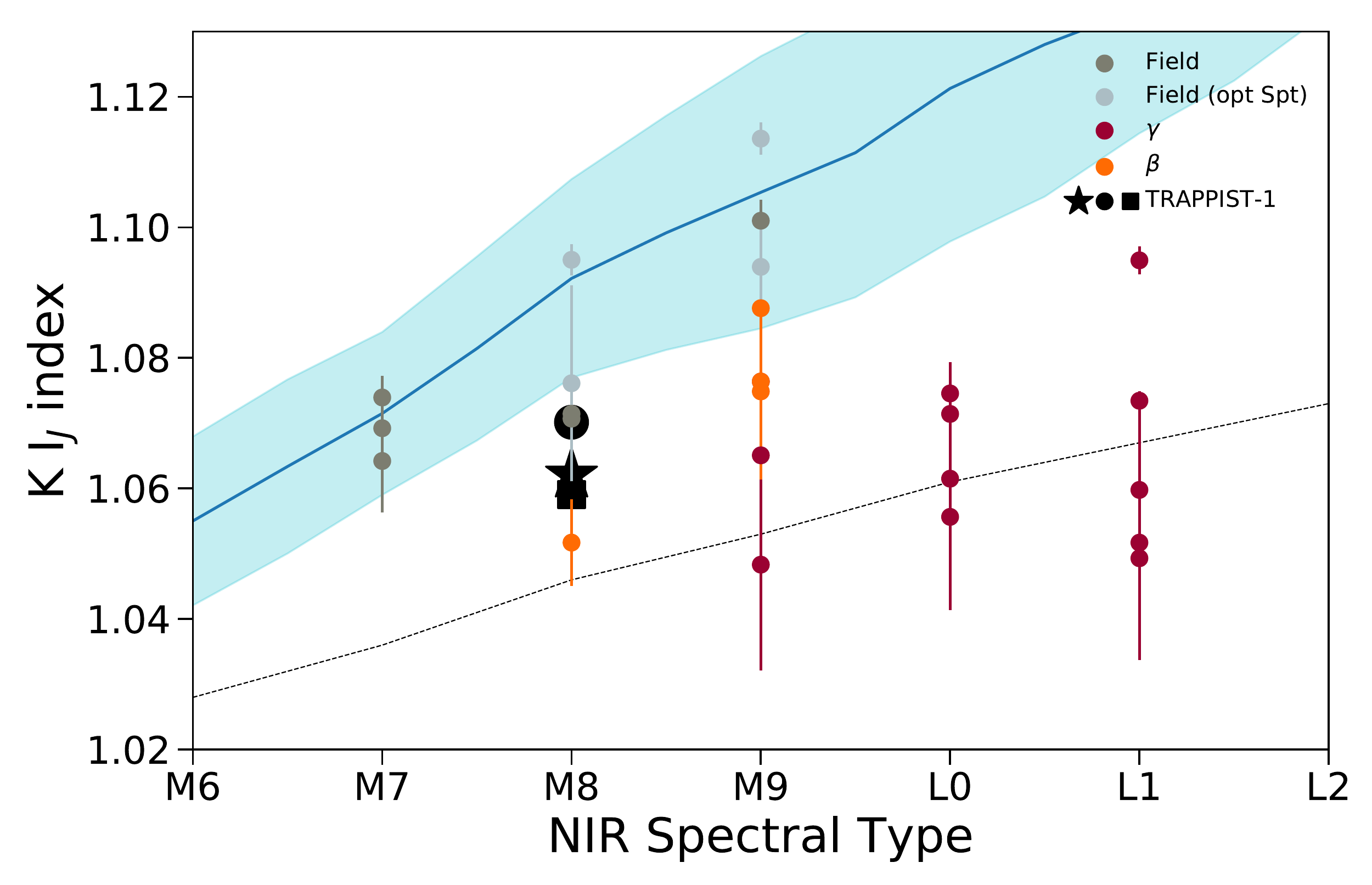}{0.48\textwidth}{\large(b)}}
\gridline{\hspace{-0.1cm}\fig{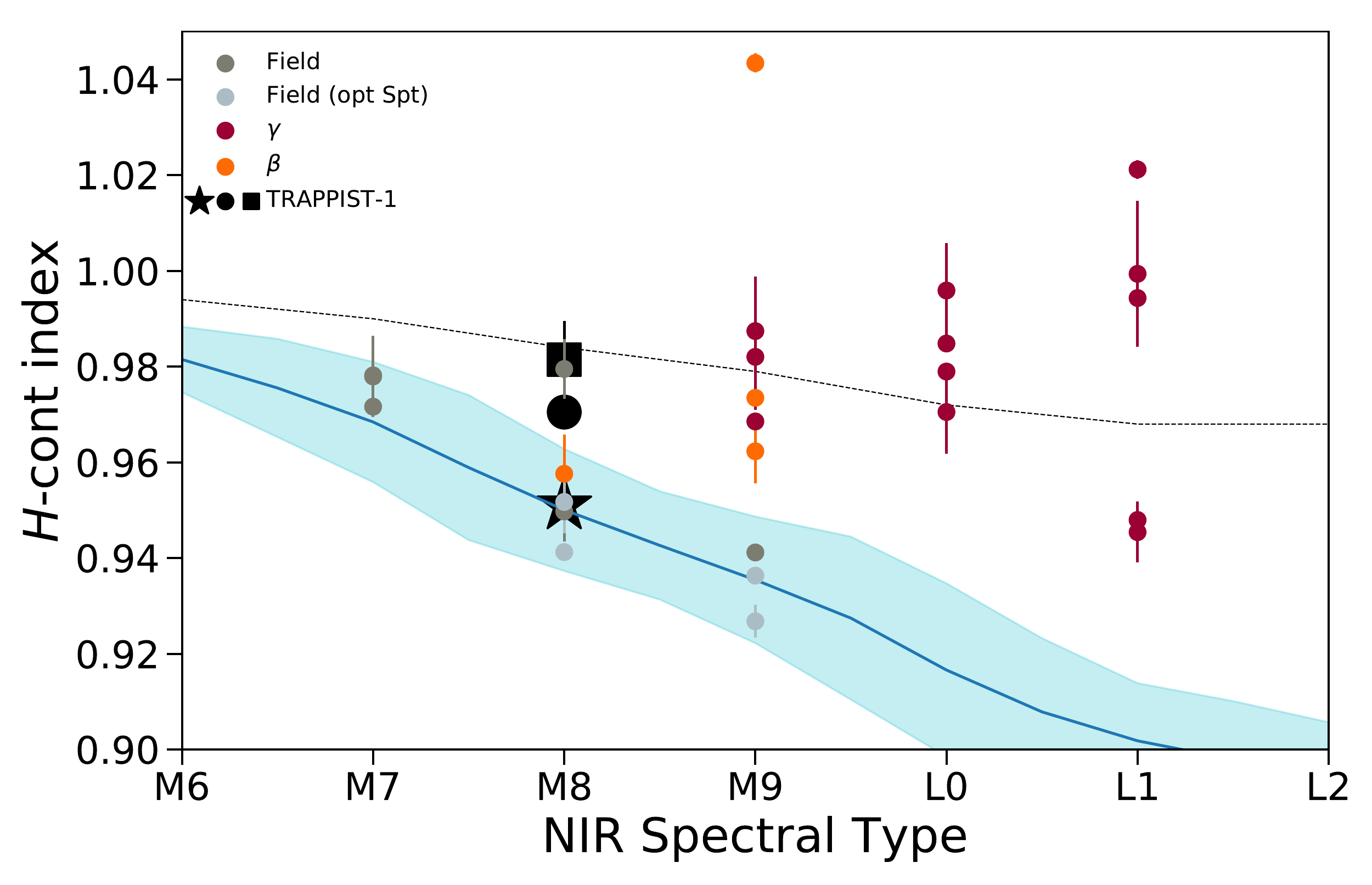}{0.48\textwidth}{\large(c)}}
\caption{\cite{Alle13} gravity-sensitive indices versus near-infrared spectral type. The field dwarfs polynomial is shown blue with the $1 \sigma$ uncertainties in aqua. The dashed line shows the boundary between a score of 1 and 2 for the \cite{Alle13} system. Sources from our comparative sample are shown as follows: field (gray), intermediate gravity (orange), low gravity (red). TRAPPIST-1 is displayed in black with the various symbols corresponding to the  different spectra, SpeX prism (square), SpeX SXD (circle), and FIRE (star). (a) NIR Spectral Type vs FeH$_z$ index (b) NIR Spectral Type vs \ion{K}{1}$_J$ index (c) NIR Spectral Type vs $H$-cont index}
\label{fig:Indices}
\end{figure}

\begin{figure}
\gridline{\hspace{-0.1cm}\fig{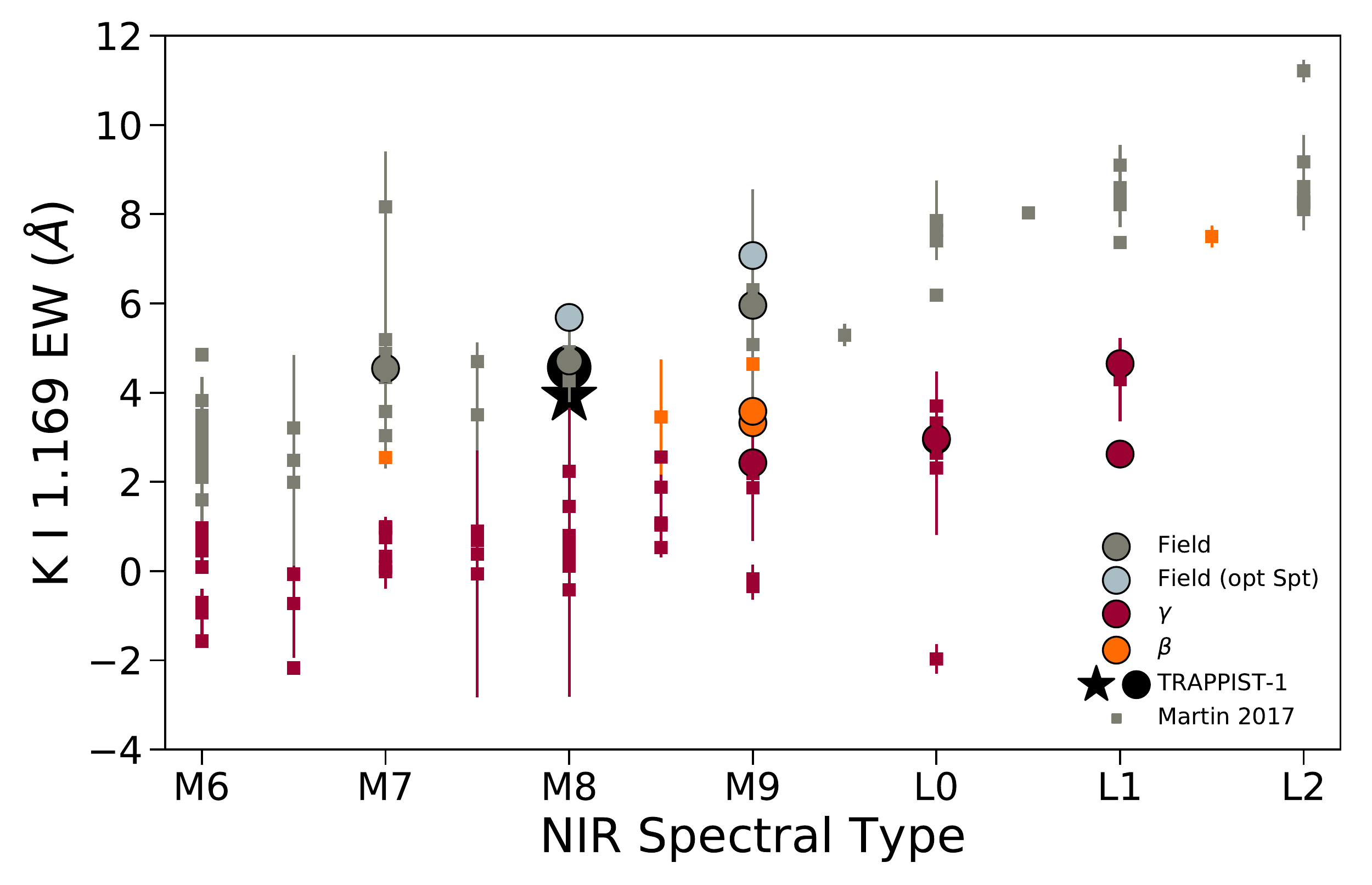}{0.48\textwidth}{\large(a)}}
\gridline{\hspace{-0.1cm}\fig{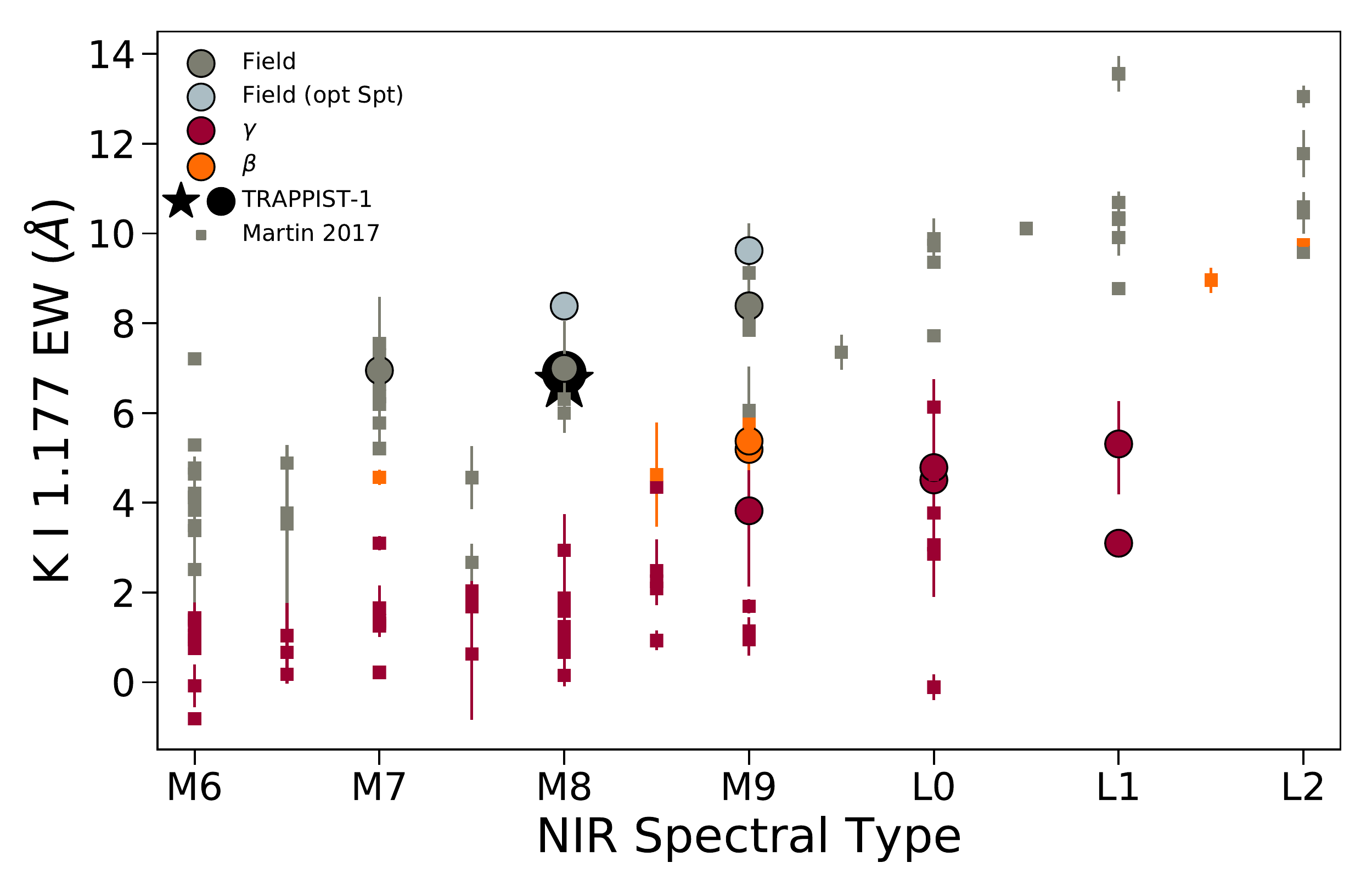}{0.48\textwidth}{\large(b)}}
\gridline{\hspace{-0.1cm}\fig{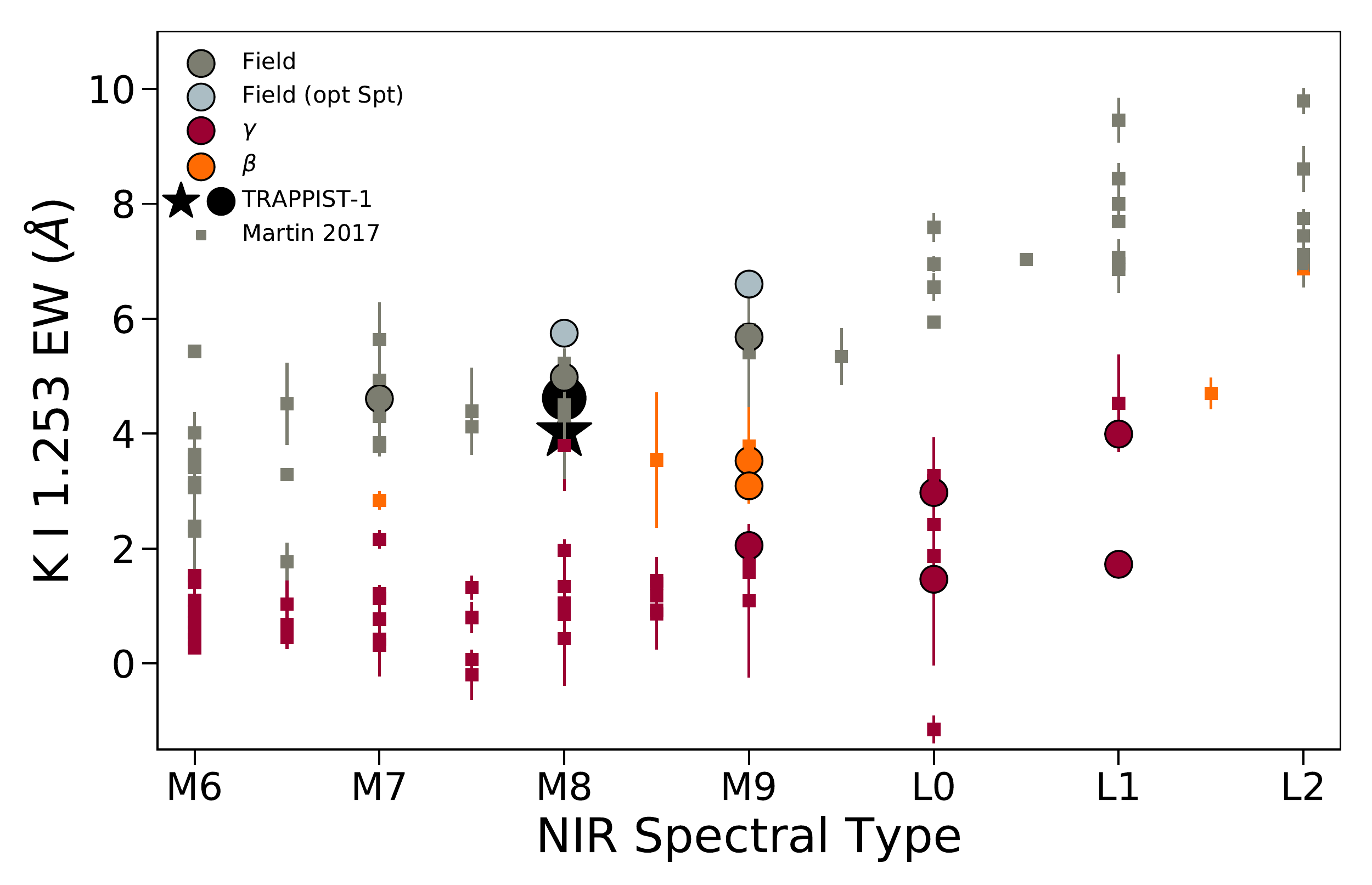}{0.48\textwidth}{\large(c)}}
\caption{Equivalent Widths for the \ion{K}{1} lines in the $J$ band. Symbols are the same as Figure \ref{fig:Indices} for our comparative sample. The \cite{Martin17} sample is shown as squares following the same color scheme as the comparative sample. (a) NIR Spectral Type vs \ion{K}{1} 1.169 line (b) NIR Spectral Type vs \ion{K}{1} 1.177 line (c) NIR Spectral Type vs \ion{K}{1} 1.253 line}
\label{fig:EquivalentWidths}
\end{figure}

Figures \ref{fig:Indices}a--c show the low-resolution spectrum index scores for the comparative sample and TRAPPIST-1 with the field dwarf polynomial and the dividing line between a score of 1 or 2 from \cite{Alle13}. A score of 1 indicates intermediate gravity, while a score of 2 indicates low-gravity for that index. Looking at the low-resolution gravity scores, TRAPPIST-1 received scores of 1, indicating intermediate gravity, in all indices using the prism and SXD data. The FIRE data however, received a score of 0 in the $H$-cont index, indicating field gravity, and a score of 1 in the other indices. All FeH$_z$ and \ion{K}{1}$_J$ index measurements (see Figure \ref{fig:Indices}a and b) for TRAPPIST-1, lie in the intermediate gravity region as defined by \cite{Alle13}. For the $H$-cont index (Figure \ref{fig:Indices}c) we see that the FIRE measurement lies in the field dwarf region, while both SpeX measurements lie in the intermediate gravity region. 

Medium resolution gravity scores of TRAPPIST-1, reveal a difference in the \ion{K}{1} line scores between the SXD and FIRE spectra. The \ion{K}{1} 1.169, 1.17, and 1.253~$\upmu$m equivalent widths are plotted for all sources in our sample with medium resolution in Figure \ref{fig:EquivalentWidths} along with the equivalent widths of all sources in the \cite{Martin17} sample. We do not show the \ion{K}{1} 1.224~$\upmu$m equivalent width plot, since as shown by \cite{Alle13} and \cite{Martin17} there is no visible trend. All three \ion{K}{1} lines for the SXD spectrum of TRAPPIST-1 received a score of 0, while the FIRE spectrum \ion{K}{1} 1.169~$\upmu$m and 1.253~$\upmu$m lines received scores of 1 and the \ion{K}{1} 1.177~$\upmu$m line received score of 0. The \ion{K}{1} 1.169~$\upmu$m and 1.253~$\upmu$m equivalent width measurements for the FIRE spectrum lies just below the field sources from \cite{Martin17} and not far from the corresponding SXD measurement. As supported by our band-by-band analysis in Sections~\ref{bandbybandField} and\ref{bandbybandYoung}, TRAPPIST-1 appears to have low surface gravity features despite our conclusion that its overall SED is best fit by a field age. 

\subsection{Comparison to Trends with Spectral Type }
\begin{figure*}
\centering
 \includegraphics[scale=.57]{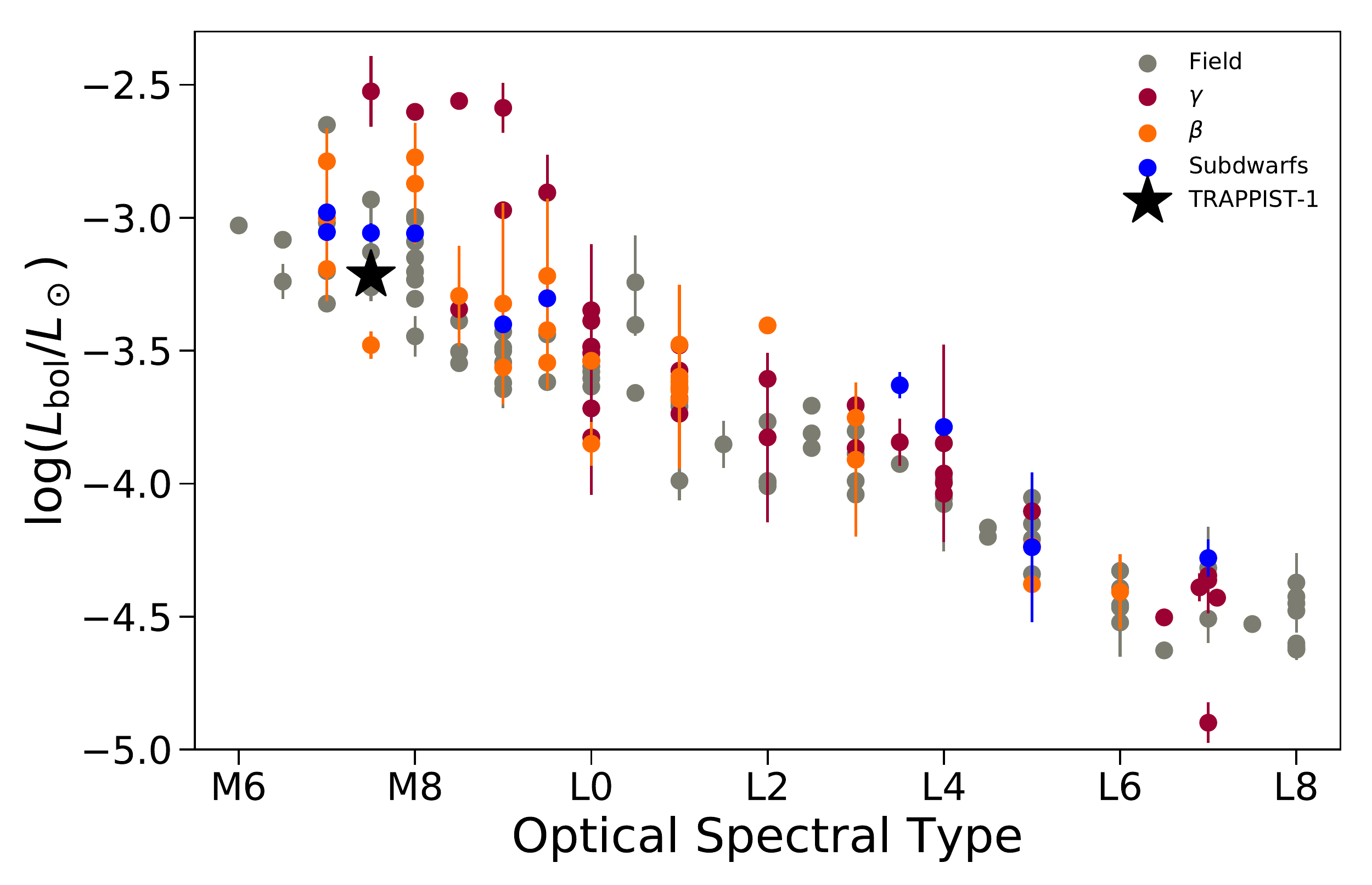} 
\caption{Optical spectral type vs $L_\mathrm{bol}$ for subdwarfs (blue), field objects (grey), and low gravity objects (red and orange). Field objects come from \cite{Fili15}, low gravity objects are from \cite{Fahe16}, and subdwarfs from \cite{Gonz18} with updates to sources in this paper. TRAPPIST-1 is shown as a black star.}
\label{fig:SptvLbol}
\end{figure*}

\begin{figure*}
\centering
 \includegraphics[scale=.57]{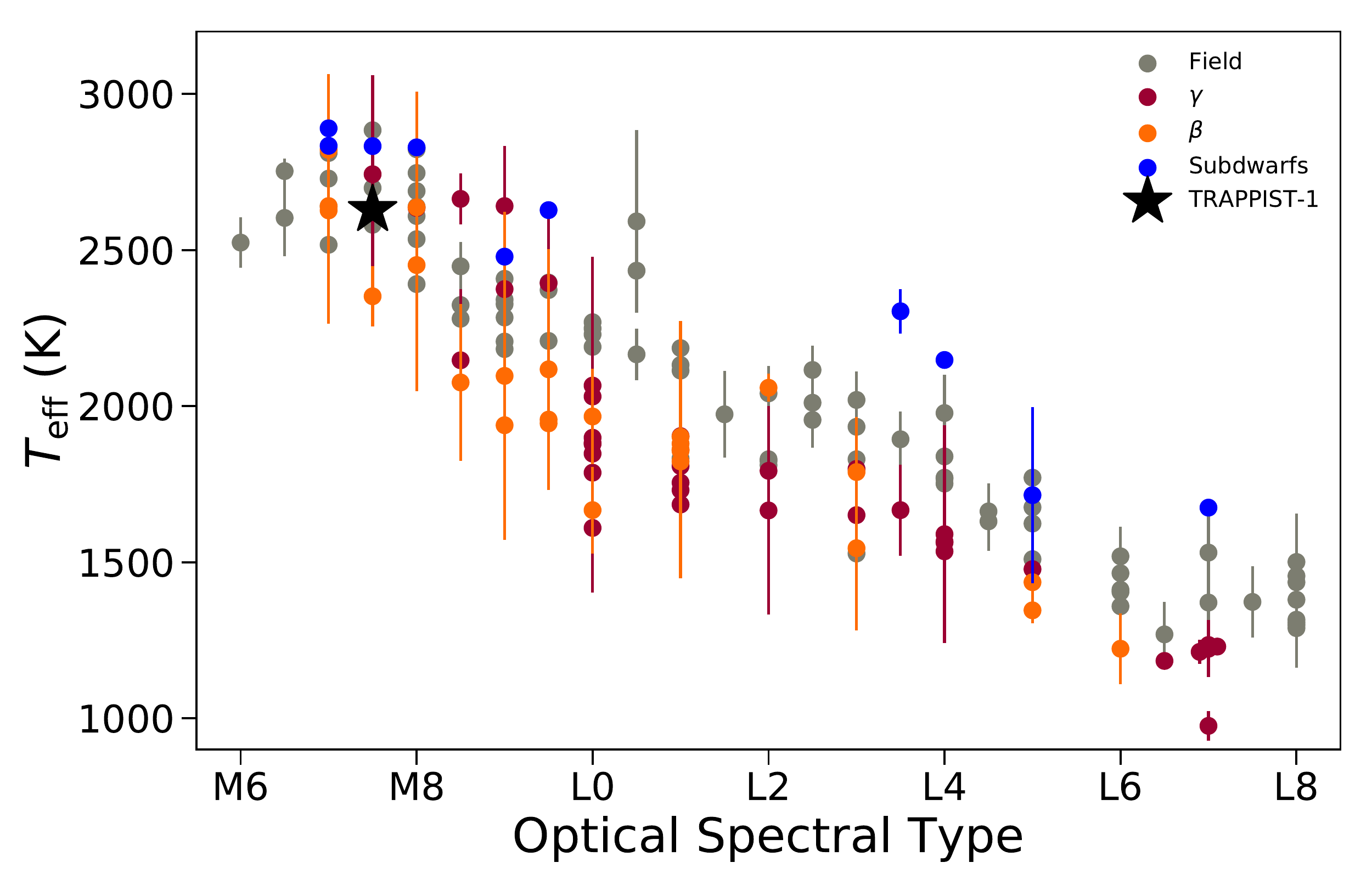}
\caption{Optical spectral type vs $T_\mathrm{eff}$. The same color coding and references as in Figure \ref{fig:SptvLbol}.}
\label{fig:SptvTeff}
\end{figure*}

\begin{figure}
\gridline{\hspace{-0.1cm}\fig{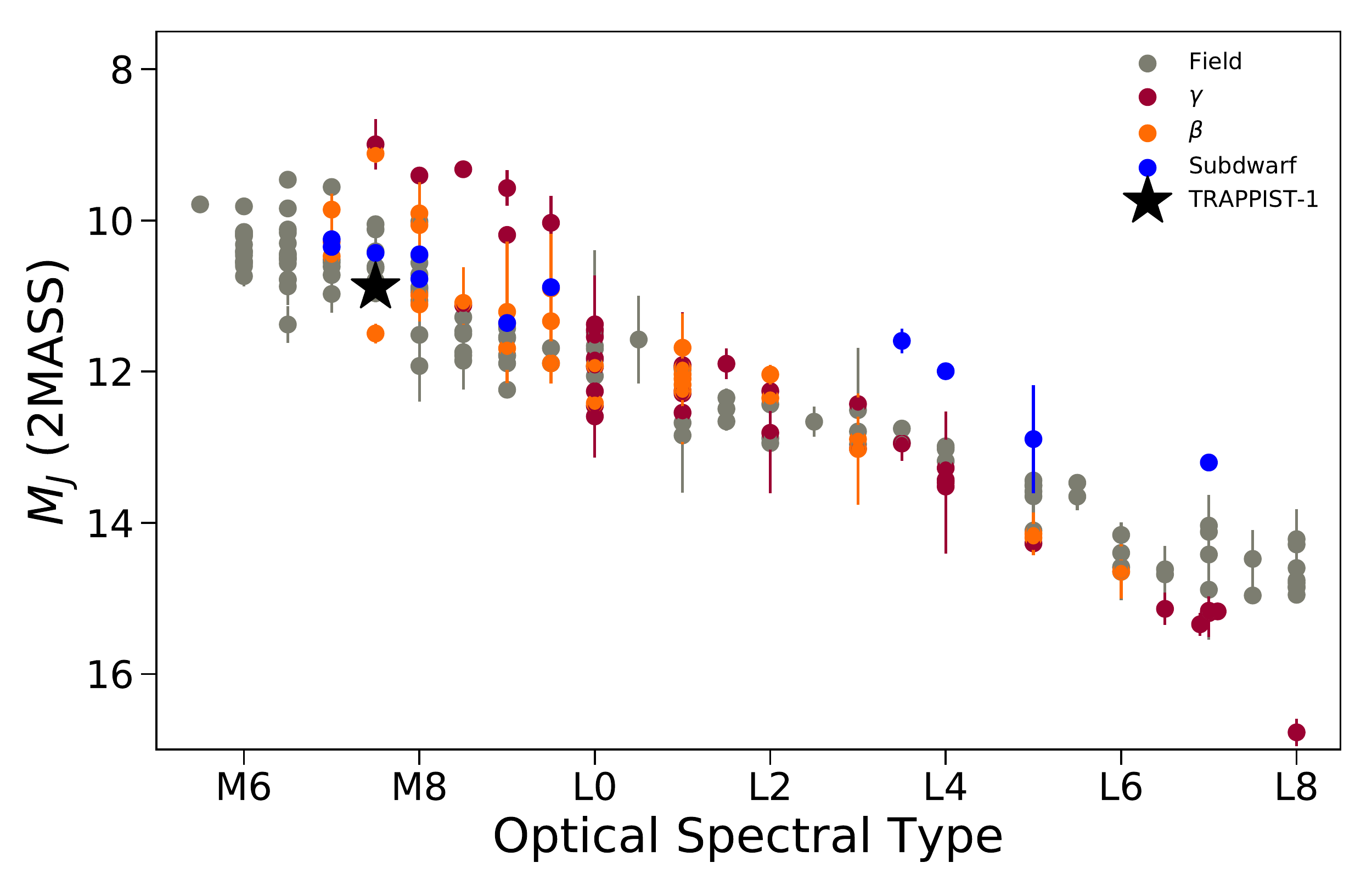}{0.48\textwidth}{\large(a)}}
\gridline{\hspace{-0.1cm}\fig{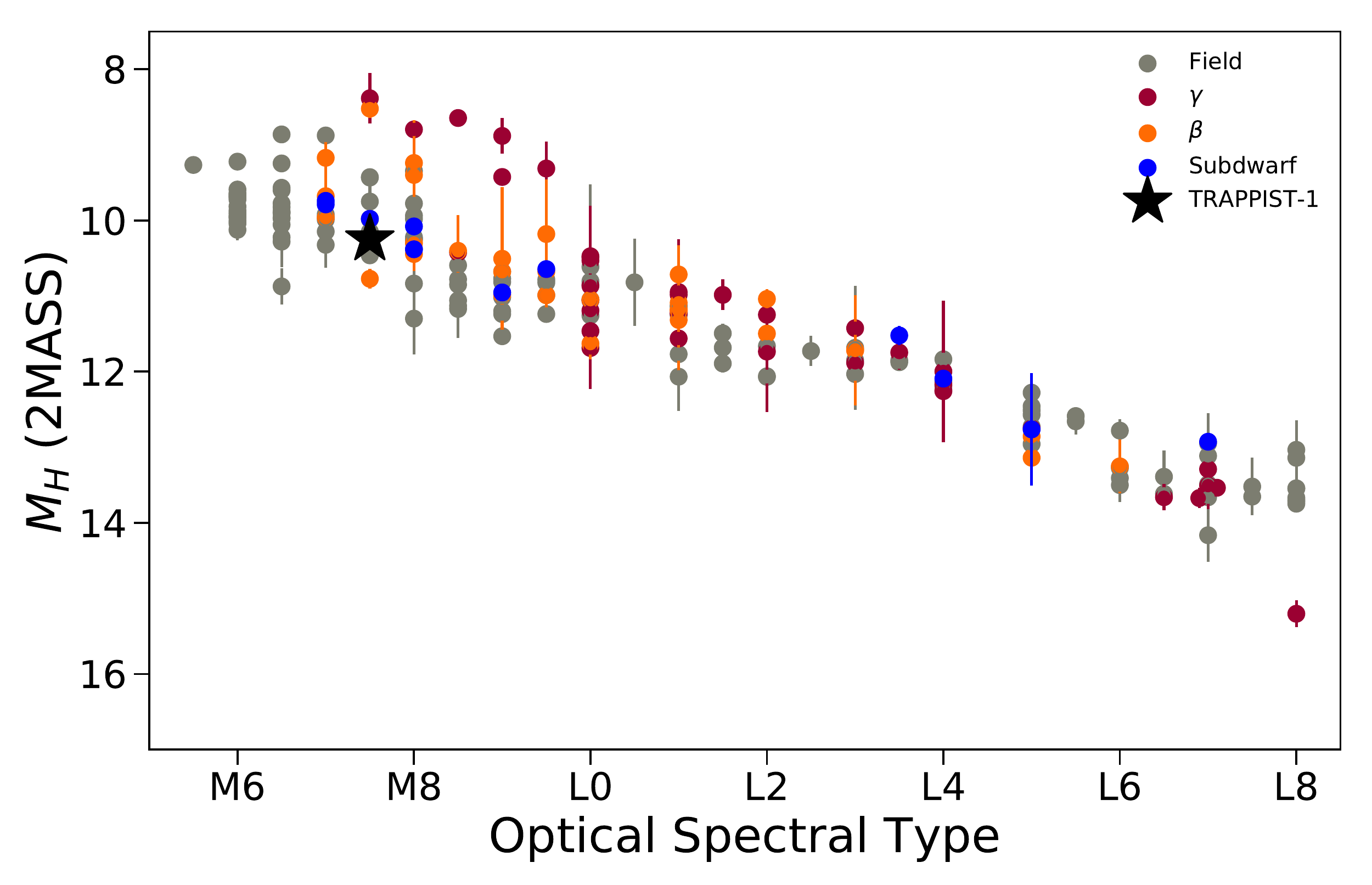}{0.48\textwidth}{\large(b)}}
\gridline{\hspace{-0.1cm}\fig{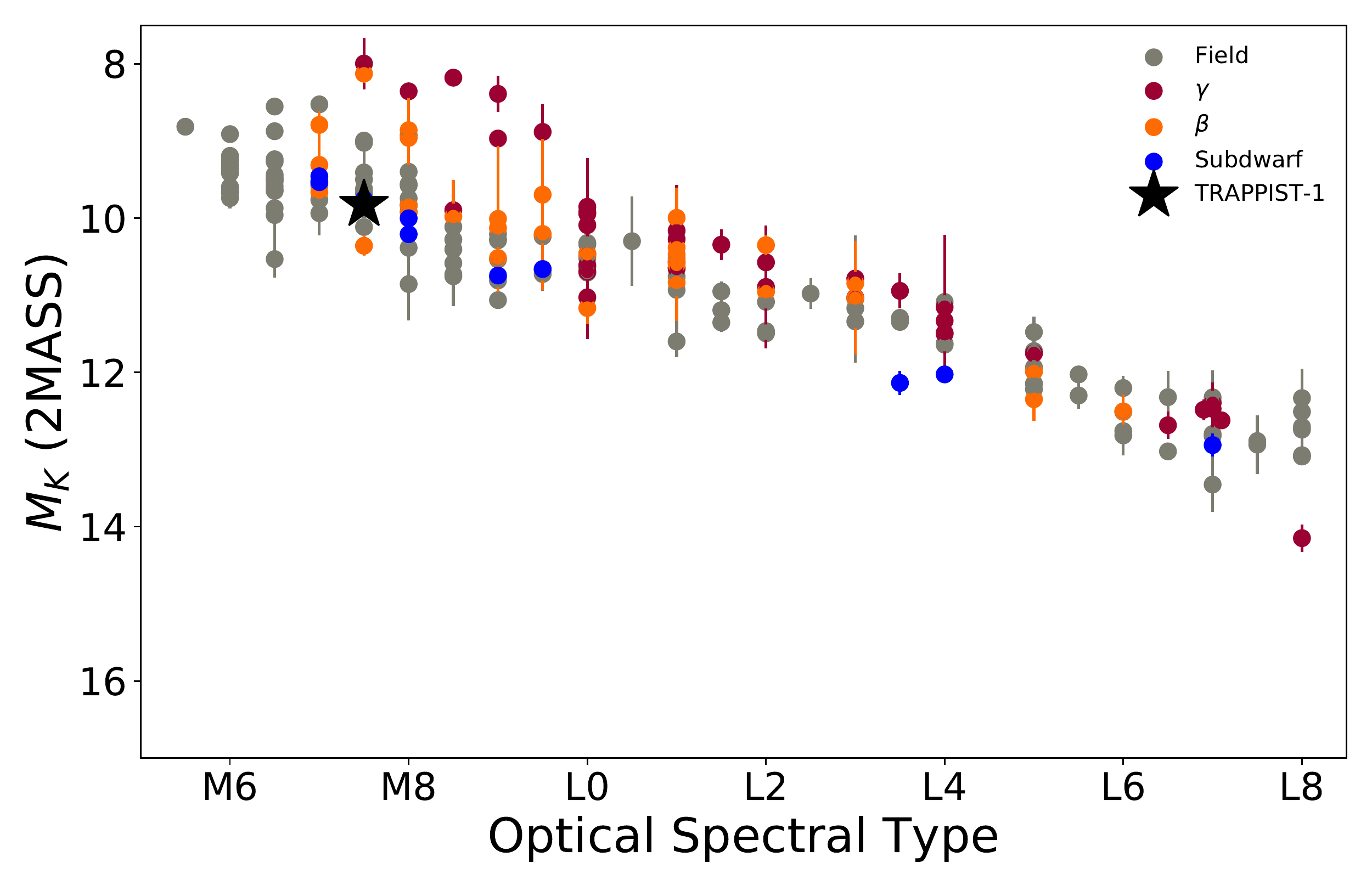}{0.48\textwidth}{\large(c)}}
\caption{Spectral type versus 2MASS absolute magnitudes. TRAPPIST-1 is shown as a black star. Same color coding and references as in Figure \ref{fig:SptvLbol}.(a) Spectral Type vs $M_J$\, (b) Spectral Type vs $M_H$ \, (c) Spectral Type vs $M_{Ks}$}
\label{fig:AbsMagsJHK}
\end{figure}

\begin{figure}
\gridline{\hspace{-0.1cm}\fig{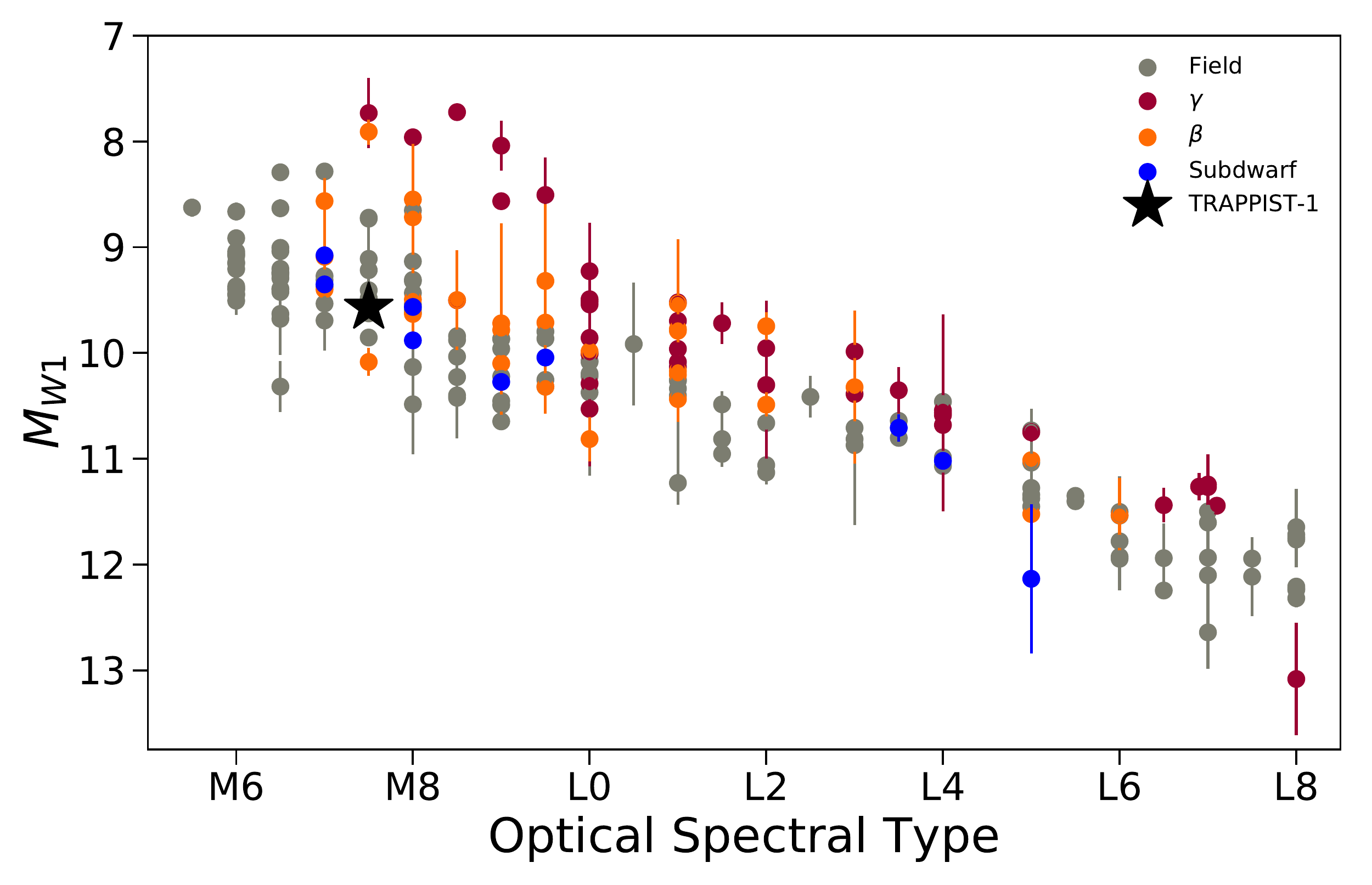}{0.48\textwidth}{\large(a)}}
\gridline{\hspace{-0.1cm}\fig{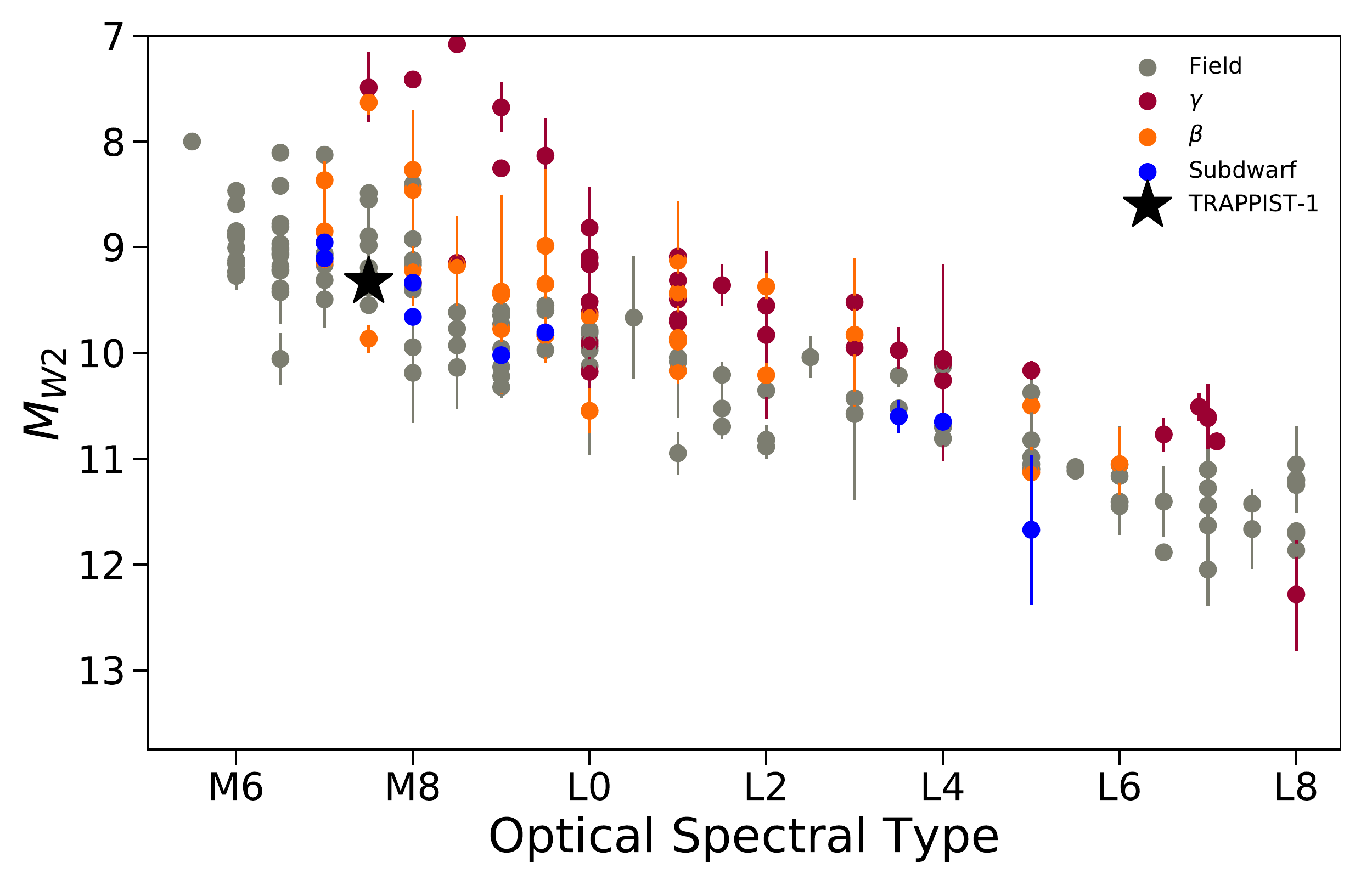}{0.48\textwidth}{\large(b)}}
\caption{Spectral type versus WISE absolute magnitudes. Same color coding and references as in Figure \ref{fig:SptvLbol}. (a) Spectral Type vs $M_{W1}$ \, (b) Spectral Type vs $M_{W2}$}
\label{fig:AbsMagsW1W2}
\vspace{0.5cm}
\end{figure}

To further examine the signatures of youth seen in the NIR spectrum of TRAPPIST-1 we place it in context with fundamnetal parameters of field sources from \cite{Fili15}, low-gravity sources from \cite{Fahe16}, and subdwarfs from \cite{Gonz18}. Figure \ref{fig:SptvLbol} shows the comparison of $L_\mathrm{bol}$ versus spectral type, Figure \ref{fig:SptvTeff} shows the comparison of $T_\mathrm{eff}$ versus spectral type, and Figures \ref{fig:AbsMagsJHK} and \ref{fig:AbsMagsW1W2} compare absolute magnitudes versus spectral type for the $J$,$H$,$K$,$W1$,and $W2$ bands.

As shown in \cite{Gonz18}, all sources are mixed when comparing $L_\mathrm{bol}$ versus optical spectral type in Figure \ref{fig:SptvLbol}. TRAPPIST-1 lands in an area where there are other field sources, however there is no visible trend for where M7.5 $\beta$ sources should be located on the diagram due to only one M7.5 $\beta$ source other than TRAPPIST-1 plotted. \cite{Fahe16} found that $\beta$ gravity sources in their sample that were not members of known moving groups fell along the field sequence, and thus not all late M dwarf $\beta$ sources are young.

Figure \ref{fig:SptvTeff} compares $T_\mathrm{eff}$ versus optical spectral type with TRAPPIST-1 again landing in an ambiguous location. TRAPPIST-1 lies near the field dwarfs, but is surrounded by low-gravity sources as well. As seen in \cite{Fahe16} the low-gravity and field dwarf polynomials overlap in the M-dwarf region, right where TRAPPIST-1 is located. 

Figures \ref{fig:AbsMagsJHK} and \ref{fig:AbsMagsW1W2} compare the absolute magnitudes in the $J$,$H$,$K$,$W1$, and $W2$ bands of field, low-gravity and subdwarfs. The M-dwarf $\beta$ gravity sources lie in the same location as the field sources or just slightly above in the $J$ band and begin to move slightly higher than the field sequence by $K$ band. By $W1$ and $W2$ the $\beta$ sources are further above the field sequence. TRAPPIST-1 remains within the field sequence from $J$ through $W2$ and does not appear to move in brightness like the other $\beta$ sources. This further supports the idea that TRAPPIST-1 is a field source requiring a different physical explanation for its low surface gravity features.

\subsection{Comparison of kinematics to other $\beta$ gravity sources}
\subsubsection{UVW}

\begin{deluxetable*}{l c c c c c c c c }
\tabletypesize{\scriptsize}
\tablecaption{Kinematics of Interesting INT-G Late-M dwarfs \label{tab:compkinematics}}
\tablehead{\colhead{Object} & \colhead{Lit. OPT SpT} & \colhead{Lit. NIR SpT} & \colhead{$U$ (km s$^{-1}$)} & \colhead{$V$ (km s$^{-1}$)} & \colhead{$W$ (km s$^{-1}$)} &  \colhead{$V_\mathrm{tan}$ (km s$^{-1}$)} & \colhead{References}} 
  \startdata
  TRAPPIST-1 & M7.5 & $\cdots$ & $-44 \pm 0.1$ & $-67.2 \pm 0.3$ & $11.7 \pm 0.4$ & $61.69 \pm 0.10$ & SpT: 1, Rest: 2\\
  Teegarden's Star & M6.5 & M7.5\,$\beta$ & $-69.46 \pm 0.31$ & $-71.17 \pm 0.15$ & $-58.68 \pm 0.25$ & $93.03 \pm 0.10$ & SpT: 3, 4, UVW: 5, Vel: 2\\
  LHS 132 & M8 & M8 & $-22.4 \pm 1.2$ & $37 \pm 1.2$ & $-68.08 \pm 0.95$ & $80.4 \pm 0.14$ & SpT: 6, 7, Rest: 2\\
  2MASS J10220489$+$0200477 & M9\,$\beta$ & M9 &$14.87 \pm 4.49$ &	$-53.28 \pm	19.96$ & $-49.14 \pm 14.81$ & $65.06 \pm 1.35$ & SpT:8, 9, UVW: 10, Vel: 11\\
  2MASS J10224821$+$5825453 & L1\,$\beta$ & L1 &$-69.35 \pm	2.74$ &	$-67.62 \pm	3.48$ &	$0.1 \pm 0.87$ &  $95.61 \pm 0.57$ & SpT: 12, 8, UVW: 10, Vel: 11 \\
  2MASS J23224684$-$3133231 & L0\,$\beta$ & L2\,$\beta$ & $40.26 \pm 2.74$ & $-30.87 \pm 3.18$ & $-24.72 \pm 1.27$ & $54.39 \pm 0.76$ & SpT: 13, 9, UVW:10 , Vel: 11 \\
  2MASS J00332386$-$1521309 & L4\,$\beta$ & L1 & $-52.85 \pm 5.68$ & $-26.91 \pm 3.93$ & $3.23 \pm 0.86$ & $34.1\pm1.5$ & SpT: 12, 9, UVW:10 , Vel: 11\\
  \enddata
  \tablecomments{All sources receive an INT-G classification using the \cite{Alle13} indices. References order: SpTs, $UVW$, Total velocity, and $V_\mathrm{tan}$. $UVW$ in this work are not with respect to LSR.}
    \tablerefs{(1) \cite{Gizi00}, (2) This Paper, (3) \cite{Teeg03}, (4) \cite{Gagn15b}, (5) \cite{Cort16} , (6) \cite{Diet14}, (7) \cite{Bard14}, (8) \cite{Fahe12}, (9) \cite{Alle13}, (10) \cite{Fahe16}, (11)Faherty et al. (in prep.), (12) \cite{Cruz09}, (13) \cite{Reid08b}}
  \tabletypesize{\small}
\end{deluxetable*}

\begin{figure*}
\centering
 \includegraphics[scale=.57]{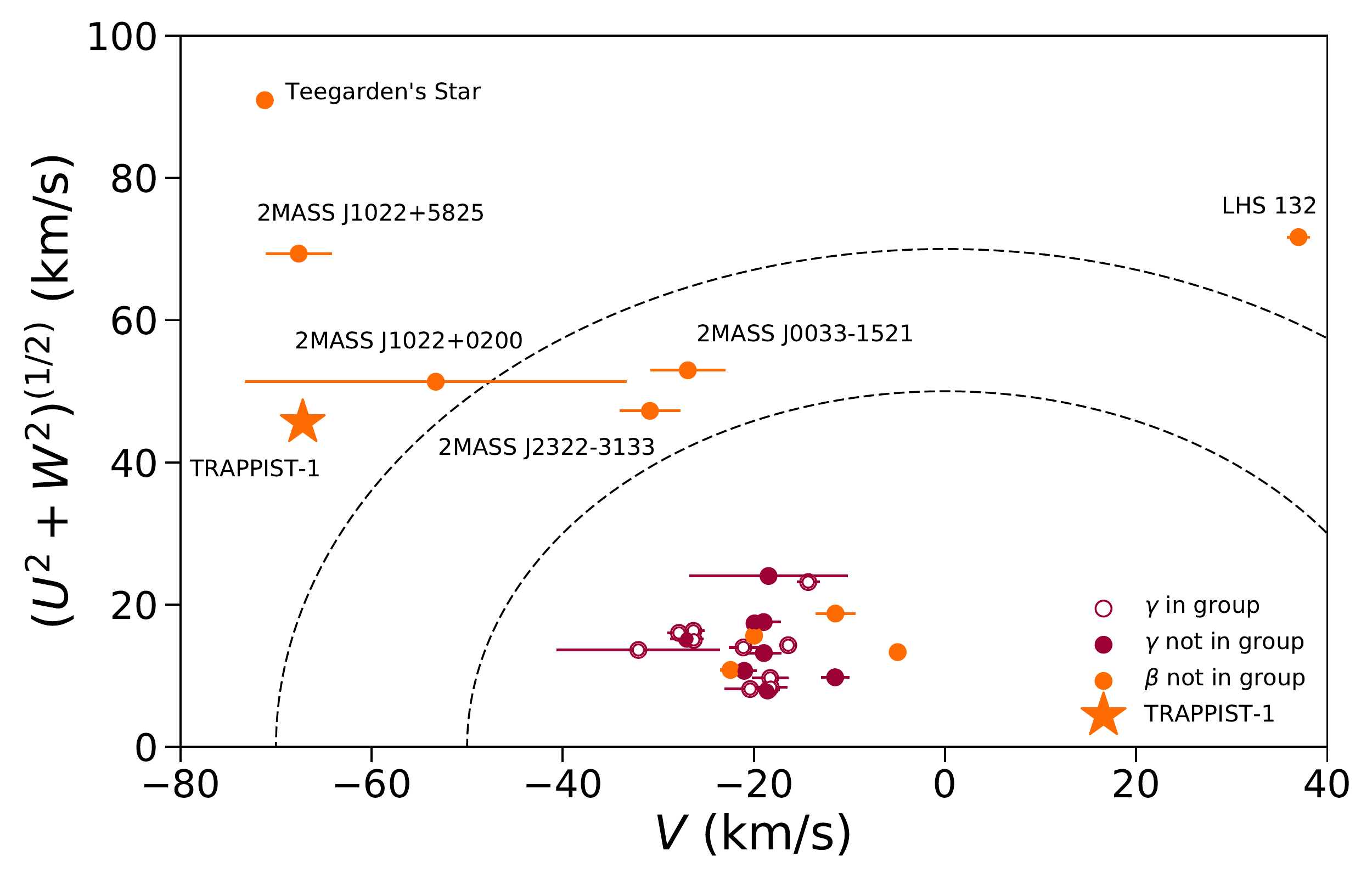}
\caption{Toomre Diagram of young sources from \cite{Fahe16} (moving group members open circles, non-members solid circles), TRAPPIST-1 (black star), Teegarden's Star (black square), and LHS 132 (black circle). The dashed circles show constant total velocity boundaries between the thin and thick disk 50 km s$^{-1}$ and 70 km s$^{-1}$, where sources with $v_{tot}<$ 50~km s$^{-1}$ are typically thin disk stars and sources with 70km s$^{-1}$ $< v_{tot} <$ 180~km s$^{-1}$ are likely thick disk stars \citep{Niss04,Bens14}. None of the $UVW$ velocities in this plot are with respect to LSR. Note that non-members in this work are objects that are lacking confirmation as bonafide moving group members.  They may have ambiguous kinematics, candidate kinematics, or they might be conclusively non-members of any currently known group.  See \citet{Fahe16} for details.}
\label{fig:Toomre}
\vspace{0.5cm}
\end{figure*}

To determine if $\beta$ sources are kinematically distinct, we examined $UVW$ velocities of all $\beta$ sources from \cite{Fahe16}. The $UVW$ velocities are displayed in a Toomre diagram (Figure \ref{fig:Toomre}) along with $\gamma$ sources from \cite{Fahe16}, Teegarden's Star, and LHS 132. For objects in this work labeled as "not in group" or "non-members", these are sources from \cite{Fahe16} that could be ambiguous members, candidate members, or true non-members of any currently known group, thus objects that are not bonafide members. 

In Figure \ref{fig:Toomre} we see all $\gamma$ sources, whether in known moving groups or not, clustered in parameter space that corresponds to the thin disk, while the $\beta$ sources are found across the thin and well into the thick disk region. Thus the $\beta$ sources with total velocities greater than 50~km s$^{-1}$ may not truly be young, but display signatures of youth for some unaccounted reason. TRAPPIST-1, LHS 132, and Teegarden's Star lie in the thick disk region, along with two intermediate gravity sources from \cite{Fahe16}- 2MASS J10220489$+$0200477 (hereafter J1022$+$0200) and 2MASS J10224821$+$5825453 (hereafter J1022$+$5825). The two intermediate gravity sources in the thin/thick disk region are 2MASS J23224684$-$3133231 (hereafter J2322$-$3133) and 2MASS J003323.86-1521309 (hereafter J0033$-$1521). Kinematics of these seven sources are listed in Table \ref{tab:compkinematics}. Since Teegarden's star also lies in the same region as TRAPPIST-1 and has recently been found to host at least two planets \citep{Zech19}, we suggest one idea for the low-gravity features may be the tug of planets on their host star. Consequently, LHS 132, J1022$+$0200, J1022$+$5825 may be ideal targets for M dwarf planet searches.

\subsubsection{Tangential Velocity}
\begin{figure}
\gridline{\hspace{-0.1cm}\fig{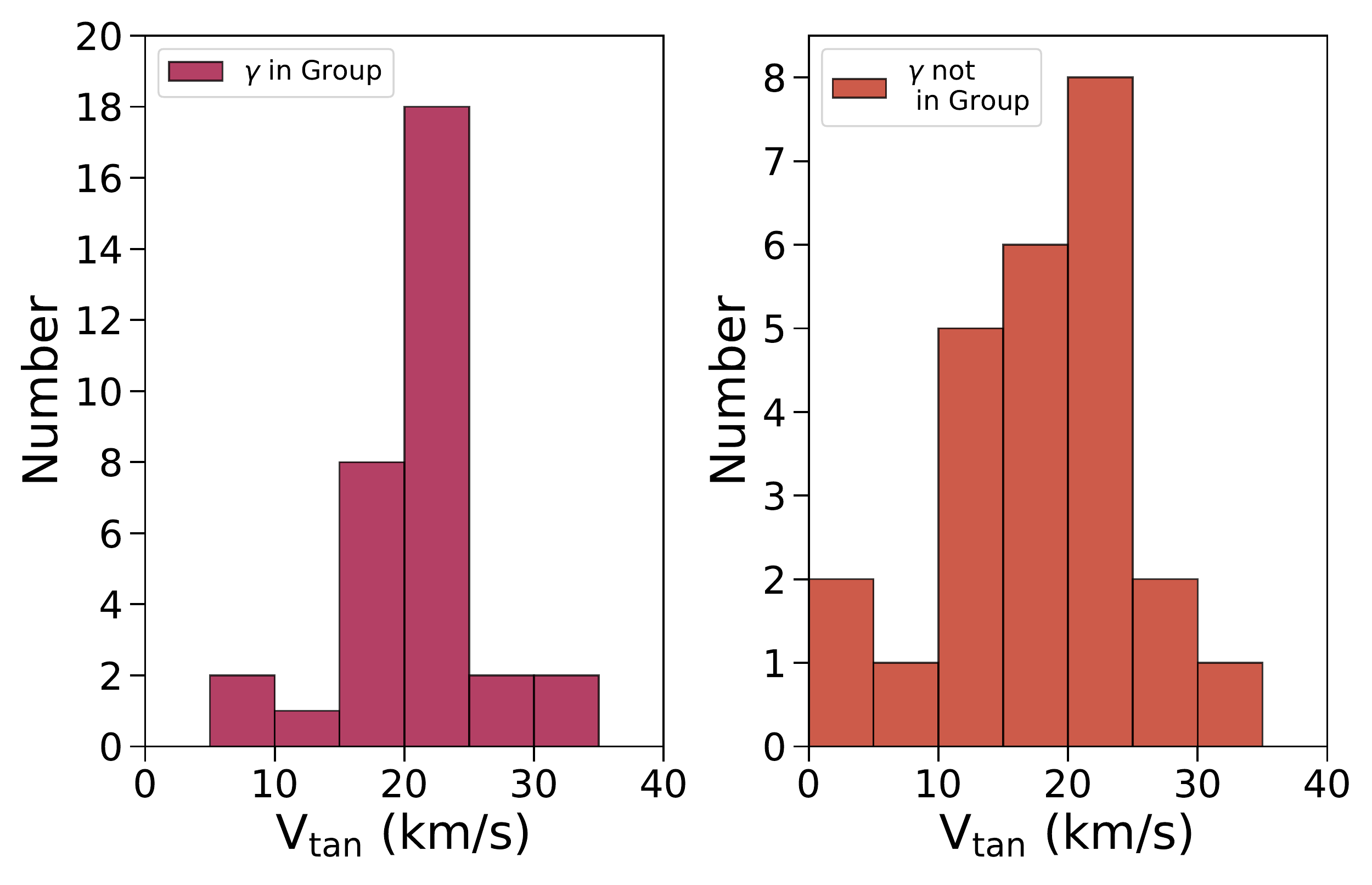}{0.48\textwidth}{\large(a)}}
\gridline{\hspace{-0.1cm}\fig{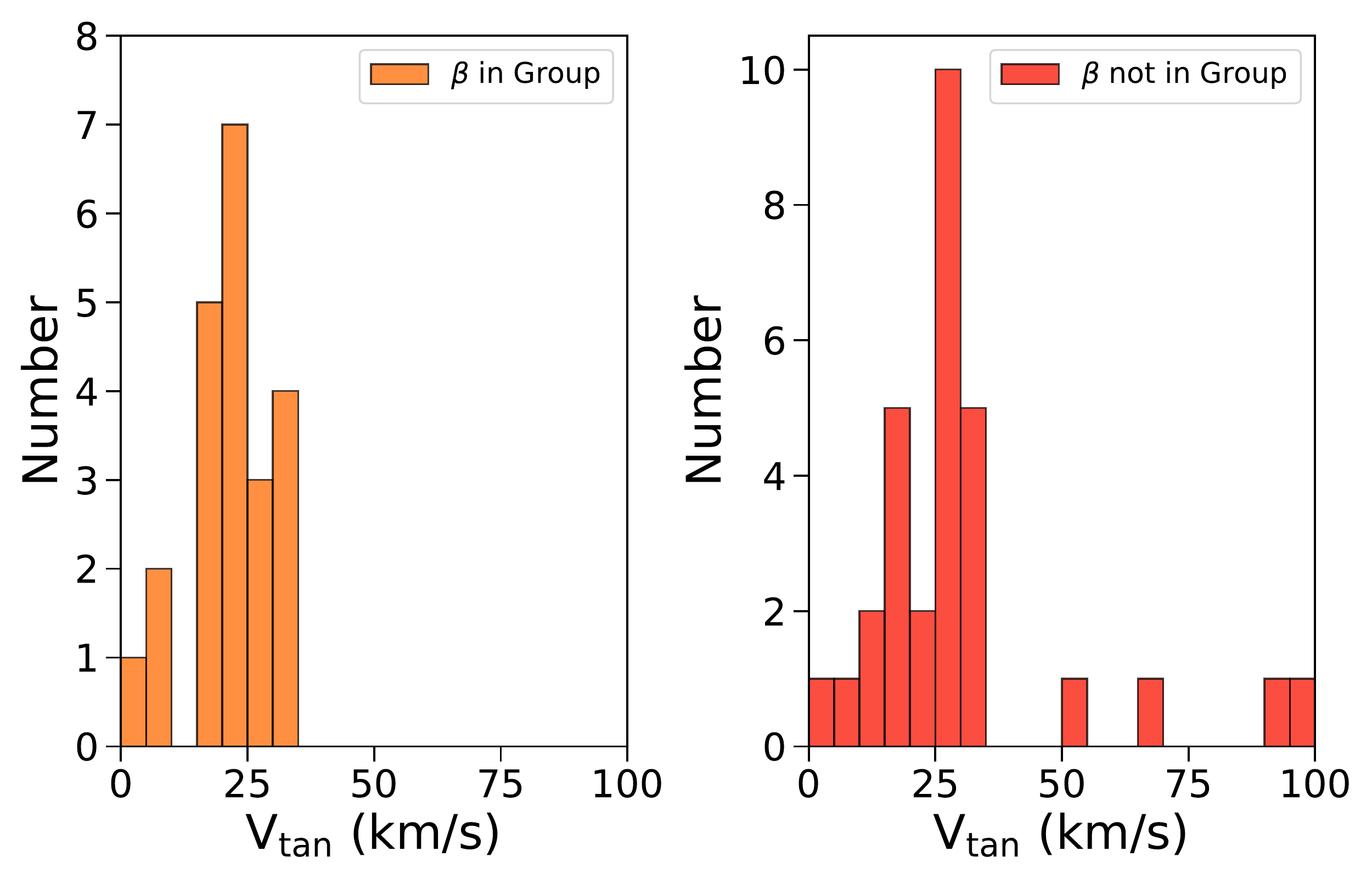}{0.48\textwidth}{\large(b)}}
\caption{Histograms of tangential velocities for low-gravity sources in moving groups and not members of groups from Faherty et al. (in prep). Members of known moving groups are shown on the left, while nonmembers are on the right. (a) $\gamma$ sources  (b) $\beta$ sources. Non-membership is defined in the same way as Figure~\ref{fig:Toomre}. In the right panel of (b), TRAPPIST-1 would lie in the long tail with a tagential velocity of $61.9 \pm 0.10$~km s$^{-1}$}
\label{fig:Vtan}
\end{figure}

Figure \ref{fig:Vtan} shows the distribution of tangential velocities for low-gravity sources from \cite{Fahe16}, with their updated membership and $V_\mathrm{tan}$ values from Faherty et al. (in prep). In this sample, any source classified as $\beta$ gravity in either optical or near-infrared receives a classification of $\beta$ (i.e. M8 in optical, but M8 $\beta$ in NIR, is designated as a $\beta$ in this sample). The same follows for a source with a $\gamma$ classification. For sources that received a $\beta$ classification in one regime, but a $\gamma$ in another, we choose the more extreme gravity classification.

Figure \ref{fig:Vtan}a shows the distribution of tangential velocities for $\gamma$ gravity sources which are members (on the right) or non-members (on the left) of known moving groups. We see that members and non-members have similar distributions of $V_\mathrm{tan}$, both peaking in the \hbox{$20-25$ km s$^{-1}$} range. The non-member distribution is not significantly different from the member distribution.

Figure \ref{fig:Vtan}b shows the distribution for $\beta$ gravity sources that are members and non-members of known moving groups. Unlike the $\gamma$ gravity sources we see that $\beta$ sources in moving groups have $V_\mathrm{tan}$ ranging from \hbox{$0-35$ km~s$^{-1}$}, whereas non-member sources have a larger range of $V_\mathrm{tan}$. However, the bulk of $\beta$ gravity non-members fall in the range seen for member sources. Our calculated $V_\mathrm{tan}$ for TRAPPIST-1 places it outside of the bulk velocity region for $\beta$ gravity sources. We also calculated $V_\mathrm{tan}$ for LHS 132, which places it in the same region as TRAPPIST-1 and the four sources plotted which are Teegarden's Star, J1022$+$0200, J1022$+$5825, and J2322$-$3133. However, J0033$-$1521 lies in the bulk region. Therefore, TRAPPIST-1 along with Teegarden's Star, J1022$+$0200, J1022$+$5825, J2322$-$3133, and LHS 132 are kinematically distinct from the other suspected young M dwarfs.

\subsection{Speculation on the $\beta$ gravity class for TRAPPIST-1}
The $\beta$ classification of TRAPPIST-1 could be due to radius inflation. Two possible causes of this could be: (1) magnetic activity and/or (2) tidal interactions of the planets with the star. In the case of the former, \cite{Chab07} state that theoretical models for sources with M$=0.08$~M$_\odot$, show that radii can range from $0.10-0.14$~R$_\odot$ for black spot coverage of up to 50\% thus offering one pathway towards a radius variation that would mimic a young M dwarf that had not contracted yet. \cite{Luge17} and \cite{Vida17} examined the $K$2 light curve and found evidence of cool, stable magnetic spots on TRAPPIST-1. However, most recently \cite{Morr18} examined the Spitzer 3.5~$\upmu$m and 4.5~$\upmu$m light curves and found no signature of cool spots leaving this line of evidence inconclusive for magnetic influence. For the second possible cause, tidal interactions of the planets with the star, could be the cause of the classification for both TRAPPIST-1 and Teegarden's Star. While it is beyond the scope of  this work to examine the full influence of the planets on their host star, we suggest that an excellent test for this theory would be to look for planets around LHS 132, the other source that matches many of TRAPPIST-1's features, as well as J1022$+$0200 and J1022$+$5825 which are kinemtically distinct from other $\beta$ sources.

\section{Conclusions}
In this work we present a distance-calibrated SED of TRAPPIST-1 using a new NIR FIRE spectrum and a new parallax from the \textit{Gaia} DR2 data release. With our new distance-calibrated SED we compare TRAPPIST-1 to objects of similar effective temperature and/or bolometric luminosity for young, old, and field-aged sources considering an age for TRAPPIST-1 of either $0.5-10$~Gyr or $<0.5$~Gyr. The $J$-band \ion{Na}{1} and \ion{K}{1} lines of TRAPPIST-1 were compared to those of the subdwarfs with medium-resolution data from \cite{Gonz18}. We also looked at TRAPPIST-1 related to objects from \cite{Burg17}. We present updated or new fundamental parameters for our comparative sample using Gaia parallaxes and Pan-Starrs photometry when available. 

Using our derived fundamental parameters we find field dwarfs of similar $T_\mathrm{eff}$ and $L_\mathrm{bol}$, and LHS 132 a M8 dwarf classified as intermediate gravity best fit the SED shape of TRAPPIST-1. From our band-by-band comparisons, TRAPPIST-1 exhibits a blend of field and young spectral features.

We measure the \cite{Alle13} indices for TRAPPIST-1, along with our entire comparative sample. TRAPPIST-1 receives a $\beta$ gravity classification when using three different spectra indicating it might be young. Examining spectral indices versus spectral type, TRAPPIST-1 lies in the $\beta$ gravity space, while when looking at equivalent width versus spectral type, TRAPPIST-1 falls with the field sources.

In an effort to better understand the $\beta$ gravity population and TRAPPIST-1, we plot $L_\mathrm{bol}$ and $T_\mathrm{eff}$ versus optical spectral type as well as $J$, $H$, $K$, $W1$, and $W2$ absolute magnitudes versus spectral type. We find TRAPPIST-1 lies in an area that has both field and $\beta$ sources when examining $L_\mathrm{bol}$ and $T_\mathrm{eff}$ versus optical spectral type and absolute magnitudes versus optical spectral type.

We present updated $UVW$ velocities for TRAPPIST-1 using the new \textit{Gaia} astrometry and compare its kinematics to $\beta$ and $\gamma$ sources which are confirmed members or not bonafide members of known moving groups, Teegarden's Star, and LHS 132. TRAPPIST-1 along with Teegarden's Star, LHS 132, J1022$+$0200, J1022$+$5825, and J2322$-$3133 fall within a subpopulation of $\beta$-gravity sources that are not bonafide members of known moving groups and have higher $UVW$ and tangential velocities.

Lastly, we present two possible causes for the $\beta$ classification of TRAPPIST-1. First, TRAPPIST-1 may have significant magnetic influence as observationally examined by observing cool stable spots. At present there are contradictory information in the literature on this topic. \cite{Luge17} and \cite{Vida17} found evidence for spots while \cite{Morr18} did not leaving this line of explanation inconclusive. Our second proposed explanation could be related to tidal interactions with planets and we suggest LHS 132, J1022$+$0200, J1022$+$5825, and J2322$-$3133 might be an excellent targets for exoplanet campaigns given their similarities to both TRAPPIST-1 and Teegarden's Star.

\acknowledgments
We thank the Magellan telescope operators for their help in collecting FIRE spectra. We thank D. Bardalez Gagliuffi and C. Galindo for their help in obtaining the SpeX spectra of J0608$-$2753. This research was supported by the NSF under Grant No. AST-1614527 and Grant No. AST-1313278, as well as by NASA under \textit{Kepler} Grant No. 80NSSC19K0106.  E.G. thanks the LSSTC Data Science Fellowship Program, which is funded by LSSTC, NSF Cybertraining Grant No. 1829740, the Brinson Foundation, and the Moore Foundation; her participation in the program has benefited this work. J.T. acknowledges support for this work provided by NASA through Hubble Fellowship grant HST-HF2-51399.001 awarded by the Space Telescope Science Institute, which is operated by the Association of Universities for Research in Astronomy, Inc., for NASA, under contract NAS5-26555.  This research has made use of the BDNYC Data Archive, an open access repository of M, L, T and Y dwarf astrometry, photometry and spectra. This paper includes data gathered with the 6.5 meter Magellan Telescopes located at Las Campanas Observatory, Chile. This publication makes use of data products from the Two Micron All Sky Survey, which is a joint project of the University of Massachusetts and the Infrared Processing and Analysis Center/California Institute of Technology, funded by the National Aeronautics and Space Administration and the National Science Foundation. This publication makes use of data products from the Wide-field Infrared Survey Explorer, which is a joint project of the University of California, Los Angeles, and the Jet Propulsion Laboratory/California Institute of Technology, funded by the National Aeronautics and Space Administration. This work has made use of data from the European Space Agency (ESA) mission {\it Gaia} (\url{https://www.cosmos.esa.int/gaia}), processed by the {\it Gaia} Data Processing and Analysis Consortium (DPAC, \url{https://www.cosmos.esa.int/web/gaia/dpac/consortium}). Funding for the DPAC has been provided by national institutions, in particular the institutions participating in the {\it Gaia} Multilateral Agreement. 

\facilities{Magellan:Baade (FIRE), IRTF:SpeX} 
\software{SEDkit (\url{https://github.com/hover2pi/SEDkit})}

\newpage
\bibliographystyle{yahapj}
\bibliography{references}

\begin{thebibliography}{}
\providecommand\natexlab[1]{#1}
\providecommand\JournalTitle[1]{#1}

\bibitem[{{Abazajian} {et~al.}(2009){Abazajian}, {Adelman-McCarthy},
  {Ag{\"u}eros}, {Allam}, {Allende Prieto}, {An}, {Anderson}, {Anderson},
  {Annis}, {Bahcall}, \& et~al.}]{Abaz09}
{Abazajian}, K.~N., {Adelman-McCarthy}, J.~K., {Ag{\"u}eros}, M.~A., {et~al.}
  2009,
  \href{http://dx.doi.org/10.1088/0067-0049/182/2/543}{\JournalTitle{\apjs},
  182, 543}

\bibitem[{{Adelman-McCarthy} {et~al.}(2008){Adelman-McCarthy}, {Ag{\"u}eros},
  {Allam}, {Allende Prieto}, {Anderson}, {Anderson}, {Annis}, {Bahcall},
  {Bailer-Jones}, {Baldry}, {Barentine}, {Bassett}, {Becker}, {Beers}, {Bell},
  {Berlind}, {Bernardi}, {Blanton}, {Bochanski}, {Boroski}, {Brinchmann},
  {Brinkmann}, {Brunner}, {Budav{\'a}ri}, {Carliles}, {Carr}, {Castander},
  {Cinabro}, {Cool}, {Covey}, {Csabai}, {Cunha}, {Davenport}, {Dilday}, {Doi},
  {Eisenstein}, {Evans}, {Fan}, {Finkbeiner}, {Friedman}, {Frieman},
  {Fukugita}, {G{\"a}nsicke}, {Gates}, {Gillespie}, {Glazebrook}, {Gray},
  {Grebel}, {Gunn}, {Gurbani}, {Hall}, {Harding}, {Harvanek}, {Hawley},
  {Hayes}, {Heckman}, {Hendry}, {Hindsley}, {Hirata}, {Hogan}, {Hogg}, {Hyde},
  {Ichikawa}, {Ivezi{\'c}}, {Jester}, {Johnson}, {Jorgensen}, {Juri{\'c}},
  {Kent}, {Kessler}, {Kleinman}, {Knapp}, {Kron}, {Krzesinski}, {Kuropatkin},
  {Lamb}, {Lampeitl}, {Lebedeva}, {Lee}, {French Leger}, {L{\'e}pine}, {Lima},
  {Lin}, {Long}, {Loomis}, {Loveday}, {Lupton}, {Malanushenko}, {Malanushenko},
  {Mandelbaum}, {Margon}, {Marriner}, {Mart{\'{\i}}nez-Delgado}, {Matsubara},
  {McGehee}, {McKay}, {Meiksin}, {Morrison}, {Munn}, {Nakajima}, {Neilsen},
  {Newberg}, {Nichol}, {Nicinski}, {Nieto-Santisteban}, {Nitta}, {Okamura},
  {Owen}, {Oyaizu}, {Padmanabhan}, {Pan}, {Park}, {Peoples}, {Pier}, {Pope},
  {Purger}, {Raddick}, {Re Fiorentin}, {Richards}, {Richmond}, {Riess}, {Rix},
  {Rockosi}, {Sako}, {Schlegel}, {Schneider}, {Schreiber}, {Schwope}, {Seljak},
  {Sesar}, {Sheldon}, {Shimasaku}, {Sivarani}, {Allyn Smith}, {Snedden},
  {Steinmetz}, {Strauss}, {SubbaRao}, {Suto}, {Szalay}, {Szapudi}, {Szkody},
  {Tegmark}, {Thakar}, {Tremonti}, {Tucker}, {Uomoto}, {Vanden Berk},
  {Vandenberg}, {Vidrih}, {Vogeley}, {Voges}, {Vogt}, {Wadadekar}, {Weinberg},
  {West}, {White}, {Wilhite}, {Yanny}, {Yocum}, {York}, {Zehavi}, \&
  {Zucker}}]{Adel08}
{Adelman-McCarthy}, J.~K., {Ag{\"u}eros}, M.~A., {Allam}, S.~S., {et~al.} 2008,
  \href{http://dx.doi.org/10.1086/524984}{\JournalTitle{\apjs}, 175, 297}

\bibitem[{{Aganze} {et~al.}(2016){Aganze}, {Burgasser}, {Faherty}, {Choban},
  {Escala}, {Lopez}, {Jin}, {Tamiya}, {Tallis}, \& {Rockward}}]{Agan16}
{Aganze}, C., {Burgasser}, A.~J., {Faherty}, J.~K., {et~al.} 2016,
  \href{http://dx.doi.org/10.3847/0004-6256/151/2/46}{\JournalTitle{\aj}, 151,
  46}

\bibitem[{{Ahn} {et~al.}(2012){Ahn}, {Alexandroff}, {Allende Prieto},
  {Anderson}, {Anderton}, {Andrews}, {Aubourg}, {Bailey}, {Balbinot}, {Barnes},
  \& et~al.}]{Ahn_12}
{Ahn}, C.~P., {Alexandroff}, R., {Allende Prieto}, C., {et~al.} 2012,
  \href{http://dx.doi.org/10.1088/0067-0049/203/2/21}{\JournalTitle{\apjs},
  203, 21}

\bibitem[{{Alam} {et~al.}(2015){Alam}, {Albareti}, {Allende Prieto}, {Anders},
  {Anderson}, {Anderton}, {Andrews}, {Armengaud}, {Aubourg}, {Bailey}, \&
  et~al.}]{Alam15}
{Alam}, S., {Albareti}, F.~D., {Allende Prieto}, C., {et~al.} 2015,
  \href{http://dx.doi.org/10.1088/0067-0049/219/1/12}{\JournalTitle{\apjs},
  219, 12}

\bibitem[{{Allers} \& {Liu}(2013)}]{Alle13ConfP}
{Allers}, K.~N., \& {Liu}, M.~C. 2013, \JournalTitle{\memsai}, 84, 1089

\bibitem[{Allers \& Liu(2013)}]{Alle13}
Allers, K.~N., \& Liu, M.~C. 2013,
  \href{http://stacks.iop.org/0004-637X/772/i=2/a=79}{\JournalTitle{The
  Astrophysical Journal}, 772, 79}

\bibitem[{{Allers} {et~al.}(2010){Allers}, {Liu}, {Dupuy}, \&
  {Cushing}}]{Alle10}
{Allers}, K.~N., {Liu}, M.~C., {Dupuy}, T.~J., \& {Cushing}, M.~C. 2010,
  \href{http://dx.doi.org/10.1088/0004-637X/715/1/561}{\JournalTitle{\apj},
  715, 561}

\bibitem[{{Anglada-Escud{\'e}} {et~al.}(2016){Anglada-Escud{\'e}}, {Amado},
  {Barnes}, {Berdi{\~n}as}, {Butler}, {Coleman}, {de La Cueva}, {Dreizler},
  {Endl}, {Giesers}, {Jeffers}, {Jenkins}, {Jones}, {Kiraga}, {K{\"u}rster},
  {L{\'o}pez-Gonz{\'a}lez}, {Marvin}, {Morales}, {Morin}, {Nelson}, {Ortiz},
  {Ofir}, {Paardekooper}, {Reiners}, {Rodr{\'{\i}}guez},
  {Rodr{\'{\i}}guez-L{\'o}pez}, {Sarmiento}, {Strachan}, {Tsapras}, {Tuomi}, \&
  {Zechmeister}}]{Angl16}
{Anglada-Escud{\'e}}, G., {Amado}, P.~J., {Barnes}, J., {et~al.} 2016,
  \href{http://dx.doi.org/10.1038/nature19106}{\JournalTitle{\nat}, 536, 437}

\bibitem[{{Ballard}(2018)}]{Ball18}
{Ballard}, S. 2018, \JournalTitle{ArXiv e-prints},
  \href{http://arxiv.org/abs/1801.04949}{{\sffamily arXiv:1801.04949
  [astro-ph.EP]}}

\bibitem[{{Bardalez Gagliuffi} {et~al.}(2014){Bardalez Gagliuffi}, {Burgasser},
  {Gelino}, {Looper}, {Nicholls}, {Schmidt}, {Cruz}, {West}, {Gizis}, \&
  {Metchev}}]{Bard14}
{Bardalez Gagliuffi}, D.~C., {Burgasser}, A.~J., {Gelino}, C.~R., {et~al.}
  2014,
  \href{http://dx.doi.org/10.1088/0004-637X/794/2/143}{\JournalTitle{\apj},
  794, 143}

\bibitem[{Barnes {et~al.}(2014)Barnes, Jenkins, Jones, Jeffers, Rojo,
  Arriagada, Jordán, Minniti, Tuomi, Pinfield, \& Anglada-Escudé}]{Barn14}
Barnes, J.~R., Jenkins, J.~S., Jones, H. R.~A., {et~al.} 2014,
  \href{http://dx.doi.org/10.1093/mnras/stu172}{\JournalTitle{Monthly Notices
  of the Royal Astronomical Society}, 439, 3094}

\bibitem[{{Batalha} {et~al.}(2010){Batalha}, {Borucki}, {Koch}, {Bryson},
  {Haas}, {Brown}, {Caldwell}, {Hall}, {Gilliland}, {Latham}, {Meibom}, \&
  {Monet}}]{Bata10}
{Batalha}, N.~M., {Borucki}, W.~J., {Koch}, D.~G., {et~al.} 2010,
  \href{http://dx.doi.org/10.1088/2041-8205/713/2/L109}{\JournalTitle{\apjl},
  713, L109}

\bibitem[{{Bell} {et~al.}(2015){Bell}, {Mamajek}, \& {Naylor}}]{Bell15}
{Bell}, C.~P.~M., {Mamajek}, E.~E., \& {Naylor}, T. 2015,
  \href{http://dx.doi.org/10.1093/mnras/stv1981}{\JournalTitle{\mnras}, 454,
  593}

\bibitem[{{Bensby} {et~al.}(2014){Bensby}, {Feltzing}, \& {Oey}}]{Bens14}
{Bensby}, T., {Feltzing}, S., \& {Oey}, M.~S. 2014,
  \href{http://dx.doi.org/10.1051/0004-6361/201322631}{\JournalTitle{\aap},
  562, A71}

\bibitem[{{Best} {et~al.}(2018){Best}, {Magnier}, {Liu}, {Aller}, {Zhang},
  {Burgett}, {Chambers}, {Draper}, {Flewelling}, {Kaiser}, {Kudritzki},
  {Metcalfe}, {Tonry}, {Wainscoat}, \& {Waters}}]{Best18}
{Best}, W.~M.~J., {Magnier}, E.~A., {Liu}, M.~C., {et~al.} 2018,
  \href{http://dx.doi.org/10.3847/1538-4365/aa9982}{\JournalTitle{\apjs}, 234,
  1}

\bibitem[{{Bochanski} {et~al.}(2010){Bochanski}, {Hawley}, {Covey}, {West},
  {Reid}, {Golimowski}, \& {Ivezi{\'c}}}]{Boch10}
{Bochanski}, J.~J., {Hawley}, S.~L., {Covey}, K.~R., {et~al.} 2010,
  \href{http://dx.doi.org/10.1088/0004-6256/139/6/2679}{\JournalTitle{\aj},
  139, 2679}

\bibitem[{{Bochanski} {et~al.}(2009){Bochanski}, {Hennawi}, {Simcoe},
  {Prochaska}, {West}, {Burgasser}, {Burles}, {Bernstein}, {Williams}, \&
  {Murphy}}]{Boch09}
{Bochanski}, J.~J., {Hennawi}, J.~F., {Simcoe}, R.~A., {et~al.} 2009,
  \href{http://dx.doi.org/10.1086/648597}{\JournalTitle{\pasp}, 121, 1409}

\bibitem[{{Borucki} {et~al.}(2010){Borucki}, {Koch}, {Basri}, {Batalha},
  {Brown}, {Caldwell}, {Caldwell}, {Christensen-Dalsgaard}, {Cochran},
  {DeVore}, {Dunham}, {Dupree}, {Gautier}, {Geary}, {Gilliland}, {Gould},
  {Howell}, {Jenkins}, {Kondo}, {Latham}, {Marcy}, {Meibom}, {Kjeldsen},
  {Lissauer}, {Monet}, {Morrison}, {Sasselov}, {Tarter}, {Boss}, {Brownlee},
  {Owen}, {Buzasi}, {Charbonneau}, {Doyle}, {Fortney}, {Ford}, {Holman},
  {Seager}, {Steffen}, {Welsh}, {Rowe}, {Anderson}, {Buchhave}, {Ciardi},
  {Walkowicz}, {Sherry}, {Horch}, {Isaacson}, {Everett}, {Fischer}, {Torres},
  {Johnson}, {Endl}, {MacQueen}, {Bryson}, {Dotson}, {Haas}, {Kolodziejczak},
  {Van Cleve}, {Chandrasekaran}, {Twicken}, {Quintana}, {Clarke}, {Allen},
  {Li}, {Wu}, {Tenenbaum}, {Verner}, {Bruhweiler}, {Barnes}, \&
  {Prsa}}]{Boru10}
{Borucki}, W.~J., {Koch}, D., {Basri}, G., {et~al.} 2010,
  \href{http://dx.doi.org/10.1126/science.1185402}{\JournalTitle{Science}, 327,
  977}

\bibitem[{{Boss} {et~al.}(2017){Boss}, {Weinberger}, {Keiser}, {Astraatmadja},
  {Anglada-Escude}, \& {Thompson}}]{Boss17}
{Boss}, A.~P., {Weinberger}, A.~J., {Keiser}, S.~A., {et~al.} 2017,
  \href{http://dx.doi.org/10.3847/1538-3881/aa84b5}{\JournalTitle{\aj}, 154,
  103}

\bibitem[{{Burgasser}(2004)}]{Burg04c}
{Burgasser}, A.~J. 2004,
  \href{http://dx.doi.org/10.1086/425418}{\JournalTitle{\apjl}, 614, L73}

\bibitem[{{Burgasser} {et~al.}(2007){Burgasser}, {Cruz}, \&
  {Kirkpatrick}}]{Burg07a}
{Burgasser}, A.~J., {Cruz}, K.~L., \& {Kirkpatrick}, J.~D. 2007,
  \href{http://dx.doi.org/10.1086/510148}{\JournalTitle{\apj}, 657, 494}

\bibitem[{{Burgasser} {et~al.}(2008{\natexlab{a}}){Burgasser}, {Liu},
  {Ireland}, {Cruz}, \& {Dupuy}}]{Burg08d}
{Burgasser}, A.~J., {Liu}, M.~C., {Ireland}, M.~J., {Cruz}, K.~L., \& {Dupuy},
  T.~J. 2008{\natexlab{a}},
  \href{http://dx.doi.org/10.1086/588379}{\JournalTitle{\apj}, 681, 579}

\bibitem[{Burgasser \& Mamajek(2017)}]{Burg17}
Burgasser, A.~J., \& Mamajek, E.~E. 2017,
  \href{http://stacks.iop.org/0004-637X/845/i=2/a=110}{\JournalTitle{The
  Astrophysical Journal}, 845, 110}

\bibitem[{{Burgasser} {et~al.}(2008{\natexlab{b}}){Burgasser}, {Vrba},
  {L{\'e}pine}, {Munn}, {Luginbuhl}, {Henden}, {Guetter}, \&
  {Canzian}}]{Burg08a}
{Burgasser}, A.~J., {Vrba}, F.~J., {L{\'e}pine}, S., {et~al.}
  2008{\natexlab{b}},
  \href{http://dx.doi.org/10.1086/523810}{\JournalTitle{\apj}, 672, 1159}

\bibitem[{{Burgasser} {et~al.}(2009){Burgasser}, {Witte}, {Helling},
  {Sanderson}, {Bochanski}, \& {Hauschildt}}]{Burg09a}
{Burgasser}, A.~J., {Witte}, S., {Helling}, C., {et~al.} 2009,
  \href{http://dx.doi.org/10.1088/0004-637X/697/1/148}{\JournalTitle{\apj},
  697, 148}

\bibitem[{{Burgasser} {et~al.}(2003){Burgasser}, {Kirkpatrick}, {Burrows},
  {Liebert}, {Reid}, {Gizis}, {McGovern}, {Prato}, \& {McLean}}]{Burg03c}
{Burgasser}, A.~J., {Kirkpatrick}, J.~D., {Burrows}, A., {et~al.} 2003,
  \href{http://dx.doi.org/10.1086/375813}{\JournalTitle{\apj}, 592, 1186}

\bibitem[{{Burgasser} {et~al.}(2015){Burgasser}, {Logsdon}, {Gagn{\'e}},
  {Bochanski}, {Faherty}, {West}, {Mamajek}, {Schmidt}, \& {Cruz}}]{Burg15}
{Burgasser}, A.~J., {Logsdon}, S.~E., {Gagn{\'e}}, J., {et~al.} 2015,
  \href{http://dx.doi.org/10.1088/0067-0049/220/1/18}{\JournalTitle{\apjs},
  220, 18}

\bibitem[{{Chabrier} \& {Baraffe}(1997)}]{Chab97}
{Chabrier}, G., \& {Baraffe}, I. 1997, \JournalTitle{\aap}, 327, 1039

\bibitem[{{Chabrier} {et~al.}(2000){Chabrier}, {Baraffe}, {Allard}, \&
  {Hauschildt}}]{Chab00}
{Chabrier}, G., {Baraffe}, I., {Allard}, F., \& {Hauschildt}, P. 2000,
  \href{http://dx.doi.org/10.1086/309513}{\JournalTitle{\apj}, 542, 464}

\bibitem[{{Chabrier} {et~al.}(2007){Chabrier}, {Gallardo}, \&
  {Baraffe}}]{Chab07}
{Chabrier}, G., {Gallardo}, J., \& {Baraffe}, I. 2007,
  \href{http://dx.doi.org/10.1051/0004-6361:20077702}{\JournalTitle{\aap}, 472,
  L17}

\bibitem[{{Chambers} {et~al.}(2016){Chambers}, {Magnier}, {Metcalfe},
  {Flewelling}, {Huber}, {Waters}, {Denneau}, {Draper}, {Farrow}, {Finkbeiner},
  {Holmberg}, {Koppenhoefer}, {Price}, {Rest}, {Saglia}, {Schlafly}, {Smartt},
  {Sweeney}, {Wainscoat}, {Burgett}, {Chastel}, {Grav}, {Heasley}, {Hodapp},
  {Jedicke}, {Kaiser}, {Kudritzki}, {Luppino}, {Lupton}, {Monet}, {Morgan},
  {Onaka}, {Shiao}, {Stubbs}, {Tonry}, {White}, {Ba{\~n}ados}, {Bell},
  {Bender}, {Bernard}, {Boegner}, {Boffi}, {Botticella}, {Calamida},
  {Casertano}, {Chen}, {Chen}, {Cole}, {Deacon}, {Frenk}, {Fitzsimmons},
  {Gezari}, {Gibbs}, {Goessl}, {Goggia}, {Gourgue}, {Goldman}, {Grant},
  {Grebel}, {Hambly}, {Hasinger}, {Heavens}, {Heckman}, {Henderson}, {Henning},
  {Holman}, {Hopp}, {Ip}, {Isani}, {Jackson}, {Keyes}, {Koekemoer}, {Kotak},
  {Le}, {Liska}, {Long}, {Lucey}, {Liu}, {Martin}, {Masci}, {McLean}, {Mindel},
  {Misra}, {Morganson}, {Murphy}, {Obaika}, {Narayan}, {Nieto-Santisteban},
  {Norberg}, {Peacock}, {Pier}, {Postman}, {Primak}, {Rae}, {Rai}, {Riess},
  {Riffeser}, {Rix}, {R{\"o}ser}, {Russel}, {Rutz}, {Schilbach}, {Schultz},
  {Scolnic}, {Strolger}, {Szalay}, {Seitz}, {Small}, {Smith}, {Soderblom},
  {Taylor}, {Thomson}, {Taylor}, {Thakar}, {Thiel}, {Thilker}, {Unger},
  {Urata}, {Valenti}, {Wagner}, {Walder}, {Walter}, {Watters}, {Werner},
  {Wood-Vasey}, \& {Wyse}}]{Cham16}
{Chambers}, K.~C., {Magnier}, E.~A., {Metcalfe}, N., {et~al.} 2016,
  \JournalTitle{arXiv e-prints},
  \href{http://arxiv.org/abs/1612.05560}{{\sffamily arXiv:1612.05560
  [astro-ph.IM]}}

\bibitem[{Cort\'es~Contreras(2016)}]{Cort16}
Cort\'es~Contreras, M. 2016, PhD thesis, Universidad Complutense de Madrid,
  Madrid, Spain

\bibitem[{{Costa} {et~al.}(2005){Costa}, {M{\'e}ndez}, {Jao}, {Henry},
  {Subasavage}, {Brown}, {Ianna}, \& {Bartlett}}]{Cost05}
{Costa}, E., {M{\'e}ndez}, R.~A., {Jao}, W.-C., {et~al.} 2005,
  \href{http://dx.doi.org/10.1086/430473}{\JournalTitle{\aj}, 130, 337}

\bibitem[{{Costa} {et~al.}(2006){Costa}, {M{\'e}ndez}, {Jao}, {Henry},
  {Subasavage}, \& {Ianna}}]{Cost06}
---. 2006, \href{http://dx.doi.org/10.1086/505706}{\JournalTitle{\aj}, 132,
  1234}

\bibitem[{{Cruz} {et~al.}(2009){Cruz}, {Kirkpatrick}, \& {Burgasser}}]{Cruz09}
{Cruz}, K.~L., {Kirkpatrick}, J.~D., \& {Burgasser}, A.~J. 2009,
  \href{http://dx.doi.org/10.1088/0004-6256/137/2/3345}{\JournalTitle{\aj},
  137, 3345}

\bibitem[{{Cruz} {et~al.}(2018){Cruz}, {N{\'u}{\~n}ez}, {Burgasser},
  {Abrahams}, {Rice}, {Reid}, \& {Looper}}]{Cruz18}
{Cruz}, K.~L., {N{\'u}{\~n}ez}, A., {Burgasser}, A.~J., {et~al.} 2018,
  \href{http://dx.doi.org/10.3847/1538-3881/aa9d8a}{\JournalTitle{\aj}, 155,
  34}

\bibitem[{{Cruz} {et~al.}(2003){Cruz}, {Reid}, {Liebert}, {Kirkpatrick}, \&
  {Lowrance}}]{Cruz03}
{Cruz}, K.~L., {Reid}, I.~N., {Liebert}, J., {Kirkpatrick}, J.~D., \&
  {Lowrance}, P.~J. 2003,
  \href{http://dx.doi.org/10.1086/378607}{\JournalTitle{\aj}, 126, 2421}

\bibitem[{{Cruz} {et~al.}(2007){Cruz}, {Reid}, {Kirkpatrick}, {Burgasser},
  {Liebert}, {Solomon}, {Schmidt}, {Allen}, {Hawley}, \& {Covey}}]{Cruz07}
{Cruz}, K.~L., {Reid}, I.~N., {Kirkpatrick}, J.~D., {et~al.} 2007,
  \href{http://dx.doi.org/10.1086/510132}{\JournalTitle{\aj}, 133, 439}

\bibitem[{{Cushing} {et~al.}(2009){Cushing}, {Looper}, {Burgasser},
  {Kirkpatrick}, {Faherty}, {Cruz}, {Sweet}, \& {Sanderson}}]{Cush09}
{Cushing}, M.~C., {Looper}, D., {Burgasser}, A.~J., {et~al.} 2009,
  \href{http://dx.doi.org/10.1088/0004-637X/696/1/986}{\JournalTitle{\apj},
  696, 986}

\bibitem[{{Cushing} {et~al.}(2005){Cushing}, {Rayner}, \& {Vacca}}]{Cush05}
{Cushing}, M.~C., {Rayner}, J.~T., \& {Vacca}, W.~D. 2005,
  \href{http://dx.doi.org/10.1086/428040}{\JournalTitle{\apj}, 623, 1115}

\bibitem[{{Cushing} \& {Vacca}(2006)}]{Cush06a}
{Cushing}, M.~C., \& {Vacca}, W.~D. 2006,
  \href{http://dx.doi.org/10.1086/499583}{\JournalTitle{\aj}, 131, 1797}

\bibitem[{{Cushing} {et~al.}(2004){Cushing}, {Vacca}, \& {Rayner}}]{Cush04}
{Cushing}, M.~C., {Vacca}, W.~D., \& {Rayner}, J.~T. 2004,
  \href{http://dx.doi.org/10.1086/382907}{\JournalTitle{\pasp}, 116, 362}

\bibitem[{{Cushing} {et~al.}(2006){Cushing}, {Roellig}, {Marley}, {Saumon},
  {Leggett}, {Kirkpatrick}, {Wilson}, {Sloan}, {Mainzer}, {Van Cleve}, \&
  {Houck}}]{Cush06b}
{Cushing}, M.~C., {Roellig}, T.~L., {Marley}, M.~S., {et~al.} 2006,
  \href{http://dx.doi.org/10.1086/505637}{\JournalTitle{\apj}, 648, 614}

\bibitem[{{Cutri} \& {et al.}(2012)}]{Cutr12}
{Cutri}, R.~M., \& {et al.} 2012, \JournalTitle{VizieR Online Data Catalog},
  2311

\bibitem[{{Cutri} \& {et al.}(2014)}]{Cutr13}
---. 2014, \JournalTitle{VizieR Online Data Catalog}, 2328

\bibitem[{{Cutri} {et~al.}(2003){Cutri}, {Skrutskie}, {van Dyk}, {Beichman},
  {Carpenter}, {Chester}, {Cambresy}, {Evans}, {Fowler}, {Gizis}, {Howard},
  {Huchra}, {Jarrett}, {Kopan}, {Kirkpatrick}, {Light}, {Marsh}, {McCallon},
  {Schneider}, {Stiening}, {Sykes}, {Weinberg}, {Wheaton}, {Wheelock}, \&
  {Zacarias}}]{Cutr03}
{Cutri}, R.~M., {Skrutskie}, M.~F., {van Dyk}, S., {et~al.} 2003, {2MASS All
  Sky Catalog of point sources.}

\bibitem[{{Dahn} {et~al.}(2008){Dahn}, {Harris}, {Levine}, {Tilleman}, {Monet},
  {Stone}, {Guetter}, {Canzian}, {Pier}, {Hartkopf}, {Liebert}, \&
  {Cushing}}]{Dahn08}
{Dahn}, C.~C., {Harris}, H.~C., {Levine}, S.~E., {et~al.} 2008,
  \href{http://dx.doi.org/10.1086/591050}{\JournalTitle{\apj}, 686, 548}

\bibitem[{{Dahn} {et~al.}(2017){Dahn}, {Harris}, {Subasavage}, {Ables},
  {Canzian}, {Guetter}, {Harris}, {Henden}, {Leggett}, {Levine}, {Luginbuhl},
  {Monet}, {Monet}, {Munn}, {Pier}, {Stone}, {Vrba}, {Walker}, \&
  {Tilleman}}]{Dahn17}
{Dahn}, C.~C., {Harris}, H.~C., {Subasavage}, J.~P., {et~al.} 2017,
  \href{http://dx.doi.org/10.3847/1538-3881/aa880b}{\JournalTitle{\aj}, 154,
  147}

\bibitem[{{DENIS Consortium}(2005)}]{DENIS}
{DENIS Consortium}. 2005, \JournalTitle{VizieR Online Data Catalog}, 2263

\bibitem[{{Dieterich} {et~al.}(2014){Dieterich}, {Henry}, {Jao}, {Winters},
  {Hosey}, {Riedel}, \& {Subasavage}}]{Diet14}
{Dieterich}, S.~B., {Henry}, T.~J., {Jao}, W.-C., {et~al.} 2014,
  \href{http://dx.doi.org/10.1088/0004-6256/147/5/94}{\JournalTitle{\aj}, 147,
  94}

\bibitem[{{Dittmann} {et~al.}(2017){Dittmann}, {Irwin}, {Charbonneau},
  {Bonfils}, {Astudillo-Defru}, {Haywood}, {Berta-Thompson}, {Newton},
  {Rodriguez}, {Winters}, {Tan}, {Almenara}, {Bouchy}, {Delfosse}, {Forveille},
  {Lovis}, {Murgas}, {Pepe}, {Santos}, {Udry}, {W{\"u}nsche}, {Esquerdo},
  {Latham}, \& {Dressing}}]{Ditt17}
{Dittmann}, J.~A., {Irwin}, J.~M., {Charbonneau}, D., {et~al.} 2017,
  \href{http://dx.doi.org/10.1038/nature22055}{\JournalTitle{\nat}, 544, 333}

\bibitem[{Dressing \& Charbonneau(2015)}]{Dres15}
Dressing, C.~D., \& Charbonneau, D. 2015,
  \href{http://stacks.iop.org/0004-637X/807/i=1/a=45}{\JournalTitle{The
  Astrophysical Journal}, 807, 45}

\bibitem[{{Dupuy} \& {Kraus}(2013)}]{Dupu13}
{Dupuy}, T.~J., \& {Kraus}, A.~L. 2013,
  \href{http://dx.doi.org/10.1126/science.1241917}{\JournalTitle{Science}, 341,
  1492}

\bibitem[{{Dupuy} \& {Liu}(2012)}]{Dupu12a}
{Dupuy}, T.~J., \& {Liu}, M.~C. 2012,
  \href{http://dx.doi.org/10.1088/0067-0049/201/2/19}{\JournalTitle{\apjs},
  201, 19}

\bibitem[{{Evans} {et~al.}(2018){Evans}, {Riello}, {De Angeli}, {Carrasco},
  {Montegriffo}, {Fabricius}, {Jordi}, {Palaversa}, {Diener}, {Busso},
  {Cacciari}, {van Leeuwen}, {Burgess}, {Davidson}, {Harrison}, {Hodgkin},
  {Pancino}, {Richards}, {Altavilla}, {Balaguer-N{\'u}{\~n}ez}, {Barstow},
  {Bellazzini}, {Brown}, {Castellani}, {Cocozza}, {De Luise}, {Delgado},
  {Ducourant}, {Galleti}, {Gilmore}, {Giuffrida}, {Holl}, {Kewley}, {Koposov},
  {Marinoni}, {Marrese}, {Osborne}, {Piersimoni}, {Portell}, {Pulone},
  {Ragaini}, {Sanna}, {Terrett}, {Walton}, {Wevers}, \& {Wyrzykowski}}]{Evan18}
{Evans}, D.~W., {Riello}, M., {De Angeli}, F., {et~al.} 2018,
  \href{http://dx.doi.org/10.1051/0004-6361/201832756}{\JournalTitle{\aap},
  616, A4}

\bibitem[{{Faherty} {et~al.}(2009){Faherty}, {Burgasser}, {Cruz}, {Shara},
  {Walter}, \& {Gelino}}]{Fahe09}
{Faherty}, J.~K., {Burgasser}, A.~J., {Cruz}, K.~L., {et~al.} 2009,
  \href{http://dx.doi.org/10.1088/0004-6256/137/1/1}{\JournalTitle{\aj}, 137,
  1}

\bibitem[{{Faherty} {et~al.}(2012){Faherty}, {Burgasser}, {Walter}, {Van der
  Bliek}, {Shara}, {Cruz}, {West}, {Vrba}, \& {Anglada-Escud{\'e}}}]{Fahe12}
{Faherty}, J.~K., {Burgasser}, A.~J., {Walter}, F.~M., {et~al.} 2012,
  \href{http://dx.doi.org/10.1088/0004-637X/752/1/56}{\JournalTitle{\apj}, 752,
  56}

\bibitem[{{Faherty} {et~al.}(2016){Faherty}, {Riedel}, {Cruz}, {Gagne},
  {Filippazzo}, {Lambrides}, {Fica}, {Weinberger}, {Thorstensen}, {Tinney},
  {Baldassare}, {Lemonier}, \& {Rice}}]{Fahe16}
{Faherty}, J.~K., {Riedel}, A.~R., {Cruz}, K.~L., {et~al.} 2016,
  \href{http://dx.doi.org/10.3847/0067-0049/225/1/10}{\JournalTitle{\apjs},
  225, 10}

\bibitem[{{Feiden} \& {Chaboyer}(2012)}]{Feid12}
{Feiden}, G.~A., \& {Chaboyer}, B. 2012,
  \href{http://dx.doi.org/10.1088/0004-637X/761/1/30}{\JournalTitle{\apj}, 761,
  30}

\bibitem[{{Filippazzo} {et~al.}(2015){Filippazzo}, {Rice}, {Faherty}, {Cruz},
  {Van Gordon}, \& {Looper}}]{Fili15}
{Filippazzo}, J.~C., {Rice}, E.~L., {Faherty}, J., {et~al.} 2015,
  \href{http://dx.doi.org/10.1088/0004-637X/810/2/158}{\JournalTitle{\apj},
  810, 158}

\bibitem[{{Gagn{\'e}} {et~al.}(2014){Gagn{\'e}}, {Faherty}, {Cruz},
  {Lafreni{\`e}re}, {Doyon}, {Malo}, \& {Artigau}}]{Gagn14b}
{Gagn{\'e}}, J., {Faherty}, J.~K., {Cruz}, K., {et~al.} 2014,
  \href{http://dx.doi.org/10.1088/2041-8205/785/1/L14}{\JournalTitle{\apjl},
  785, L14}

\bibitem[{Gagn\'e {et~al.}(2015)Gagn\'e, Lambrides, Faherty, \&
  Simcoe}]{zenodofirehose}
Gagn\'e, J., Lambrides, E., Faherty, J.~K., \& Simcoe, R. 2015, {Firehose v2.0,
  Zenodo}

\bibitem[{{Gagn{\'e}} {et~al.}(2015){Gagn{\'e}}, {Faherty}, {Cruz},
  {Lafreni{\'e}re}, {Doyon}, {Malo}, {Burgasser}, {Naud}, {Artigau},
  {Bouchard}, {Gizis}, \& {Albert}}]{Gagn15b}
{Gagn{\'e}}, J., {Faherty}, J.~K., {Cruz}, K.~L., {et~al.} 2015,
  \href{http://dx.doi.org/10.1088/0067-0049/219/2/33}{\JournalTitle{\apjs},
  219, 33}

\bibitem[{{Gagn{\'e}} {et~al.}(2017){Gagn{\'e}}, {Faherty}, {Burgasser},
  {Artigau}, {Bouchard}, {Albert}, {Lafreni{\`e}re}, {Doyon}, \& {Bardalez
  Gagliuffi}}]{Gagn17}
{Gagn{\'e}}, J., {Faherty}, J.~K., {Burgasser}, A.~J., {et~al.} 2017,
  \href{http://dx.doi.org/10.3847/2041-8213/aa70e2}{\JournalTitle{\apjl}, 841,
  L1}

\bibitem[{{Gaia Collaboration} {et~al.}(2018){Gaia Collaboration}, {Brown, A.
  G. A.}, {Vallenari, A.}, {Prusti, T.}, {de Bruijne, J. H. J.}, \& {et
  al.}}]{GaiaDR2}
{Gaia Collaboration}, {Brown, A. G. A.}, {Vallenari, A.}, {et~al.} 2018,
  \href{http://dx.doi.org/10.1051/0004-6361/201833051}{\JournalTitle{A\&A}}

\bibitem[{{Gaia Collaboration} {et~al.}(2016){Gaia Collaboration}, {Brown},
  {Vallenari}, {Prusti}, {de Bruijne}, {Mignard}, {Drimmel}, {Babusiaux},
  {Bailer-Jones}, {Bastian}, \& et~al.}]{GaiaDR1}
{Gaia Collaboration}, {Brown}, A.~G.~A., {Vallenari}, A., {et~al.} 2016,
  \href{http://dx.doi.org/10.1051/0004-6361/201629512}{\JournalTitle{\aap},
  595, A2}

\bibitem[{{Gaidos} {et~al.}(2016){Gaidos}, {Mann}, {Kraus}, \&
  {Ireland}}]{Gaid16}
{Gaidos}, E., {Mann}, A.~W., {Kraus}, A.~L., \& {Ireland}, M. 2016,
  \href{http://dx.doi.org/10.1093/mnras/stw097}{\JournalTitle{\mnras}, 457,
  2877}

\bibitem[{{Geballe} {et~al.}(2002){Geballe}, {Knapp}, {Leggett}, {Fan},
  {Golimowski}, {Anderson}, {Brinkmann}, {Csabai}, {Gunn}, {Hawley},
  {Hennessy}, {Henry}, {Hill}, {Hindsley}, {Ivezi{\'c}}, {Lupton}, {McDaniel},
  {Munn}, {Narayanan}, {Peng}, {Pier}, {Rockosi}, {Schneider}, {Smith},
  {Strauss}, {Tsvetanov}, {Uomoto}, {York}, \& {Zheng}}]{Geba02}
{Geballe}, T.~R., {Knapp}, G.~R., {Leggett}, S.~K., {et~al.} 2002,
  \href{http://dx.doi.org/10.1086/324078}{\JournalTitle{\apj}, 564, 466}

\bibitem[{{Gillon} {et~al.}(2016){Gillon}, {Jehin}, {Lederer}, {Delrez}, {de
  Wit}, {Burdanov}, {Van Grootel}, {Burgasser}, {Triaud}, {Opitom}, {Demory},
  {Sahu}, {Bardalez Gagliuffi}, {Magain}, \& {Queloz}}]{Gill16}
{Gillon}, M., {Jehin}, E., {Lederer}, S.~M., {et~al.} 2016,
  \href{http://dx.doi.org/10.1038/nature17448}{\JournalTitle{\nat}, 533, 221}

\bibitem[{{Gillon} {et~al.}(2017){Gillon}, {Triaud}, {Demory}, {Jehin}, {Agol},
  {Deck}, {Lederer}, {de Wit}, {Burdanov}, {Ingalls}, {Bolmont}, {Leconte},
  {Raymond}, {Selsis}, {Turbet}, {Barkaoui}, {Burgasser}, {Burleigh}, {Carey},
  {Chaushev}, {Copperwheat}, {Delrez}, {Fernandes}, {Holdsworth}, {Kotze}, {Van
  Grootel}, {Almleaky}, {Benkhaldoun}, {Magain}, \& {Queloz}}]{Gill17}
{Gillon}, M., {Triaud}, A.~H.~M.~J., {Demory}, B.-O., {et~al.} 2017,
  \href{http://dx.doi.org/10.1038/nature21360}{\JournalTitle{\nat}, 542, 456}

\bibitem[{{Gizis}(1997)}]{Gizi97}
{Gizis}, J.~E. 1997,
  \href{http://dx.doi.org/10.1086/118302}{\JournalTitle{\aj}, 113, 806}

\bibitem[{{Gizis}(2002)}]{Gizi02b}
---. 2002, \href{http://dx.doi.org/10.1086/341259}{\JournalTitle{\apj}, 575,
  484}

\bibitem[{{Gizis} {et~al.}(2000){Gizis}, {Monet}, {Reid}, {Kirkpatrick},
  {Liebert}, \& {Williams}}]{Gizi00}
{Gizis}, J.~E., {Monet}, D.~G., {Reid}, I.~N., {et~al.} 2000,
  \href{http://dx.doi.org/10.1086/301456}{\JournalTitle{\aj}, 120, 1085}

\bibitem[{{Golimowski} {et~al.}(2004){Golimowski}, {Leggett}, {Marley}, {Fan},
  {Geballe}, {Knapp}, {Vrba}, {Henden}, {Luginbuhl}, {Guetter}, {Munn},
  {Canzian}, {Zheng}, {Tsvetanov}, {Chiu}, {Glazebrook}, {Hoversten},
  {Schneider}, \& {Brinkmann}}]{Goli04a}
{Golimowski}, D.~A., {Leggett}, S.~K., {Marley}, M.~S., {et~al.} 2004,
  \href{http://dx.doi.org/10.1086/420709}{\JournalTitle{\aj}, 127, 3516}

\bibitem[{{Gonzales} {et~al.}(2018){Gonzales}, {Faherty}, {Gagn{\'e}},
  {Artigau}, \& {Bardalez Gagliuffi}}]{Gonz18}
{Gonzales}, E.~C., {Faherty}, J.~K., {Gagn{\'e}}, J., {Artigau}, {\'E}., \&
  {Bardalez Gagliuffi}, D. 2018,
  \href{http://stacks.iop.org/0004-637X/864/i=1/a=100}{\JournalTitle{\apj},
  864, 100}

\bibitem[{{Hawley} {et~al.}(2002){Hawley}, {Covey}, {Knapp}, {Golimowski},
  {Fan}, {Anderson}, {Gunn}, {Harris}, {Ivezi{\'c}}, {Long}, {Lupton},
  {McGehee}, {Narayanan}, {Peng}, {Schlegel}, {Schneider}, {Spahn}, {Strauss},
  {Szkody}, {Tsvetanov}, {Walkowicz}, {Brinkmann}, {Harvanek}, {Hennessy},
  {Kleinman}, {Krzesinski}, {Long}, {Neilsen}, {Newman}, {Nitta}, {Snedden}, \&
  {York}}]{Hawl02}
{Hawley}, S.~L., {Covey}, K.~R., {Knapp}, G.~R., {et~al.} 2002,
  \href{http://dx.doi.org/10.1086/340697}{\JournalTitle{\aj}, 123, 3409}

\bibitem[{{Henry} \& {Kirkpatrick}(1990)}]{Henr90}
{Henry}, T.~J., \& {Kirkpatrick}, J.~D. 1990,
  \href{http://dx.doi.org/10.1086/185715}{\JournalTitle{\apjl}, 354, L29}

\bibitem[{{Kane} {et~al.}(2016){Kane}, {Hill}, {Kasting}, {Kopparapu},
  {Quintana}, {Barclay}, {Batalha}, {Borucki}, {Ciardi}, {Haghighipour},
  {Hinkel}, {Kaltenegger}, {Selsis}, \& {Torres}}]{Kane16}
{Kane}, S.~R., {Hill}, M.~L., {Kasting}, J.~F., {et~al.} 2016,
  \href{http://dx.doi.org/10.3847/0004-637X/830/1/1}{\JournalTitle{\apj}, 830,
  1}

\bibitem[{{Kellogg} {et~al.}(2016){Kellogg}, {Metchev}, {Gagn{\'e}}, \&
  {Faherty}}]{Kell16}
{Kellogg}, K., {Metchev}, S., {Gagn{\'e}}, J., \& {Faherty}, J. 2016,
  \href{http://dx.doi.org/10.3847/2041-8205/821/1/L15}{\JournalTitle{\apjl},
  821, L15}

\bibitem[{{Kirkpatrick} {et~al.}(2006){Kirkpatrick}, {Barman}, {Burgasser},
  {McGovern}, {McLean}, {Tinney}, \& {Lowrance}}]{Kirk06}
{Kirkpatrick}, J.~D., {Barman}, T.~S., {Burgasser}, A.~J., {et~al.} 2006,
  \href{http://dx.doi.org/10.1086/499622}{\JournalTitle{\apj}, 639, 1120}

\bibitem[{{Kirkpatrick} {et~al.}(1991){Kirkpatrick}, {Henry}, \&
  {McCarthy}}]{Kirk91}
{Kirkpatrick}, J.~D., {Henry}, T.~J., \& {McCarthy}, Jr., D.~W. 1991,
  \href{http://dx.doi.org/10.1086/191611}{\JournalTitle{\apjs}, 77, 417}

\bibitem[{{Kirkpatrick} {et~al.}(1995){Kirkpatrick}, {Henry}, \&
  {Simons}}]{Kirk95}
{Kirkpatrick}, J.~D., {Henry}, T.~J., \& {Simons}, D.~A. 1995,
  \href{http://dx.doi.org/10.1086/117323}{\JournalTitle{\aj}, 109, 797}

\bibitem[{{Kirkpatrick} {et~al.}(2008){Kirkpatrick}, {Cruz}, {Barman},
  {Burgasser}, {Looper}, {Tinney}, {Gelino}, {Lowrance}, {Liebert},
  {Carpenter}, {Hillenbrand}, \& {Stauffer}}]{Kirk08}
{Kirkpatrick}, J.~D., {Cruz}, K.~L., {Barman}, T.~S., {et~al.} 2008,
  \href{http://dx.doi.org/10.1086/592768}{\JournalTitle{\apj}, 689, 1295}

\bibitem[{{Kirkpatrick} {et~al.}(2010){Kirkpatrick}, {Looper}, {Burgasser},
  {Schurr}, {Cutri}, {Cushing}, {Cruz}, {Sweet}, {Knapp}, {Barman},
  {Bochanski}, {Roellig}, {McLean}, {McGovern}, \& {Rice}}]{Kirk10}
{Kirkpatrick}, J.~D., {Looper}, D.~L., {Burgasser}, A.~J., {et~al.} 2010,
  \href{http://dx.doi.org/10.1088/0067-0049/190/1/100}{\JournalTitle{\apjs},
  190, 100}

\bibitem[{{Koren} {et~al.}(2016){Koren}, {Blake}, {Dahn}, \& {Harris}}]{Kore16}
{Koren}, S.~C., {Blake}, C.~H., {Dahn}, C.~C., \& {Harris}, H.~C. 2016,
  \href{http://dx.doi.org/10.3847/0004-6256/151/3/57}{\JournalTitle{\aj}, 151,
  57}

\bibitem[{{Lawrence} {et~al.}(2012){Lawrence}, {Warren}, {Almaini}, {Edge},
  {Hambly}, {Jameson}, {Lucas}, {Casali}, {Adamson}, {Dye}, {Emerson},
  {Foucaud}, {Hewett}, {Hirst}, {Hodgkin}, {Irwin}, {Lodieu}, {McMahon},
  {Simpson}, {Smail}, {Mortlock}, \& {Folger}}]{Lawr12}
{Lawrence}, A., {Warren}, S.~J., {Almaini}, O., {et~al.} 2012,
  \JournalTitle{VizieR Online Data Catalog}, 2314

\bibitem[{{Leggett} {et~al.}(1998){Leggett}, {Allard}, \&
  {Hauschildt}}]{Legg98}
{Leggett}, S.~K., {Allard}, F., \& {Hauschildt}, P.~H. 1998,
  \href{http://dx.doi.org/10.1086/306517}{\JournalTitle{\apj}, 509, 836}

\bibitem[{{Leggett} {et~al.}(2002){Leggett}, {Golimowski}, {Fan}, {Geballe},
  {Knapp}, {Brinkmann}, {Csabai}, {Gunn}, {Hawley}, {Henry}, {Hindsley},
  {Ivezi{\'c}}, {Lupton}, {Pier}, {Schneider}, {Smith}, {Strauss}, {Uomoto}, \&
  {York}}]{Legg02a}
{Leggett}, S.~K., {Golimowski}, D.~A., {Fan}, X., {et~al.} 2002,
  \href{http://dx.doi.org/10.1086/324037}{\JournalTitle{\apj}, 564, 452}

\bibitem[{{L{\'e}pine} {et~al.}(2003{\natexlab{a}}){L{\'e}pine}, {Rich}, \&
  {Shara}}]{Lepi03c}
{L{\'e}pine}, S., {Rich}, R.~M., \& {Shara}, M.~M. 2003{\natexlab{a}},
  \href{http://dx.doi.org/10.1086/377069}{\JournalTitle{\apjl}, 591, L49}

\bibitem[{{L{\'e}pine} {et~al.}(2003{\natexlab{b}}){L{\'e}pine}, {Rich}, \&
  {Shara}}]{Lepi03a}
---. 2003{\natexlab{b}},
  \href{http://dx.doi.org/10.1086/345972}{\JournalTitle{\aj}, 125, 1598}

\bibitem[{{L{\'e}pine} {et~al.}(2002){L{\'e}pine}, {Shara}, \& {Rich}}]{Lepi02}
{L{\'e}pine}, S., {Shara}, M.~M., \& {Rich}, R.~M. 2002,
  \href{http://dx.doi.org/10.1086/341783}{\JournalTitle{\aj}, 124, 1190}

\bibitem[{{Liebert} {et~al.}(1979){Liebert}, {Dahn}, {Gresham}, \&
  {Strittmatter}}]{Lieb79}
{Liebert}, J., {Dahn}, C.~C., {Gresham}, M., \& {Strittmatter}, P.~A. 1979,
  \href{http://dx.doi.org/10.1086/157384}{\JournalTitle{\apj}, 233, 226}

\bibitem[{{Lindegren, L.} {et~al.}(2018){Lindegren, L.}, {Hernandez, J.},
  {Bombrun, A.}, {Klioner, S.}, {Bastian, U.}, \& {Ramos-Lerate, M.}}]{Lind18}
{Lindegren, L.}, {Hernandez, J.}, {Bombrun, A.}, {et~al.} 2018,
  \href{http://dx.doi.org/10.1051/0004-6361/201832727}{\JournalTitle{A\&A}}

\bibitem[{{Liu} {et~al.}(2016){Liu}, {Dupuy}, \& {Allers}}]{Liu_16}
{Liu}, M.~C., {Dupuy}, T.~J., \& {Allers}, K.~N. 2016,
  \href{http://dx.doi.org/10.3847/1538-4357/833/1/96}{\JournalTitle{\apj}, 833,
  96}

\bibitem[{{Liu} {et~al.}(2013){Liu}, {Magnier}, {Deacon}, {Allers}, {Dupuy},
  {Kotson}, {Aller}, {Burgett}, {Chambers}, {Draper}, {Hodapp}, {Jedicke},
  {Kaiser}, {Kudritzki}, {Metcalfe}, {Morgan}, {Price}, {Tonry}, \&
  {Wainscoat}}]{Liu_13}
{Liu}, M.~C., {Magnier}, E.~A., {Deacon}, N.~R., {et~al.} 2013,
  \href{http://dx.doi.org/10.1088/2041-8205/777/2/L20}{\JournalTitle{\apjl},
  777, L20}

\bibitem[{{Luger} {et~al.}(2017){Luger}, {Sestovic}, {Kruse}, {Grimm},
  {Demory}, {Agol}, {Bolmont}, {Fabrycky}, {Fernandes}, {Van Grootel},
  {Burgasser}, {Gillon}, {Ingalls}, {Jehin}, {Raymond}, {Selsis}, {Triaud},
  {Barclay}, {Barentsen}, {Howell}, {Delrez}, {de Wit}, {Foreman-Mackey},
  {Holdsworth}, {Leconte}, {Lederer}, {Turbet}, {Almleaky}, {Benkhaldoun},
  {Magain}, {Morris}, {Heng}, \& {Queloz}}]{Luge17}
{Luger}, R., {Sestovic}, M., {Kruse}, E., {et~al.} 2017,
  \href{http://dx.doi.org/10.1038/s41550-017-0129}{\JournalTitle{Nature
  Astronomy}, 1, 0129}

\bibitem[{{Luhman} {et~al.}(2009){Luhman}, {Mamajek}, {Allen}, \&
  {Cruz}}]{Luhm09b}
{Luhman}, K.~L., {Mamajek}, E.~E., {Allen}, P.~R., \& {Cruz}, K.~L. 2009,
  \href{http://dx.doi.org/10.1088/0004-637X/703/1/399}{\JournalTitle{\apj},
  703, 399}

\bibitem[{{Luyten}(1979)}]{Luyt79a}
{Luyten}, W.~J. 1979, {LHS catalogue. A catalogue of stars with proper motions
  exceeding 0``5 annually}

\bibitem[{{Martin} {et~al.}(2017){Martin}, {Mace}, {McLean}, {Logsdon}, {Rice},
  {Kirkpatrick}, {Burgasser}, {McGovern}, \& {Prato}}]{Martin17}
{Martin}, E.~C., {Mace}, G.~N., {McLean}, I.~S., {et~al.} 2017,
  \href{http://dx.doi.org/10.3847/1538-4357/aa6338}{\JournalTitle{\apj}, 838,
  73}

\bibitem[{{Ment} {et~al.}(2018){Ment}, {Dittmann}, {Astudillo-Defru},
  {Charbonneau}, {Irwin}, {Bonfils}, {Murgas}, {Almenara}, {Forveille}, {Agol},
  {Ballard}, {Berta-Thompson}, {Bouchy}, {Cloutier}, {Delfosse}, {Doyon},
  {Dressing}, {Esquerdo}, {Haywood}, {Kipping}, {Latham}, {Lovis}, {Newton},
  {Pepe}, {Rodriguez}, {Santos}, {Tan}, {Udry}, {Winters}, \&
  {W{\"u}nsche}}]{Ment18}
{Ment}, K., {Dittmann}, J.~A., {Astudillo-Defru}, N., {et~al.} 2018,
  \JournalTitle{ArXiv e-prints},
  \href{http://arxiv.org/abs/1808.00485}{{\sffamily arXiv:1808.00485
  [astro-ph.EP]}}

\bibitem[{{Montes} {et~al.}(2001){Montes}, {L{\'o}pez-Santiago}, {G{\'a}lvez},
  {Fern{\'a}ndez-Figueroa}, {De Castro}, \& {Cornide}}]{Mont01}
{Montes}, D., {L{\'o}pez-Santiago}, J., {G{\'a}lvez}, M.~C., {et~al.} 2001,
  \href{http://dx.doi.org/10.1046/j.1365-8711.2001.04781.x}{\JournalTitle{\mnras},
  328, 45}

\bibitem[{{Morris} {et~al.}(2018){Morris}, {Agol}, {Hebb}, {Hawley}, {Gillon},
  {Ducrot}, {Delrez}, {Ingalls}, \& {Demory}}]{Morr18}
{Morris}, B.~M., {Agol}, E., {Hebb}, L., {et~al.} 2018,
  \href{http://dx.doi.org/10.3847/2041-8213/aad8aa}{\JournalTitle{\apjl}, 863,
  L32}

\bibitem[{{Mulders} {et~al.}(2015){Mulders}, {Pascucci}, \& {Apai}}]{Muld15c}
{Mulders}, G.~D., {Pascucci}, I., \& {Apai}, D. 2015,
  \href{http://dx.doi.org/10.1088/0004-637X/814/2/130}{\JournalTitle{\apj},
  814, 130}

\bibitem[{{Nissen}(2004)}]{Niss04}
{Nissen}, P.~E. 2004, in Origin and Evolution of the Elements, ed.
  A.~{McWilliam} \& M.~{Rauch}, 154

\bibitem[{{Patten} {et~al.}(2006){Patten}, {Stauffer}, {Burrows}, {Marengo},
  {Hora}, {Luhman}, {Sonnett}, {Henry}, {Raghavan}, {Megeath}, {Liebert}, \&
  {Fazio}}]{Patt06}
{Patten}, B.~M., {Stauffer}, J.~R., {Burrows}, A., {et~al.} 2006,
  \href{http://dx.doi.org/10.1086/507264}{\JournalTitle{\apj}, 651, 502}

\bibitem[{{Phan-Bao} {et~al.}(2003){Phan-Bao}, {Crifo}, {Delfosse},
  {Forveille}, {Guibert}, {Borsenberger}, {Epchtein}, {Fouqu{\'e}}, {Simon}, \&
  {Vetois}}]{Phan03}
{Phan-Bao}, N., {Crifo}, F., {Delfosse}, X., {et~al.} 2003,
  \href{http://dx.doi.org/10.1051/0004-6361:20030188}{\JournalTitle{\aap}, 401,
  959}

\bibitem[{{Rajpurohit} {et~al.}(2016){Rajpurohit}, {Reyl{\'e}}, {Allard},
  {Homeier}, {Bayo}, {Mousis}, {Rajpurohit}, \&
  {Fern{\'a}ndez-Trincado}}]{Rajp16}
{Rajpurohit}, A.~S., {Reyl{\'e}}, C., {Allard}, F., {et~al.} 2016,
  \href{http://dx.doi.org/10.1051/0004-6361/201526776}{\JournalTitle{\aap},
  596, A33}

\bibitem[{{Rayner} {et~al.}(2009){Rayner}, {Cushing}, \& {Vacca}}]{Rayn09}
{Rayner}, J.~T., {Cushing}, M.~C., \& {Vacca}, W.~D. 2009,
  \href{http://dx.doi.org/10.1088/0067-0049/185/2/289}{\JournalTitle{\apjs},
  185, 289}

\bibitem[{{Reid} \& {Cruz}(2002)}]{Reid02b}
{Reid}, I.~N., \& {Cruz}, K.~L. 2002,
  \href{http://dx.doi.org/10.1086/339699}{\JournalTitle{\aj}, 123, 2806}

\bibitem[{{Reid} {et~al.}(2008){Reid}, {Cruz}, {Kirkpatrick}, {Allen},
  {Mungall}, {Liebert}, {Lowrance}, \& {Sweet}}]{Reid08b}
{Reid}, I.~N., {Cruz}, K.~L., {Kirkpatrick}, J.~D., {et~al.} 2008,
  \href{http://dx.doi.org/10.1088/0004-6256/136/3/1290}{\JournalTitle{\aj},
  136, 1290}

\bibitem[{{Reid} \& {Gizis}(2005)}]{Reid05}
{Reid}, I.~N., \& {Gizis}, J.~E. 2005,
  \href{http://dx.doi.org/10.1086/430462}{\JournalTitle{\pasp}, 117, 676}

\bibitem[{{Reid} {et~al.}(2003){Reid}, {Cruz}, {Laurie}, {Liebert}, {Dahn},
  {Harris}, {Guetter}, {Stone}, {Canzian}, {Luginbuhl}, {Levine}, {Monet}, \&
  {Monet}}]{Reid03b}
{Reid}, I.~N., {Cruz}, K.~L., {Laurie}, S.~P., {et~al.} 2003,
  \href{http://dx.doi.org/10.1086/344946}{\JournalTitle{\aj}, 125, 354}

\bibitem[{{Reid}(1987)}]{Reid87}
{Reid}, N. 1987,
  \href{http://dx.doi.org/10.1093/mnras/225.4.873}{\JournalTitle{\mnras}, 225,
  873}

\bibitem[{Reiners \& Basri(2009)}]{Rein09}
Reiners, A., \& Basri, G. 2009,
  \href{http://stacks.iop.org/0004-637X/705/i=2/a=1416}{\JournalTitle{The
  Astrophysical Journal}, 705, 1416}

\bibitem[{{Reiners} \& {Basri}(2010)}]{Rein10}
{Reiners}, A., \& {Basri}, G. 2010,
  \href{http://dx.doi.org/10.1088/0004-637X/710/2/924}{\JournalTitle{\apj},
  710, 924}

\bibitem[{{Ricker} {et~al.}(2015){Ricker}, {Winn}, {Vanderspek}, {Latham},
  {Bakos}, {Bean}, {Berta-Thompson}, {Brown}, {Buchhave}, {Butler}, {Butler},
  {Chaplin}, {Charbonneau}, {Christensen-Dalsgaard}, {Clampin}, {Deming},
  {Doty}, {De Lee}, {Dressing}, {Dunham}, {Endl}, {Fressin}, {Ge}, {Henning},
  {Holman}, {Howard}, {Ida}, {Jenkins}, {Jernigan}, {Johnson}, {Kaltenegger},
  {Kawai}, {Kjeldsen}, {Laughlin}, {Levine}, {Lin}, {Lissauer}, {MacQueen},
  {Marcy}, {McCullough}, {Morton}, {Narita}, {Paegert}, {Palle}, {Pepe},
  {Pepper}, {Quirrenbach}, {Rinehart}, {Sasselov}, {Sato}, {Seager},
  {Sozzetti}, {Stassun}, {Sullivan}, {Szentgyorgyi}, {Torres}, {Udry}, \&
  {Villasenor}}]{Rick15}
{Ricker}, G.~R., {Winn}, J.~N., {Vanderspek}, R., {et~al.} 2015,
  \href{http://dx.doi.org/10.1117/1.JATIS.1.1.014003}{\JournalTitle{Journal of
  Astronomical Telescopes, Instruments, and Systems}, 1, 014003}

\bibitem[{{Riello} {et~al.}(2018){Riello}, {De Angeli}, {Evans}, {Busso},
  {Hambly}, {Davidson}, {Burgess}, {Montegriffo}, {Osborne}, {Kewley},
  {Carrasco}, {Fabricius}, {Jordi}, {Cacciari}, {van Leeuwen}, \&
  {Holland}}]{Riel18}
{Riello}, M., {De Angeli}, F., {Evans}, D.~W., {et~al.} 2018,
  \href{http://dx.doi.org/10.1051/0004-6361/201832712}{\JournalTitle{\aap},
  616, A3}

\bibitem[{{Saumon} {et~al.}(1996){Saumon}, {Hubbard}, {Burrows}, {Guillot},
  {Lunine}, \& {Chabrier}}]{Saum96}
{Saumon}, D., {Hubbard}, W.~B., {Burrows}, A., {et~al.} 1996,
  \href{http://dx.doi.org/10.1086/177027}{\JournalTitle{\apj}, 460, 993}

\bibitem[{{Saumon} \& {Marley}(2008)}]{Saum08}
{Saumon}, D., \& {Marley}, M.~S. 2008,
  \href{http://dx.doi.org/10.1086/592734}{\JournalTitle{\apj}, 689, 1327}

\bibitem[{{Schmidt} {et~al.}(2007){Schmidt}, {Cruz}, {Bongiorno}, {Liebert}, \&
  {Reid}}]{Schm07}
{Schmidt}, S.~J., {Cruz}, K.~L., {Bongiorno}, B.~J., {Liebert}, J., \& {Reid},
  I.~N. 2007, \href{http://dx.doi.org/10.1086/512158}{\JournalTitle{\aj}, 133,
  2258}

\bibitem[{{Schneider} {et~al.}(2011){Schneider}, {Melis}, {Song}, \&
  {Zuckerman}}]{Schn11}
{Schneider}, A., {Melis}, C., {Song}, I., \& {Zuckerman}, B. 2011,
  \href{http://dx.doi.org/10.1088/0004-637X/743/2/109}{\JournalTitle{\apj},
  743, 109}

\bibitem[{{Schneider} {et~al.}(2016){Schneider}, {Windsor}, {Cushing},
  {Kirkpatrick}, \& {Wright}}]{Schn16}
{Schneider}, A.~C., {Windsor}, J., {Cushing}, M.~C., {Kirkpatrick}, J.~D., \&
  {Wright}, E.~L. 2016,
  \href{http://dx.doi.org/10.3847/2041-8205/822/1/L1}{\JournalTitle{\apjl},
  822, L1}

\bibitem[{{Scholz} {et~al.}(2004{\natexlab{a}}){Scholz}, {Lehmann}, {Matute},
  \& {Zinnecker}}]{Scho04a}
{Scholz}, R.-D., {Lehmann}, I., {Matute}, I., \& {Zinnecker}, H.
  2004{\natexlab{a}},
  \href{http://dx.doi.org/10.1051/0004-6361:20041059}{\JournalTitle{\aap}, 425,
  519}

\bibitem[{{Scholz} {et~al.}(2004{\natexlab{b}}){Scholz}, {Lodieu}, \&
  {McCaughrean}}]{Scho04c}
{Scholz}, R.-D., {Lodieu}, N., \& {McCaughrean}, M.~J. 2004{\natexlab{b}},
  \href{http://dx.doi.org/10.1051/0004-6361:200400098}{\JournalTitle{\aap},
  428, L25}

\bibitem[{{Simcoe} {et~al.}(2013){Simcoe}, {Burgasser}, {Schechter}, {Fishner},
  {Bernstein}, {Bigelow}, {Pipher}, {Forrest}, {McMurtry}, {Smith}, \&
  {Bochanski}}]{Simcoe13}
{Simcoe}, R.~A., {Burgasser}, A.~J., {Schechter}, P.~L., {et~al.} 2013,
  \href{http://dx.doi.org/10.1086/670241}{\JournalTitle{\pasp}, 125, 270}

\bibitem[{{Sivarani} {et~al.}(2009){Sivarani}, {L{\'e}pine}, {Kembhavi}, \&
  {Gupchup}}]{Siva09}
{Sivarani}, T., {L{\'e}pine}, S., {Kembhavi}, A.~K., \& {Gupchup}, J. 2009,
  \href{http://dx.doi.org/10.1088/0004-637X/694/2/L140}{\JournalTitle{\apjl},
  694, L140}

\bibitem[{{Tanner} {et~al.}(2012){Tanner}, {White}, {Bailey}, {Blake}, {Blake},
  {Cruz}, {Burgasser}, \& {Kraus}}]{Tann12}
{Tanner}, A., {White}, R., {Bailey}, J., {et~al.} 2012,
  \href{http://dx.doi.org/10.1088/0067-0049/203/1/10}{\JournalTitle{\apjs},
  203, 10}

\bibitem[{{Teegarden} {et~al.}(2003){Teegarden}, {Pravdo}, {Hicks}, {Lawrence},
  {Shaklan}, {Covey}, {Fraser}, {Hawley}, {McGlynn}, \& {Reid}}]{Teeg03}
{Teegarden}, B.~J., {Pravdo}, S.~H., {Hicks}, M., {et~al.} 2003,
  \href{http://dx.doi.org/10.1086/375803}{\JournalTitle{\apjl}, 589, L51}

\bibitem[{{van Biesbroeck}(1961)}]{vanB61}
{van Biesbroeck}, G. 1961,
  \href{http://dx.doi.org/10.1086/108457}{\JournalTitle{\aj}, 66, 528}

\bibitem[{{Van Grootel} {et~al.}(2018){Van Grootel}, {Fernandes}, {Gillon},
  {Jehin}, {Manfroid}, {Scuflaire}, {Burgasser}, {Barkaoui}, {Benkhaldoun},
  {Burdanov}, {Delrez}, {Demory}, {de Wit}, {Queloz}, \& {Triaud}}]{vanG18}
{Van Grootel}, V., {Fernandes}, C.~S., {Gillon}, M., {et~al.} 2018,
  \href{http://dx.doi.org/10.3847/1538-4357/aaa023}{\JournalTitle{\apj}, 853,
  30}

\bibitem[{{Vida} {et~al.}(2017){Vida}, {K{\H{o}}v{\'a}ri}, {P{\'a}l},
  {Ol{\'a}h}, \& {Kriskovics}}]{Vida17}
{Vida}, K., {K{\H{o}}v{\'a}ri}, Z., {P{\'a}l}, A., {Ol{\'a}h}, K., \&
  {Kriskovics}, L. 2017,
  \href{http://dx.doi.org/10.3847/1538-4357/aa6f05}{\JournalTitle{\apj}, 841,
  124}

\bibitem[{{Weinberger} {et~al.}(2016){Weinberger}, {Boss}, {Keiser},
  {Anglada-Escud{\'e}}, {Thompson}, \& {Burley}}]{Wein16}
{Weinberger}, A.~J., {Boss}, A.~P., {Keiser}, S.~A., {et~al.} 2016,
  \href{http://dx.doi.org/10.3847/0004-6256/152/1/24}{\JournalTitle{\aj}, 152,
  24}

\bibitem[{{Zechmeister} {et~al.}(2019){Zechmeister}, {Dreizler}, {Ribas},
  {Reiners}, {Caballero}, {Bauer}, {B{\'e}jar}, {Gonz{\'a}lez-Cuesta},
  {Herrero}, \& {Lalitha}}]{Zech19}
{Zechmeister}, M., {Dreizler}, S., {Ribas}, I., {et~al.} 2019,
  \href{http://dx.doi.org/10.1051/0004-6361/201935460}{\JournalTitle{\aap},
  627, A49}

\end{thebibliography}

\appendix
\restartappendixnumbering
\section{Photometry and Spectra Tables for SEDs}

\movetabledown=4cm
\begin{rotatetable*}
\begin{deluxetable*}{l c c c c c c c c c c c c c c c}
\tablecaption{Optical Pan STARRS and SDSS photometry used for the construction of SEDs \label{tab:opticalphot}}
\tabletypesize{\scriptsize}
\tablehead{\colhead{Name} & \colhead{Pan STARRS} &\colhead{Pan STARRS} & \colhead{Pan STARRS} & \colhead{Pan STARRS} & \colhead{Pan STARRS} & \colhead{SDSS} &\colhead{SDSS} & \colhead{SDSS} & \colhead{SDSS} & \colhead{SDSS}&  \colhead{Ref.}\\
 \colhead{} & \colhead{$g$} & \colhead{$r$} & \colhead{$i$} & \colhead{$z$}& \colhead{$y$} & \colhead{$u$} & \colhead{$g$} & \colhead{$r$} & \colhead{$i$} & \colhead{$z$} & \colhead{}}
   \startdata
  TRAPPIST-1 & $19.35\pm0.02$ & $17.87\pm0.01 $ & $15.13\pm0.01$ & $13.73\pm0.01$ & $12.97\pm0.01$ & $\cdots$ & $\cdots$ & $\cdots$ & $\cdots$ &$\cdots$ & 1 \\ \hline
  J0320$+$1854 & $19.8\pm0.03$ & $18.2\pm0.01$ & $15.6\pm0.01 $ & $14.24\pm0.01$ &$13.43\pm0.01$ &$20.411\pm 0.093$ &$20.021\pm 0.016$ & $17.996\pm 0.007$ & $15.303 \pm 0.004$ & $13.448 \pm 0.004$ & 1, 2  \\
  J0443$+$0002 & $21.51\pm0.07$ & $19.66\pm0.02$ & $16.96\pm0.01 $& $15.38\pm0.01 $& $14.4\pm0.01$ & $\cdots$ & $21.99\pm 0.07$ & $19.76\pm0.02$ & $17.0048\pm 0.0055$ & $15.0886 \pm 0.0053$ & 1, 2\\ 
  J0518$-$2756 & $\cdots$ & $\cdots$ & $19.99\pm0.02$ & $18.49\pm0.01$ & $17.5\pm0.01$ & $\cdots$  & $\cdots$ & $\cdots$ & $\cdots$ & $\cdots$ & 1\\
  J0532+8246 & $\cdots$ & $\cdots$ & $\cdots$ &$18.07\pm0.01$ & $16.92\pm0.03 $& $\cdots$ & $\cdots$ & $22.59\pm0.26$ & $20.366\pm0.054 $ & $ 17.581\pm0.02$ & 1, 3\\
  J0608$-$2753 & $\cdots$ & $20.94\pm0.02$ & $18.14\pm0.01$ & $16.55\pm0.01$ & $15.54\pm0.01$ & $\cdots$  & $\cdots$ & $\cdots$ & $\cdots$ & $\cdots$ & 1\\
  J0853$-$0329 & $19.8\pm0.04$ & $17.85\pm0.01$ & $15.5\pm0.01$ & $14.06\pm0.01$ & $13.1\pm0.01 $ & $\cdots$  & $\cdots$ & $\cdots$ & $\cdots$ & $\cdots$ & 1\\
  J1013$-$1356 & $20.99\pm0.07$ & $19.25\pm0.01$ & $17.18\pm0.01$ & $ 16.25\pm0.01$ & $15.9\pm0.01$ & $\cdots$ & $\cdots$ & $\cdots$ & $\cdots$& $\cdots$  & 1\\ 
  J1256$-$0224 & $\cdots$ & $\cdots$ & $19.47\pm0.04$ & $18.0\pm0.03$ & $17.54\pm0.01$& $\cdots$ & $\cdots$ & $\cdots$ &$19.41 \pm 0.02$ & $17.71 \pm 0.02$ & 1, 4 \\ 
  LHS 377 & $18.99\pm0.02$ & $17.69\pm0.01$ &$15.68\pm0.01$ &$14.84\pm0.01$ &$ 14.48\pm0.01$ &$21.89 \pm 0.13$ & $19.384 \pm 0.011$ & $17.681 \pm 0.006$ & $15.696 \pm 0.005$ & $14.707 \pm 0.006$ & 1, 2 \\  
  J1444$-$2019 & $\cdots$ & $19.31\pm0.03$ & $16.09\pm0.05$ & $\cdots$ & $14.06\pm0.02$ & $\cdots$  & $\cdots$ & $\cdots$ & $\cdots$ & $\cdots$ & 5\\
  LHS 3003 & $17.65\pm0.01$ & $16.21\pm0.01$ & $\cdots$ & $\cdots$ & $\cdots$ & $\cdots$ & $\cdots$ & $\cdots$ & $\cdots$ & $\cdots$ & 5 \\  
  J1610$-$0040 & $20.12\pm0.1$ & $18.08\pm0.01$ & $15.9\pm0.01$ & $14.86\pm.01$ & $ 14.41\pm0.02$ & $\cdots$ & $\cdots$ & $17.976 \pm 0.007$ & $15.903 \pm 0.004$ & $14.663 \pm 0.005$ & 1,6 \\ 
  vB 8 & $17.43\pm0.01$ & $16.02\pm0.01$ & $\cdots$ & $\cdots$& $\cdots$ & $19.201 \pm 0.035$ &$17.738 \pm 0.006 $ & $16.046 \pm 0.003$ & $13.221 \pm 0.002$ & $11.766 \pm 0.003 $  & 1, 2 \\ 
  GJ 660.1B & $\cdots$ & $\cdots$ & $\cdots$ & $\cdots$ & $14.4\pm0.18$& $\cdots$ & $\cdots$ & $\cdots$ & $\cdots$ & $\cdots$ & 5\\ 
  J1835$+$3295 & $19.02\pm0.01$ & $17.18\pm0.01$ & $\cdots$ & $\cdots$ & $\cdots$ & $\cdots$ & $\cdots$ & $\cdots$ & $\cdots$ & $\cdots$ & 1 \\
  vB 10 & $\cdots$ & $16.59\pm0.01$ & $\cdots$ & $\cdots$ & $\cdots$ & $\cdots$ & $\cdots$ & $\cdots$ & $\cdots$ & $\cdots$ & 1 \\ 
  J2036+5059 & $19.61\pm0.02$ & $18.14\pm0.01$ & $16.13\pm0.01$ & $15.24\pm0.01$ & $14.89\pm0.01$ & $\cdots$ & $\cdots$ & $\cdots$ & $\cdots$ & $\cdots$ & 1 \\ 
  J2341$-$1133 & $21.1\pm0.03$ & $19.84\pm0.02$ & $17.13\pm0.01$ & $15.86\pm0.01$ & $15.13\pm0.01$ & $\cdots$ & $\cdots$ & $\cdots$ & $\cdots$ & $\cdots$ & 1 \\ 
  J2352$-$1100 & $20.57\pm0.03$& $19.18\pm0.01$ & $16.49\pm0.01$ & $15.23\pm0.01$ & $14.49\pm0.01$ & $\cdots$ & $20.930\pm 0.042$ & $19.250 \pm 0.015$ & $16.4830 \pm 0.0044$  &$14.9416 \pm 0.0047$ & 1,2\\
  \enddata
  \tablecomments{Photometric points with uncertainties greater than 0.5 magnitudes were excluded from the construction of the SED.}
  \tablerefs{(1) \cite{Cham16}, (2) \cite{Ahn_12}, (3) \cite{Alam15}, (4) \cite{Adel08}, (5) \cite{Best18}, (6) \cite{Abaz09}}   
  \tabletypesize{\small} 
\end{deluxetable*}
\end{rotatetable*}

\begin{deluxetable*}{l c c c c c c c}[h]
\tablecaption{Other Optical Photometry: Gaia, Johnson-Cousins and DENIS used for the construction of SEDs \label{tab:opticalphot2}}
\tabletypesize{\small}
\tablehead{\colhead{Name} & \colhead{\textit{Gaia}} & \colhead{\textit{Gaia}} & \colhead{Johnson} & \colhead{Cousins} & \colhead{Cousins} & \colhead{DENIS} & \colhead{Ref.}\\
  &\colhead{BP} & \colhead{RP}&\colhead{$V$} & \colhead{$R$} & \colhead{$I$} & \colhead{$I$} & \colhead{}}
  \startdata
  TRAPPIST-1 & $18.998 \pm 0.048$ & $14.1 \pm 0.01$ & $\cdots$ & $\cdots$ & $\cdots$ & $\cdots$ & 1 \\ \hline
  LHS 132 & $\cdots$ & $\cdots$ & $ 18.53 \pm 0.021$ & $16.3 \pm 0.008$ & $13.88 \pm 0.012$ & $13.83 \pm 0.03$ & 2, 3 \\ 
  J0853$-$0329 & $\cdots$ & $\cdots$ & $18.94\pm0.032$ & $\cdots$ & $\cdots$ & $\cdots$ & 2\\
  J1048$-$3956 & $\cdots$ & $\cdots$ & $17.532\pm0.057$ & $15.051\pm0.014$ & $\cdots$ & $\cdots$ & 4\\
  J1247$-$3816 & $\cdots$ & $\cdots$ & $\cdots$ & $\cdots$ & $\cdots$ & $17.85 \pm 0.16$ & 3\\
  LHS 3003  & $\cdots$ & $\cdots$ & $16.95 \pm 0.014$ & $14.9 \pm 0.006$ & $12.53\pm0.008$ & $\cdots$ & 2 \\  
  vB 8 & $\cdots$ & $\cdots$ & $16.85 \pm 0.059$ & $\cdots$ & $12.25\pm0.015$ & $\cdots$& 2 \\
  J2000$-$7523 & $\cdots$ & $\cdots$ & $21.157\pm0.008$ & $18.379\pm0.001$ & $16.119\pm0.024$ & $\cdots$ & 5\\
  \enddata
  \tablecomments{Photometric points with uncertainties greater than 0.5 magnitudes were excluded from the construction of the SED.}
  \tablerefs{(1) \cite{GaiaDR1, GaiaDR2, Riel18, Evan18}, (2) \cite{Diet14}, (3) \cite{DENIS}, (4) \cite{Cost05}, (5) \cite{Cost06}}    
  \tabletypesize{\small} 
\end{deluxetable*}

\begin{deluxetable*}{l c c c c c c c c c c c c}
\tabletypesize{\footnotesize}
\tablecaption{NIR Photometry used for the construction of SEDs \label{tab:NIRphot}}
\tablehead{\colhead{Name} & \colhead{2MASS $J$} & \colhead{2MASS $H$} & \colhead{2MASS $K_\mathrm{s}$} & \colhead{DENIS $J$} & \colhead{DENIS $K_\mathrm{s}$} & \colhead{MKO $J$} & \colhead{MKO $H$} & \colhead{MKO $K$} & \colhead{Reference}}
  \startdata
  TRAPPIST-1 & $11.354 \pm 0.022$ & $10.718 \pm 0.021$ & $10.296 \pm 0.023$ & $\cdots$ & $\cdots$ & $\cdots$ & $\cdots$ & $\cdots$ &1\\ \hline
  LHS 132 & $11.13 \pm 0.023$ & $10.479 \pm 0.024$ & $10.069 \pm 0.021 $ & $11.165 \pm 0.07 $ & $10.03 \pm 0.07$ & $\cdots$ & $\cdots$ & $\cdots$ &1, 2\\
  J0320$+$1854 & $11.759 \pm 0.021$ & $11.066 \pm 0.022$ & $10.639 \pm 0.018$ & $\cdots$ & $\cdots$ & $\cdots$ & $\cdots$ & $\cdots$ &1\\
  J0436$-$4114 & $13.097 \pm 0.026$ & $12.43 \pm 0.022$ & $12.05 \pm 0.024$ & $\cdots$ & $\cdots$ & $\cdots$ & $\cdots$ & $\cdots$ &1\\
  J0443$+$0002 & $ 12.507 \pm 0.026 $ & $11.804 \pm 0.024$ & $11.216 \pm 0.021$ & $\cdots$ & $\cdots$ & $\cdots$ & $\cdots$ & $\cdots$ &1\\
  J0518$-$2756 & $15.262 \pm 0.043$ & $ 14.295 \pm 0.046$ & $ 13.615 \pm 0.04$ & $\cdots$ & $\cdots$ & $\cdots$ & $\cdots$ & $\cdots$ &1\\
  J0532+8246 & $15.179 \pm 0.058$ & $14.904 \pm 0.091$ & $14.92 \pm 0.15$ & $\cdots$ & $\cdots$ & $\cdots$ & $\cdots$ & $\cdots$ &1\\
  J0608$-$2753 & $13.595 \pm 0.028$ & $12.897 \pm 0.026$ & $12.37 \pm 0.024$ & $\cdots$ & $\cdots$ & $\cdots$ & $\cdots$ & $\cdots$ &1\\
  J0853$-$0329 & $11.212\pm0.026$ & $10.469\pm0.026$ & $9.942\pm0.024$ & $\cdots$ & $\cdots$ & $11.18\pm0.05$ & $10.48\pm0.05$ & $9.91\pm0.05$ & 1, 3\\
  J1013$-$1356 & $14.621 \pm 0.032$ & $14.382 \pm 0.049$ & $14.398 \pm 0.078$ & $\cdots$ & $\cdots$ & $\cdots$ & $\cdots$ & $\cdots$ & 1\\
  J1048$-$3956 & $9.538\pm0.022$ & $8.905\pm0.044$ & $8.447\pm0.023$ & $\cdots$ & $\cdots$ & $\cdots$ & $\cdots$ & $\cdots$ & 1\\
  J1207$-$3900 & $15.494 \pm 0.058$ & $14.608 \pm 0.04$ & $ 14.04 \pm 0.059$ & $\cdots$ & $\cdots$ & $\cdots$ & $\cdots$ & $\cdots$ & 1\\
  J1247$-$3816 & $14.785 \pm 0.031$ & $14.096 \pm 0.035$ & $13.573 \pm 0.038$ & $\cdots$ & $\cdots$ & $\cdots$ & $\cdots$ & $\cdots$ & 1\\
  J1256$-$0224 & $16.10 \pm 0.11$ & $15.79\pm0.15$ & $\cdots$ & $\cdots$ & $\cdots$ & $\cdots$ & $16.078 \pm 0.016$ & $16.605 \pm 0.099$ & 1, 4, 5\\
  LHS 377 & $13.194 \pm 0.029$ & $12.73 \pm 0.03$ & $12.479 \pm 0.025$ & $\cdots$ & $\cdots$ & $\cdots$ & $\cdots$ & $\cdots$ &1\\
  J1444$-$2019 & $12.546 \pm 0.026$ & $12.142 \pm 0.026$ & $11.933 \pm 0.026$ & $\cdots$ & $\cdots$ & $\cdots$ & $\cdots$ & $\cdots$ &1\\
  LHS 3003 & $9.965 \pm 0.026$ & $9.315 \pm 0.022$& $8.928 \pm 0.027$ & $\cdots$ & $\cdots$ & $ 9.94 \pm 0.05$ & $9.43 \pm 0.05$ & $8.93 \pm 0.05$ & 1,6\\
  J1610$-$0040 & $12.911 \pm 0.022$ & $12.302 \pm 0.022$ & $12.302 \pm 0.022$ & $\cdots$ & $\cdots$ & $\cdots$ & $\cdots$ & $\cdots$ &1\\
  vB 8 & $9.776 \pm 0.029$ & $9.201 \pm 0.024$ & $8.816 \pm 0.023$ & $9.737 \pm 0.04$ & $8.819 \pm 0.06$ & $\cdots$ & $\cdots$ & $\cdots$ &1, 2\\
  GJ 660.1B & $ 13.052 \pm 0.045$ & $12.565 \pm 0.023$ & $ 12.227 \pm 0.027$ & $\cdots$ & $\cdots$ & $\cdots$ & $\cdots$ & $\cdots$ &1\\ 
  J1835$+$3295 & $10.27\pm0.022$ & $9.617\pm0.021$ & $9.171\pm0.018$ & $\cdots$ & $\cdots$ & $\cdots$ & $\cdots$ & $\cdots$ &1\\ 
  vB 10 & $9.908 \pm 0.025$ & $9.226 \pm 0.026$ & $8.765 \pm 0.022$ & $\cdots$ & $\cdots$ & $\cdots$ & $\cdots$ & $\cdots$ &1\\
  J2000$-$7523 &  $15.07 \pm 0.048$ &$14.003 \pm 0.036$& $13.42 \pm 0.042$ &$\cdots$ & $\cdots$ & $\cdots$ & $\cdots$ & $\cdots$ &1\\
  J2036+5059 & $13.611 \pm 0.029$ & $13.160 \pm 0.036$ & $12.936 \pm 0.033$ & $\cdots$ & $\cdots$ & $\cdots$ & $\cdots$ & $\cdots$ &1\\
  J2341$-$1133 & $13.546 \pm 0.023$ & $12.939 \pm 0.03$ & $12.546 \pm 0.033$ & $\cdots$ & $\cdots$ & $\cdots$ & $\cdots$ & $\cdots$ &1\\
  J2352$-$1100 & $12.84 \pm 0.022$ & $12.166\pm0.021$ & $11.742 \pm 0.02$ & $\cdots$ & $\cdots$ & $\cdots$ & $\cdots$ & $\cdots$ &1\\
 \enddata
  \tablecomments{Photometric points with uncertainties greater than 0.5 magnitudes were excluded from the construction of the SED.}
  \tablerefs{(1) \cite{Cutr03}, (2) \cite{DENIS}, (3)\cite{Goli04a}, (4) \cite{Lawr12}, (5) \cite{Gonz18}, (6) \cite{Legg02a}}
  \tabletypesize{\small}
\end{deluxetable*}


\begin{deluxetable*}{l c c c c c c c c c c c}
\tabletypesize{\scriptsize} 
\tablecaption{MIR Photometry used for construction of SEDs \label{tab:MIRphot}}
\tablehead{\colhead{Name} &  \colhead{MKO} &\colhead{MKO} &\colhead{WISE} & \colhead{WISE} & \colhead{WISE} & \colhead{WISE} & \colhead{IRAC} & \colhead{IRAC} & \colhead{IRAC} & \colhead{IRAC} & \colhead{Ref.} \\
\colhead{} & \colhead{$L^\prime$} & \colhead{$M^\prime$} &\colhead{$W1$} & \colhead{$W2$} & \colhead{$W3$} & \colhead{$W4$} & \colhead{[3.6 $\upmu$m]} & \colhead{[4.5 $\upmu$m]} & \colhead{[5.8 $\upmu$m]} & \colhead{[8.0 $\upmu$m]} & \colhead{}}
  \startdata
  TRAPPIST-1 &$\cdots$ &  $\cdots$ & $10.042 \pm 0.023$ & $9.80 \pm 0.02$ & $9.528 \pm 0.041$ & $\cdots$ & $\cdots$ & $\cdots$ & $\cdots$ & $\cdots$ & 1 \\ \hline
  LHS 132 &$\cdots$ &  $\cdots$ &  $\cdots$ & $\cdots$ & $\cdots$ & $\cdots$ & $9.64 \pm 0.02 $ & $9.62 \pm 0.02$ & $9.52 \pm 0.02$ & $9.48 \pm 0.01$ & 2 \\ 
  J0320$+$1854 & $\cdots$ &  $\cdots$ & $10.347 \pm 0.023$ & $10.148 \pm 0.02 $ & $9.874 \pm 0.048$ & $\cdots$ & $\cdots$ & $\cdots$ & $\cdots$ & $\cdots$ & 1 \\ 
  J0436$-$4114 & $\cdots$ &  $\cdots$ & $11.74 \pm 0.023$ & $11.46 \pm 0.021$ & $11.111 \pm 0.082$ & $\cdots$ & $\cdots$ & $\cdots$ & $\cdots$ & $\cdots$ &1\\
  J0443$+$0002 & $\cdots$ & $\cdots$ & $10.826 \pm 0.024$ & $ 10.476 \pm 0.021$ & $ 10.031 \pm 0.054$ & $\cdots$ & $ 10.55 \pm 0.02$ & $ 10.45 \pm 0.02$ & $ 10.35 \pm 0.03 $ & $ 10.22 \pm 0.03$ & 1, 3\\
  J0518$-$2756 & $\cdots$ & $\cdots$ & $ 13.045 \pm 0.024$ & $12.661 \pm 0.026$& $12.581 \pm 0.349$ & $\cdots$ & $\cdots$ & $\cdots$ & $\cdots$ & $\cdots$ & 1\\
  J0532$+$8246 & $\cdots$ &  $\cdots$ & $\cdots$ & $\cdots$ & $\cdots$ & $\cdots$ & $13.37 \pm 0.03$ & $13.22 \pm 0.02$ & $13.23 \pm 0.1 $ & $13.03 \pm 0.1$ & 1 \\ 
  J0608$-$2753 & $\cdots$ &  $\cdots$ & $11.976 \pm 0.024$ & $11.623 \pm 0.021$ & $11.31 \pm 0.11$ & $\cdots$ & $11.75 \pm 0.02$ & $11.62 \pm 0.02$ & $11.52 \pm 0.03$ & $11.44 \pm 0.03$ & 1, 3 \\ 
  J0853$-$0329 & $9.39\pm0.07$ & $9.62\pm0.1$ & $9.624\pm0.023$ & $9.381\pm0.02$ & $8.967\pm0.028$ & $8.756\pm0.428$ & $9.41\pm0.02$ & $9.39\pm0.03 $ & $9.22\pm0.01$ & $9.13\pm0.01$ & 2, 4\\
  J1013$-$1356 & $\cdots$ & $\cdots$ & $13.782 \pm 0.028$ & $13.545 \pm 0.035$ & $12.68 \pm 0.51$ & $\cdots$ & $\cdots$ & $\cdots$ & $\cdots$ & $\cdots$ & 1 \\ 
  J1048$-$3956 & $\cdots$ & $\cdots$ & $ 8.103\pm0.024$ & $7.814\pm0.021$ & $7.462\pm0.018$ & $7.226\pm0.087$ & $\cdots$ & $\cdots$ & $\cdots$ & $\cdots$ & 5 \\ 
  J1207$-$3900 & $\cdots$ & $\cdots$ & $13.64 \pm 0.024$ & $13.204 \pm 0.027$ & $\cdots$ & $\cdots$ & $\cdots$ & $\cdots$ & $\cdots$ & $\cdots$ & 5\\
  J1247$-$3816 & $\cdots$ & $\cdots$ & $13.119 \pm 0.024$ & $12.532 \pm 0.024$ & $10.953 \pm 0.077$ & $8.84 \pm 0.29$ & $\cdots$ & $\cdots$ & $\cdots$ & $\cdots$ & 1,5\\
  J1256$-$0224 & $\cdots$ & $\cdots$ & $15.214 \pm 0.038$ & $15.106 \pm 0.098$ & $\cdots$ & $\cdots$ & $\cdots$ & $\cdots$ & $\cdots$ & $\cdots$ & 1 \\
  LHS 377 & $\cdots$ & $\cdots$ & $12.298 \pm 0.027$ & $12.051 \pm 0.025$ & $11.67 \pm 0.11$ & $\cdots$ & $\cdots$ & $\cdots$ & $\cdots$ & $\cdots$ & 1 \\ 
  J1444$-$2019 & $\cdots$ & $\cdots$ & $11.464 \pm 0.024$ & $11.211 \pm 0.022$ & $10.967 \pm 0.09$ & $\cdots$ & $\cdots$ & $\cdots$ & $\cdots$ & $\cdots$ & 5 \\
  LHS 3003 & $8.43 \pm 0.03$ &$\cdots$ & $\cdots$ & $\cdots$ & $\cdots$ & $\cdots$ & $8.47\pm0.02$ & $8.49\pm0.01$ & $8.39\pm0.02$ & $8.36\pm0.01$ & 2,6  \\
  J1610$-$0040 & $\cdots$ &  $\cdots$ & $11.639 \pm 0.025$ & $11.639 \pm 0.025$ & $11.639 \pm 0.025$ & $\cdots$ & $\cdots$ & $\cdots$ & $\cdots$ & $\cdots$ & 1 \\ 
  vB 8 & $\cdots$ &  $\cdots$ & $8.588 \pm 0.023$ & $8.365 \pm 0.021$ & $8.132 \pm 0.022$ & $7.857 \pm 0.181$ & $8.37 \pm 0.02$ & $8.38 \pm 0.01$ & $8.28 \pm 0.02$ & $ 8.24 \pm 0.02$ & 1, 2 \\ 
  GJ 660.1B\tablenotemark{a} & $\cdots$ & $\cdots$ & $11.689\pm0.24$ & $11.496\pm0.318$ &$\cdots$ & $\cdots$ & $\cdots$ & $\cdots$ & $\cdots$ & $\cdots$  & 7 \\
  J1835$+$3295 & $\cdots$ & $\cdots$ &$8.803\pm0.022$ & $8.539\pm0.019$ & $8.16\pm0.019$ & $7.886\pm0.132$ & $8.55\pm0.02$ &$8.55\pm0.01$ &$8.39\pm0.01$ & $8.29\pm0.01$ & 1, 2\\
  vB 10 & $\cdots$ &  $\cdots$ & $8.465 \pm 0.023$ & $8.249 \pm 0.02$ & $8.08 \pm 0.022$ & $\cdots$ & $8.29 \pm 0.02$ & $8.3 \pm 0.03$ & $8.15 \pm 0.01$ & $8.14 \pm 0.01$ & 1, 2 \\ 
  J2000$-$7523 & $\cdots$ & $\cdots$ &$12.819 \pm 0.024$ &$12.431 \pm 0.024$ &$11.64 \pm 0.15$ &$\cdots$ &$\cdots$ &$\cdots$ &$\cdots$ & $\cdots$& 1\\
  J2036$+$5059\tablenotemark{a}  & $\cdots$ & $\cdots$& $12.667\pm0.24$ & $12.436\pm0.318$ &$\cdots$ & $\cdots$ & $\cdots$ & $\cdots$ & $\cdots$ & $\cdots$  & 7 \\
  J2341$-$1133 & $\cdots$ & $\cdots$ & $12.224 \pm 0.025$ & $12.001 \pm 0.024$ & $11.721 \pm 0.218$ & $\cdots$ & $\cdots$ & $\cdots$ & $\cdots$ & $\cdots$ & 5 \\ 
  J2352$-$1100 & $\cdots$ & $\cdots$ & $11.44 \pm 0.025$ & $11.146 \pm 0.022$ & $10.849 \pm 0.109$ & $\cdots$ & $\cdots$ & $\cdots$ & $\cdots$ & $\cdots$ & 5 \\
 \enddata
  \tablecomments{Photometric points with uncertainties greater than 0.5 magnitudes were excluded from the construction of the SED.}
  \tablenotetext{a}{Magnitudes were estimated from 2MASS $K_\mathrm{s}$ to properly append the Rayleigh-Jeans tail.}
  \tablerefs{(1) \cite{Cutr12}, (2) \cite{Patt06}, (3) \cite{Luhm09b}, (4) \cite{Reid02b}, (5) \cite{Cutr13}, (6)\cite{Legg98}, (7) This paper}
  \tabletypesize{\small} 
\end{deluxetable*}

\begin{deluxetable*}{l c c c c c c c c c}
\tablecaption{Spectra used to construct SEDs \label{tab:SpectraReferences}}
\tabletypesize{\footnotesize}
\tablehead{\colhead{Name} & \colhead{OPT} & \colhead{OPT} & \colhead{OPT} & \colhead{NIR} & \colhead{NIR} & \colhead{ NIR} & \colhead{MIR} & \colhead{MIR} & \colhead{MIR} \vspace{-.1cm}\\
 & & \colhead{Obs. Date} & \colhead{Ref.} & & \colhead{Obs. Date} & \colhead{Ref.} & &\colhead{Obs. Date} & \colhead{Ref.}}   
  \startdata
  TRAPPIST-1 & KPNO 4m: R--C Spec & 2003--07--10 & 1 & FIRE & 2017--07--28 & 2 & $\cdots$ & $\cdots$ & $\cdots$\\
  LHS 132 & CTIO 1.5m: R--C Spec & 2003--11--09 & 1 & SpeX Prism & 2008--09--07 & 3 & $\cdots$ & $\cdots$ & $\cdots$\\ 
  J0320$+$1854 & GoldCam & 2000--10--01 & 4 & SpeX SXD, LXD1.9 & 2009--12--01 & 5 & $\cdots$ & $\cdots$ & $\cdots$\\ 
  J0436$-$4114 & LRIS & 2009--10--11 & 6 & SpeX SXD & 2012--09--20 & 7 & $\cdots$ & $\cdots$ & $\cdots$\\ 
  J0443$+$0002 & CTIO 1.5m: R--C Spec & 2002--01--28 & 4 & SpeX SXD & 2012--02--05 & 8 & $\cdots$ & $\cdots$ & $\cdots$\\ 
  J0518$-$2756 & LRIS & 2009--02--17 & 9 & FIRE & 2015--12--20 & 11 & $\cdots$ & $\cdots$ & $\cdots$\\ 
  J0532+8246 & LRIS & 2003--01--03 & 11 & NIRSPEC & 2002--12--24 & 11 & IRS & 2005--03--23& 12\\
  J0608$-$2753 & CTIO 4m: R--C Spec & 2002--01--25 & 4 & SpeX SXD & 2007--11--13 & 13 & $\cdots$ & $\cdots$ & $\cdots$ \\ 
  J0853$-$0329 & CTIO 4m: R--C Spec & 2003--04--21& 14 & SpeX SXD & 2009--12--01 & 5 & $\cdots$ & $\cdots$ & $\cdots$\\
  J1013$-$1356 & GMOS-N & 2004--11--21 & 15 & SpeX Prism & 2004--03--12& 16 & IRS & 2005--06--07& 17\\
  J1048$-$3956 & CTIO 1.5m: R--C Spec& 2003--05--15 & 14 & SpeX SXD & 2009--12--01 & 5 & $\cdots$ & $\cdots$ & $\cdots$\\ 
  J1207$-$3900 & MagE & 2013--05--14 & 18 & FIRE & 2015--12--22 & 10 & $\cdots$ & $\cdots$ & $\cdots$\\
  J1247$-$3816 & $\cdots$ & $\cdots$ & $\cdots$ & SpeX Prism & 2013--05--10 & 18 &$\cdots$ & $\cdots$ & $\cdots$\\
  J1256$-$0224 & LDSS3 & 2006--05--07 & 19 & FIRE & 2016--08--13 & 20 & $\cdots$ & $\cdots$ & $\cdots$\\
  LHS 377 & X-Shooter (UVB,VIS) & 2014--02--20 & 21 & X-Shooter & 2014--02--20 & 21 & IRS & 2005--07--01 & 12 \\
  J1444$-$2019 & $\cdots$ & $\cdots$ & $\cdots$ & SpeX Prism & 2005--03--23 & 3 & $\cdots$ & $\cdots$ & $\cdots$\\ 
  LHS 3003 & GoldCam & 2003--03--13 & 14 & SpeX Prism & 2008--07--29 & 3 & IRS & 2006--01--14 & 22\\
  J1610$-$0040 & MkIII & 2003-02-19 & 23 & SpeX Prism & 2003--07--06 & 24 & $\cdots$ & $\cdots$ & $\cdots$\\ 
  vB 8 & KPNO 4m: R--C Spec & 2002--09--25 & 14 & SpeX SXD, LXD1.9 & 2001--07--12 & 5 & $\cdots$ & $\cdots$ & $\cdots$\\
  GJ 660.1B & $\cdots$ & $\cdots$ & $\cdots$ & SpeX Prism & 2011--03--09 & 25 & $\cdots$ & $\cdots$ & $\cdots$\\ 
  J1835$+$3295 & KPNO 4m: R--C Spec & 2001--07--22 & 1 & SpeX Prism & 2003--09--05 & 9 & $\cdots$ & $\cdots$ & $\cdots$\\
  vB 10 & KPNO 4m: R--C Spec & 2002--09--26 & 14 & SpeX SXD, LXD1.9 & 2001--06--13 & 5, 26 & IRS & 2005-10-11 & 27\\
  J2000$-$7523 & CTIO 4m: R--C Spec & 2003--04--23 & 14 & FIRE & 2013--07--28 & 28 & $\cdots$ & $\cdots$ & $\cdots$\\ 
  J2036+5059 & KAST &2001-12-09 & 29 & SpeX Prism & 2003--10--06 & 24 & $\cdots$ & $\cdots$ & $\cdots$ \\ 
  J2341$-$1133 & GoldCam & 2002--07--06 & 1 & SpeX Prism & 2010--07--07 & 3 & $\cdots$ & $\cdots$ & $\cdots$\\ 
  J2352$-$1100 & GoldCam & 2002--07--06 & 1 & SpeX Prism & 2010-07--07 & 3 & $\cdots$ & $\cdots$ & $\cdots$\\ 
  \enddata
  \tablerefs{(1) \cite{Cruz07}, (2) This Paper, (3) \cite{Bard14}, (4) \cite{Cruz03}, (5) \cite{Rayn09}, (6) \cite{Phan03}, (7) \cite{Alle13}, (8) \cite{Gagn15b}, (9) \cite{Cruz18}, (10) Faherty et al. in prep, (11)\cite{Burg03c}, (12) Spitzer PID51, (13) CruzUnpub (Zendodo),  (14) \cite{Reid08b}, (15)\cite{Burg07a}, (16) \cite{Burg04c}, (17) Spitzer PID251, (18) \cite{Gagn14b}, (19) \cite{Burg09a}, (20) \cite{Gonz18}, (21) \cite{Rajp16}, (22)\cite{Cush06b}, (23) \cite{Lepi03c}, (24) \cite{Cush06a}, (25) \cite{Agan16}, (26) \cite{Cush05}, (27) Spitzer PID29, (28) \cite{Fahe16}, (29) \cite{Lepi03a}}
  \tabletypesize{\small}
\end{deluxetable*} 

\begin{deluxetable*}{l c c c c c c c c c c}
\tabletypesize{\scriptsize}
\tablecaption{Allers \& Liu Gravity indices for low-resolution spectra from all comparison samples\label{tab:gravityindiceslow}}
\tablehead{\colhead{} & \colhead{} & \colhead{Literature}& \colhead{Literature} & \colhead{} & \colhead{} & \colhead{} & \colhead{} & \colhead{Gravity} & \colhead{Gravity} &\colhead{Spectrum} \\ 
\colhead{Object} & \colhead{Spectrum} & \colhead{Opt SpT}& \colhead{NIR SpT} & \colhead{FeH$_z$} & \colhead{VO$_z$} & \colhead{\ion{K}{1}$_\mathrm{J}$} & \colhead{$H$-cont} & \colhead{Scores} & \colhead{Class} &\colhead{Reference}} 
  \startdata
  TRAPPIST-1   & prism & M7.5 & $\cdots$ &	$1.078\pm0.008$	&	$1.054\pm0.003$	&	$1.059\pm0.010$	&	$0.981	\pm	0.008$ & 1n11 & INT-G & 1\\
               & SXD   & M7.5 & $\cdots$ &    $1.119	\pm	0.001$	&	$1.070	\pm	0.002$	&	$1.070	\pm	0.001$	&	$0.971	\pm	0.001$ & 1n11 & INT-G & 2\\
               & FIRE  & M7.5 & $\cdots$ &    $1.105	\pm	0.001$	&	$1.084	\pm	0.001$	&	$1.062	\pm	0.001$	&	$0.951	\pm	0.001$ & 1n10 & INT-G & 3\\ \hline
  LHS 132      & prism & M8   & M8       &    $1.141	\pm	0.005$	&	$1.055	\pm	0.010$	&	$1.071	\pm	0.009$	&	$0.980	\pm	0.006$ & 1n11 & INT-G & 4\\
  J0320$+$1854 & prism & M8 & $\cdots$ & $1.244	\pm	0.009$	&	$1.056	\pm	0.009$	&	$1.076	\pm	0.015$	&	$0.952	\pm	0.007$ & 0n10 & FLD-G & 1\\
               & SXD   & M8 & $\cdots$ & $1.228	\pm	0.003$	&	$1.049	\pm	0.003$	&	$1.095	\pm	0.002$	&	$0.941	\pm	0.001$ & 0n00 & FLD-G & 6\\
  J0436$-$4114 & prism & M8\,$\beta$  & M9\,$\gamma$	& $1.076	\pm	0.015$	&	$1.104	\pm	0.015$	&	$1.065	\pm	0.013$	&	$0.982	\pm	0.011$	& 2n12 & VL-G & 1\\
  J0443$+$0002 & prism & M9\,$\gamma$ & L1\,$\gamma$ & $	1.109	\pm	0.011	$	&	$	1.163	\pm	0.016	$	&	$	1.056	\pm	0.014	$	&	$	0.971	\pm	0.009$ & 2122 & VL-G & 7\\
  	           & FIRE  & M9\,$\gamma$ & L1\,$\gamma$ & $	1.117	\pm	0.001	$	&	$	1.204	\pm	0.002	$	&	$	1.075	\pm	0.001	$	&	$	0.979	\pm	0.001$ & 2112 & VL-G & 3\\ 
  J0518$-$2756 & prism & 11 & 11 &	$	1.141	\pm	0.019	$	&	$	1.278	\pm	0.043	$	&	$	1.060	\pm	0.015	$	&	$	0.945	\pm	0.006$	& 1221 & VL-G & 1 \\ 
               & FIRE  &	11 & 11	& $	1.097	\pm	0.001	$	&	$	1.300	\pm	0.003	$	&	$	1.073	\pm	0.001	$	&	$	0.948	\pm	0.001$ &	2211 & VL-G	& 3\\ 
    J0532$+$8246 & NIRSPEC &	sdL7	& 	$\cdots$	&	$\cdots$		&		$\cdots$	& $\cdots$		&	$	1.134	\pm	0.002	$	&		$\cdots$	&	$\cdots$		& 8\\
  J0608$-$2753 & prism	 & M8.5\,$\gamma$ & L0\,$\gamma$	&	$	1.066	\pm	0.005	$	&	$	1.195	\pm	0.011	$	&	$	1.062	\pm	0.018	$	&	$	0.996	\pm	0.010	$	&	2112	&	VL-G	& 1\\ 
               & SXD     &  M8.5\,$\gamma$ & L0\,$\gamma$	&	$	1.047	\pm	0.003	$	&	$	1.173	\pm	0.002	$	&	$	1.071	\pm	0.003	$	&	$	0.985	\pm	0.002	$	&	2112	&	VL-G	& 9\\
  J0853$-$0329 & SXD     & M9 & M9 & $	1.201	\pm	0.003	$	&	$	1.119	\pm	0.004	$	&	$	1.101	\pm	0.003	$	&	$	0.941	\pm	0.001	$& 0n00 &	FLD-G &6\\
  J1013$-$1356 & prism	 & sdM9.5 &	$\cdots$	&$	1.133	\pm	0.014	$	&	$	0.987	\pm	0.008	$	&	$	1.024	\pm	0.003	$	&	$	0.957	\pm	0.013	$	&	1021	&	INT-G	& 10\\
  J1048$-$3956 & SXD     & M9 & $\cdots$ & $	1.250	\pm	0.003	$	&	$	1.054	\pm	0.003	$	&	$	1.114	\pm	0.002	$	&	$	0.936	\pm	0.002	$ &0n00 &	FLD-G & 6\\
  J1207$-$3900 & prism	 & L0\,$\gamma$ & L1\,$\gamma$	&   $	1.016	\pm	0.016	$	&	$	1.264	\pm	0.019	$	&	$	1.049	\pm	0.016	$	&	$	0.999	\pm	0.015	$	&	2222	&	VL-G	& 11\\ 
               & SXD	 & L0\,$\gamma$ & L1\,$\gamma$	&	$	1.019	\pm	0.005	$	&	$	1.517	\pm	0.018	$	&	$	1.095	\pm	0.002	$	&	$	1.021	\pm	0.002	$	&	2212	&	VL-G	& 11\\ 
               & FIRE	 & L0\,$\gamma$ & L1\,$\gamma$	&	$	1.052	\pm	0.001	$	&	$	1.388	\pm	0.003	$	&	$	1.052	\pm	0.002	$	&	$	0.994	\pm	0.001	$	&	2222	&	VL-G & 3 \\ 
  J1247$-$3816 & prism	 & $\cdots$	& M9\,$\gamma$	&$	1.047	\pm	0.010	$	&	$	1.102	\pm	0.007	$	&	$	1.048	\pm	0.016	$	&	$	0.987	\pm	0.011	$	&	2n22	&	VL-G	& 11\\ 
  J1256$-$0224 & prism	 & sdL3.5	&	$\cdots$	&	$	1.341	\pm	0.052	$	&	$	0.977	\pm	0.017	$	&	$	1.049	\pm	0.008	$	&	$	0.919	\pm	0.015	$	&	1021	&	INT-G	&12 \\
               & FIRE    & sdL3.5 & $\cdots$	&	$	1.336	\pm	0.005	$	&	$	0.982	\pm	0.001	$	&	$	1.090	\pm	0.001	$	&	$	1.082	\pm	0.004	$	&	1012	&	INT-G	& 13\\ 
  LHS 377      & X-Shooter &	sdM7	&		$\cdots$	&	$\cdots$		&	$	0.991	\pm	0.001	$	&	$	0.982	\pm	0.002	$	&	$	0.824	\pm	0.001	$	&	$\cdots$	&	$\cdots$	& 14\\
  J1444$-$2019 & prism & sdM9 & $\cdots$ &$	1.256	\pm	0.019	$	&	$	0.985	\pm	0.013	$	&	$	1.067	\pm	0.007	$	&	$	0.922	\pm	0.006	$ & 1n10	& INT-G & 4\\
  LHS3003      & prism & M7	&	M7	&	$	1.121	\pm	0.006	$	&	$	1.027	\pm	0.007	$	&	$	1.069	\pm	0.008	$	&	$	0.978	\pm	0.007	$	&	0n00	&	FLD-G	& 4\\
  J1610$-$0040 & SXD   & sdM7	&	$\cdots$	&	$	1.102	\pm	0.002	$	&	$	0.984	\pm	0.001	$	&	$	1.058	\pm	0.001	$	&	$	0.958	\pm	0.001	$	&	1n00	&	FLD-G	& 15\\
  vB 8         & prism & M7	&	M7	&	$	1.155	\pm	0.008	$	&	$	1.015	\pm	0.007	$	&	$	1.064	\pm	0.008	$	&	$	0.978	\pm	0.009	$	&	0n00	&	FLD-G	&5\\
               & SXD   & M7	&	M7	&	$	1.155	\pm	0.002	$	&	$	1.014	\pm	0.002	$	&	$	1.074	\pm	0.001	$	&	$	0.972	\pm	0.001	$	&	0n00	&	FLD-G	& 6\\
  GJ 660.1B    & prism & $\cdots$ & d/sdM7 &$	1.200	\pm	0.017	$	&	$	1.012	\pm	0.011	$	&	$	1.087	\pm	0.006	$	&	$	0.958	\pm	0.006	$& 0n00 &	FLD-G& 16\\
  J1835$+$3295 & prism & M8.5 & $\cdots$ &$	1.142	\pm	0.049	$	&	$	1.057	\pm	0.012	$	&	$	1.094	\pm	0.006	$	&	$	0.927	\pm	0.003	$ &1n00&	FLD-G & 1\\
  vB 10        & SXD   & M8	&	M8	&$	1.148	\pm	0.012	$	&	$	1.052	\pm	0.002	$	&	$	1.071	\pm	0.002	$	&	$	0.950	\pm	0.006	$	&	0n10	&	FLD-G	&6, 17\\
  J2000$-$7523 & FIRE  & M9\,$\gamma$ &  M9\,$\gamma$ & $	1.094	\pm	0.001	$	&	$	1.195	\pm	0.001	$	&	$	1.076	\pm	0.002	$	&	$	0.969	\pm	0.001	$& 1n11 &	INT-G &18\\
  J2036$+$5059 & prism & sdM7.5 & $\cdots$ &$	1.090	\pm	0.011	$	&	$	0.977	\pm	0.009	$	&	$	1.029	\pm	0.003	$	&	$	0.976	\pm	0.007	$	&	1n21	&	INT-G	& 4\\
               & SXD   & sdM7.5	&	$\cdots$	&	$	1.095	\pm	0.002	$	&	$	0.977	\pm	0.001	$	&	$	1.032	\pm	0.001	$	&	$	0.969	\pm	0.001	$	&	1n21	&	INT-G	& 15\\
  J2341$-$1133 & prism & M8 & $\cdots$ & $	1.105	\pm	0.010	$	&	$	1.059	\pm	0.012	$	&	$	1.069	\pm	0.012	$	&	$	0.973	\pm	0.006$ & 1n11 & INT-G & 4\\ 
  J2352$-$1100 & prism & M7 & M8\,$\beta$ & $1.110	\pm	0.004$	&	$1.054	\pm	0.008$	&	$1.066	\pm	0.009$	&	$0.979	\pm	0.006$  & 1n11 & INT-G & 4\\ 
  \enddata
  \tablecomments{When determining the gravity scores, the literature near-infrared spectral type was used. In cases where there is no NIR spectral type, we used the optical spectral type. Half spectral types were rounded to the nearest whole type. Gravity Scores are listed for each index in order, where scores are as follows: 0- field gravity (FLD-G), 1- intermediate gravity(INT-G), 2- low gravity (VL-G). The appropriate combinations of scores are used to get the \cite{Alle13} gravity class designations. To receive the following gravity classifications a median score of such is needed- FLD-G: $\leq0.5$, INT-G: 1, VL-G: $\geq1.5$. Again in this paper we use $\beta$ and INT-G interchangeably as well as $\gamma$ and VL-G, however we choose to list the final gravity class in this table following the \cite{Alle13} notation.}
  \tablerefs{(1) \cite{Cruz18}, (2) \cite{Gill16}, (3)This Paper, (4) \cite{Bard14}, (5) \cite{Burg08d} , (6) \cite{Rayn09}, (7) \cite{Alle13}, (8) \cite{Burg03c}, (9) CruzUnpub (Zenodo), (10) \cite{Burg04c}, (11) \cite{Burg09a}, (12) \cite{Gagn14b}, (13) \cite{Gonz18}, (14) \cite{Rajp16}, (15)\cite{Cush06a} , (16)\cite{Agan16} , (17)\cite{Cush05} , (18) \cite{Fahe16} }
  \tabletypesize{\small}
\end{deluxetable*}

\begin{deluxetable*}{l c c c c c c c c c c}
\tabletypesize{\scriptsize}
\tablecaption{Allers \& Liu Gravity indices for medium-resolution spectra from all comparison samples \label{tab:gravityindicesmed}}
\tablehead{\colhead{} & \colhead{} & \colhead{Literature} & \colhead{Literature} & \colhead{} & \colhead{} & \colhead{} & \colhead{} & \colhead{} & \colhead{Gravity} & \colhead{Gravity} \\
\colhead{Object} & \colhead{Spectrum} & \colhead{Opt SpT} & \colhead{NIR SpT} & \colhead{FeH$_z$} & \colhead{FeH$_J$} & \colhead{VO$_z$} & \colhead{\ion{K}{1}$_\mathrm{J}$} & \colhead{$H$-cont} & \colhead{Scores} & \colhead{Class}} 
  \startdata
  TRAPPIST-1 & SXD & M7.5 & $\cdots$	& $	1.119	\pm	0.001	$	&	$	1.093	\pm	0.010	$	&	$	1.070	\pm	0.002	$	&	$	1.070	\pm	0.001	$	&	$	0.971	\pm	0.001	$	& 1n01 & INT-G \\
             & FIRE & M7.5 & $\cdots$ & $	1.105	\pm	0.001	$	&	$	1.110	\pm	0.009	$	&	$	1.084	\pm	0.001	$	&	$	1.062	\pm	0.001	$	&	$	0.951	\pm	0.001	$ & 1n10 & INT-G \\ \hline
  J0320$+$1854 & SXD & M8 &	$\cdots$ & $	1.228	\pm	0.003	$	&	$	1.168	\pm	0.015	$	&	$	1.049	\pm	0.003	$	&	$	1.095	\pm	0.002	$	&	$	0.941	\pm	0.001	$ &	0n00	&	FLD-G \\
  J0443$+$0002 & FIRE & M9\,$\gamma$ & L0\,$\gamma$ & $	1.117	\pm	0.001	$	&	$	1.113	\pm	0.010	$	&	$	1.204	\pm	0.002	$	&	$	1.075	\pm	0.001	$	&	$	0.979	\pm	0.001	$	& 2122	&	VL-G \\
  J0518$-$2756 & FIRE & L1\,$\gamma$ & L1\,$\gamma$ &	$	1.097	\pm	0.001	$	&	$	1.137	\pm	0.011	$	&	$	1.300	\pm	0.003	$	&	$	1.073	\pm	0.001	$	&	$	0.948	\pm	0.001	$	&	2221	& VL-G \\
  J0532$+$8246 & NIRSPEC & sdL7 & $\cdots$	&	$\cdots$ &	$1.239	\pm	0.026$	&	$\cdots$	&	$1.134	\pm	0.002$	&	$\cdots$ & $\cdots$ & $\cdots$ \\
  J0608$-$2753 & SXD & M8.5\,$\gamma$ & L0\,$\gamma$ &	$	1.047	\pm	0.003	$	&	$	1.081	\pm	0.012	$	&	$	1.173	\pm	0.002	$	&	$	1.071	\pm	0.003	$	&	$	0.985	\pm	0.002	$	& 2122	&	VL-G \\
  J0853$-$0329 & SXD & M9 & M9 &$	1.201	\pm	0.003	$	&	$	1.119	\pm	0.004	$	&	$	1.192	\pm	0.016	$	&	$	1.101	\pm	0.003	$	&	$	0.941	\pm	0.001	$ & 0n00 & FLD-G\\
  J1048$-$3956 & SXD & M9 & $\cdots$ &$	1.250	\pm	0.003	$	&	$	1.190	\pm	0.022	$	&	$	1.054	\pm	0.003	$	&	$	1.114	\pm	0.002	$	&	$	0.936	\pm	0.002	$ & 0n00 &FLD-G\\
  J1207$-$3900 & SXD & L0\,$\gamma$ & L1\,$\gamma$	&	$	1.019	\pm	0.005	$	&	$	1.178	\pm	0.040	$	&	$	1.517	\pm	0.018	$	&	$	1.095	\pm	0.002	$	&	$	1.021	\pm	0.002	$	& 2222	&	VL-G \\
               & FIRE & L0\,$\gamma$ & L1\,$\gamma$ & $	1.052	\pm	0.001	$	&	$	1.119	\pm	0.011	$	&	$	1.388	\pm	0.003	$	&	$	1.052	\pm	0.002	$	&	$	0.994	\pm	0.001	$	&	2222	& VL-G \\
  J1256$-$0224 & FIRE & sdL3.5 &	$\cdots$ & $	1.336	\pm	0.005	$	&	$	1.161	\pm	0.020	$	&	$	0.982	\pm	0.001	$	&	$	1.090	\pm	0.001	$	&	$	1.082	\pm	0.004	$	& 1012	&	INT-G \\
  LHS 377 & X-Shooter &	sdM7 & $\cdots$ & $\cdots$ &	$	1.047	\pm	0.003	$	&	$	0.991	\pm	0.001	$	&	$	0.982	\pm	0.002	$	&	$	0.824	\pm	0.001	$ &	$\cdots$ & $\cdots$\\
  J1610$-$0040 & SXD & sdM7 & $\cdots$ & $	1.102	\pm	0.002	$	&	$	1.073	\pm	0.010	$	&	$	0.984	\pm	0.001	$	&	$	1.058	\pm	0.001	$	&	$	0.958	\pm	0.001	$	& 1n00	&	FLD-G \\
  vB 8 & SXD & M7 & M7 & $	1.155	\pm	0.002	$	&	$	1.111	\pm	0.011	$	&	$	1.014	\pm	0.002	$	&	$	1.074	\pm	0.001	$	&	$	0.972	\pm	0.001	$	& 0n00	&	FLD-G \\
  vB 10 & SXD & M8 & M8 & $	1.148	\pm	0.012	$	&	$	1.036	\pm	0.002	$	&	$	1.052	\pm	0.002	$	&	$	1.071	\pm	0.002	$	&	$	0.950	\pm	0.006	$	& 2n00	&	INT-G	\\ 
  J2000$-$7523 & FIRE & M9\,$\gamma$ &  M9\,$\gamma$ &$	1.094	\pm	0.001	$	&	$	1.195	\pm	0.001	$	&	$	1.102	\pm	0.007	$	&	$	1.076	\pm	0.002	$	&	$	0.969	\pm	0.001	$& 1n21&	INT-G\\
  J2036$+$5059 & SXD & sdM7.5 & $\cdots$ &	$	1.095	\pm	0.002	$	&	$	1.055	\pm	0.009	$	&	$	0.977	\pm	0.001	$	&	$	1.032	\pm	0.001	$	&	$	0.969	\pm	0.001	$ & 2n21	&	INT-G \\
  \enddata
  \tablecomments{Gravity Scores are listed for each index in order as follows: FeH (score based on the FeH$_z$ and FeH$_J$ scores), VO$_z$, alkali lines scores (combination of the \ion{K}{1} 1.169, 1.17, and 1.253 equivalent width scores), and $H$-cont. The appropriate combinations of scores are used to get the \cite{Alle13} gravity class designations and follow the same median scores needed as in Table \ref{tab:gravityindiceslow}. For M8 dwarfs the VO$_z$ value has no index score, thus is labeled as "n". For medium resolution data the FeH indices are combined to get a final FeH score. Scores correspond to gravities as follows: 0- field gravity (FLD-G), 1- intermediate gravity (INT-G), 2- low gravity (VL-G). Again in this paper we use $\beta$ and INT-G interchangeably as well as $\gamma$ and VL-G, however we choose to list the final gravity class in this table following the \cite{Alle13} notation.}
  \tabletypesize{\small}
  \tablerefs{Spectrum references are the same as those listed in Table \ref{tab:gravityindiceslow}}
\end{deluxetable*}

\begin{deluxetable*}{l c c c c c c c c}
\tabletypesize{\footnotesize}
\tablecaption{Equivalent widths from Medium Resolution Spectra \label{tab:eqw}}
\tablehead{\colhead{Object} & \colhead{Spectrum} & \colhead{Lit. Opt SpT} & \colhead{Lit. NIR SpT} & \colhead{\ion{Na}{1} 1.138 $\upmu$m} & \colhead{\ion{K}{1} 1.169 $\upmu$m} & \colhead{\ion{K}{1} 1.177 $\upmu$m} & \colhead{\ion{K}{1} 1.224 $\upmu$m} & \colhead{\ion{K}{1} 1.253 $\upmu$m}} 
  \startdata
  TRAPPIST-1 &	SXD	& M7.5 & $\cdots$ & $11.762 \pm 0.095$ & $4.566 \pm 0.084$ & $6.891 \pm 0.073$ & $4.124 \pm 0.12$ & $4.618 \pm 0.067$ \\
             & FIRE & M7.5 & $\cdots$ & $12.049 \pm 0.015$ & $3.913 \pm 0.019$ & $6.690 \pm 0.014$ & $4.043 \pm	0.024$ & $4.027 \pm 0.014 $ \\  \hline
  J0320$+$1854	& SXD &	M8 &	$\cdots$ & $14.426 \pm	0.074$ & $5.685 \pm	0.064 $	& $8.38 \pm 0.068$ &	$5.04 \pm	0.10$ &	$5.749 \pm	0.063$	\\
  J0443$+$0002	& FIRE & M9\,$\gamma$ & L0\,$\gamma$	&	$	9.223	\pm	0.065 $	&	$	2.941	\pm	0.059$	&	$	4.509	\pm	0.054$	&	$	3.629	\pm	0.091$	&	$	2.974	\pm	0.052$	\\
  J0518$-$2756	& FIRE & L1\,$\gamma$ & L1\,$\gamma$	&	$	6.25	\pm	0.18$	&	$	4.65	\pm	0.16$	&	$	5.31	\pm	0.14$	&	$	4.56\pm	0.22$	&	$3.99\pm 0.12$	\\
  J0532$+$8246	& NIRSPEC &	sdL7	&	$\cdots$	&	$\cdots$	&	$	12.8	\pm	1.8$	&	$	16.0	\pm	1.7$	&	$	0.6	\pm	1.2$	&	$	6.66	\pm	0.88$	\\
  J0608$-$2753	& SXD &	M8.5\,$\gamma$ & L0\,$\gamma$	&	$	7.12	\pm	0.45$	&	$	2.97	\pm	0.36$	&	$	4.78	\pm	0.33$	&	$	1.42\pm	0.52$	&	$	1.46	\pm	0.27$	\\
  J0853$-$0329 & SXD & M9 & M9 & $13.645\pm0.063$ &	$5.956\pm0.057$ &	$8.389\pm0.057$ &$4.640\pm	0.092$	&$5.682\pm0.051$\\ 
  J1048$-$3956 & SXD & M9 & $\cdots$ &$13.739\pm	0.044$&	$7.073\pm	0.043$&	$9.620\pm	0.038$& $5.205\pm	0.074$ &	$6.604\pm	0.036$\\
  J1207$-$3900 & SXD &	L0\,$\gamma$ & L1\,$\gamma$	&	$	12.0	\pm	3.4	$	&	$	3.6\pm	2.4$	&	$	2.5	\pm	1.8$	&	$	-10.0	\pm	6.8$	&	$	6.6	\pm	3.3$	\\
             & FIRE & L0\,$\gamma$ & L1\,$\gamma$	&	$	6.83	\pm	0.14$	&	$	2.62	\pm	0.13$	&	$	3.10 \pm	0.12$	&	$	3.39	\pm	0.18$	&	$	1.73\pm	0.11$	\\
  J1256$-$0224	& FIRE & sdL3.5	&	$\cdots$	&	$	12.56\pm	0.62$	&	$	6.72	\pm	0.42$	&	$	9.09	\pm	0.33$	&	$	2.45	\pm	0.61$	&	$	4.30	\pm	0.34$ \\
  LHS 377	& X-Shooter & sdM7	&	$\cdots$	&	$	3.0	\pm	1.0$	&	$	1.11 \pm	0.20$	&	$	3.53	\pm	0.33$	&	$	1.33	\pm	0.12$	&	$	1.31	\pm	0.15$	\\
  J1610$-$0040	& SXD &	sdM7	&	$\cdots$	&	$	13.08	\pm	0.18$	&	$	4.93	\pm	0.16$	&	$	7.15	\pm	0.13$	&	$	3.43\pm	0.23$	&	$	5.01	\pm	0.13$	\\
  vB 8	& SXD &	M7	& M7	&	$	12.169	\pm	0.073$	&	$	4.543	\pm	0.065$	&	$	6.947 \pm	0.056$	&	$	3.87\pm	0.10$	&	$	4.608\pm	0.052$	\\  
  vB 10	& SXD &	M8 &	M8 &	$\cdots$ & $4.71 \pm 0.27$ & $6.99 \pm 0.26$ &	$5.43 \pm 0.25$	&	$4.98 \pm 0.30$ \\ 
  J2000$-$7523 & FIRE & M9\,$\gamma$ &  M9\,$\gamma$ &$7.60\pm0.11$&	$2.428\pm0.088$&	$3.822\pm	0.071$&$2.59\pm0.12$ &	$2.053\pm	0.064$ \\
  J2036$+$5059	& SXD &	sdM7.5	& $\cdots$ & $	7.86	\pm	0.32$	&	$	2.18\pm	0.29$	&	$	4.36	\pm	0.27$	&	$	0.71	\pm	0.39$	&	$2.20 \pm	0.24$	\\
  \enddata
  \tablecomments{Equivalent width measurements of the \ion{Na}{1} $1.138\,\upmu$m line and the \ion{K}{1} $1.169\,\upmu$m, $1.177\,\upmu$, and $1.244\,\upmu$m lines. Table 10 of \cite{Alle13} shows the equivalent width measurements cut off to translate to a score of a 0 or 1 for the alkali lines.}
  \tabletypesize{\small}
\end{deluxetable*}

\end{document}